\documentclass[useAMS,usenatbib]{mn2e}
\usepackage{epsfig,rotate,graphicx}
\usepackage[fleqn]{amsmath}
\usepackage{subfigure}
\usepackage{lscape}
\usepackage{bm}

\newcommand{\p}{\partial}
\newcommand{\mnras}{MNRAS}
\newcommand{\apj}{ApJ}
\newcommand{\aap}{A\&A}
\newcommand{\apjl}{ApJL}
\newcommand{\dd}{\delta}

\newcommand{\be}{\begin{equation}}
\newcommand{\ee}{\end{equation}}

\newcommand{\apjs}{{\it ApJS, }}

\newcommand{\nat}{{\it Nature, }}

\newcommand{\aaps}{{\it A\&AS, }}

\newcommand{\sbar}{\bar{\sigma}}

\title[Edge modes]{Edge modes in
  self-gravitating disc-planet interactions}

\author[Lin and Papaloizou]{ Min-Kai Lin$^1$
  \thanks{E-mail: mkl23@cam.ac.uk} and John C. B. Papaloizou$^1$ \thanks{E-mail:
    J.C.B.Papaloizou@damtp.cam.ac.uk} \\ 
$^1$ Department of Applied Mathematics and Theoretical Physics,
University of Cambridge, Centre for Mathematical Sciences,\\
\  \ Wilberforce  Road, Cambridge, CB3 0WA, UK \\}

\begin{document}

\maketitle

\begin{abstract}
We study the stability of gaps opened by a giant planet in a
self-gravitating protoplanetary disc. We find a
linear instability associated with both the self-gravity of the disc
and  local vortensity maxima which coincide with  gap edges.
For our models, these edge modes
develop and extend to twice the orbital radius
of a Saturn mass  planet in discs with
total masses $M_d\ga 0.06M_*$ where $M_*$ is the central stellar
mass, corresponding to a Toomre $Q\la 1.5$ at twice the 
planet's orbital radius.
The disc models, although massive,  are such that they are stable
 in the absence of the planet.
 Unlike the previously studied local  vortex forming
instabilities associated with gap edges in  weakly
or non-self-gravitating discs with low viscosity,
 the edge modes we consider are global and exist 
 only in sufficiently massive discs, but for the
typical  viscosity values adopted for protoplanetary discs.

It is shown through analytic modelling and linear calculations
 that edge modes may be
interpreted as a localised disturbance associated with a  gap edge
 inducing  activity in the extended disc, through the launching of density
waves excited through gravitational potential perturbation
at Lindblad resonances.
We also perform hydrodynamic simulations in order  to investigate
the evolution of  edge modes in the linear and  nonlinear regimes in
disc-planet systems. The form and growth rates of developing
unstable modes  are found to be consistent with linear theory.
Their
dependence  on viscosity and  gravitational softening is also explored.

We also performed a first study of the effect of
edge modes on disc-planet torques and the orbital migration 
of the planet. We
found that if edge modes develop, then the average torque on the
planet becomes more positive with increasing disc mass. In simulations
where the planet was  allowed to migrate,
although a  fast type III migration  could be seen 
that was similar  to that  seen in non self-gravitating discs,
 we found that it was possible
for the planet to interact gravitationally with the spiral arms associated
with an  edge mode and that this could result in the planet being 
scattered  outwards. Thus orbital migration is likely
to be complex and non monotonic  in massive discs of the type we consider.  


\end{abstract}


\section{Introduction}
Understanding the interaction between gaseous protoplanetary discs and
 embedded planets is important for  planet formation theory. Such interaction
 may lead to inward orbital migration and account for
at least some of the observed class of exo-planets called `hot Jupiters'
\citep{mayor95} which orbit  close  to their central stars. Analytical
studies of disc-planet interaction began well before the discovery of extrasolar planets
\citep{goldreich79}. It is known that giant planets with masses  comparable  to or
exceeding  that of Saturn can open gaps in  standard model discs \citep{papaloizoulin84}
subsequent to which they may migrate inwards.

The stability of protoplanetary discs with gaps opened by  interaction with a planet has
also been the subject of study \citep{koller03,li05,valborro07}, as well
as the consequences of instability on planetary migration
\citep{ou07,li09,lin10,yu10}. These works focused on low viscosity
discs where  gap edges become unstable with the result that  vortices form.
 Such  instabilities are
known to be associated with steep  surface  density gradients or narrow rings 
\citep{papaloizou85, lovelace99,li00,li01}. These works, like most disc-planet
studies, either ignore self-gravity  completely or at best  consider weak self-gravity.     
We remark that partial disc  gaps induced by a Saturn mass planet may be
 associated with rapid type III migration \citep{masset03,peplinski08b} in a massive enough disc
which is viscous enough to be stable against vortex formation.
\citet{lin10} found that type III migration, although unsteady, could on average persist
when the disc viscosity was small enough for vortex instability to occur.

However, as type III migration occurs in massive discs  the effects of self-gravity
need to be properly considered for consistency.
The stability of a disc gap induced by interaction
with a giant planet  in   massive, self-gravitating discs has
not yet been studied. 
The stability of structured self-gravitating
discs without planets  was  explored for particle discs by \cite{sellwood91} and
for gaseous  discs by \cite{papaloizou89} and \cite{papaloizou91}. Similarly, more recently
\cite{meschiari08} also demonstrated gravitational instabilities associated
with prescribed surface density profiles that model gap structure.

 In
this paper, we extend  the above works by studying the gravitational
stability of gaps self-consistently opened by a
planet. In this respect the disc models are stable 
if the planet is not introduced and there is thus no gap. 
 We extend the analytic discussion of neutral modes
associated with  surface density gap edges given by \citet{sellwood91} 
to gaseous  discs. In addition we develop the  physical interpretation of the associated  gap
edge instabilities  in massive discs as
disturbances localised around a  vortensity maximum
 that further perturb gravitationally
the smooth parts of the disc by exciting waves at Lindblad resonances. 
The angular momentum carried away reacts back on the edge disturbance so
as to destabilise it.

We perform both linear calculations and nonlinear simulations
for  discs with a range of masses that consistently identify the growth of
low azimuthal mode number,  $m,$ edge modes as being the dominant  form of instability.
In addition we evaluate the effect of these on the disc torques acting on the planet.
We find that these torques become unsteady and oscillate in time.
First estimates of the effect on the planetary migration show 
that although fast type III-like inward migration may occur
in the unstable massive discs we study, scattering by spiral arms may also occur leading to
short periods of significant outward migration.

This paper is organised as follows. We present the governing equations
and basic model in \S\ref{basic_equations}. In \S\ref{motivation}
 we then present the results of an illustrative simulation 
of a self-gravitating disc with an embedded Saturn mass
planet that produces a dip/gap in the surface density profile,
and subsequently undergoes an instability in which the gap
edges play an important role. In the absence of the planet and  gap,
the Toomre $Q$ value is high enough for the disc model to remain stable. 

In  \S\ref{analytic1}  we present an analytic discussion 
of the linearly unstable modes. We derive the general wave action
conservation law for these modes and apply it to study their angular momentum
balance. In particular we show that, when the equation of state is  barotropic,
at marginal stability the angular momentum loss through waves propagating out
of the system is balanced by corotation torques exerted on the disc.
These are expected to be strongest near an edge. 
In  \S\ref{analytic2} we  consider modes localised
around  a vortensity maximum near an edge for which the self-gravity response  balances
the potentially  singular response at corotation, showing that such disturbances may be driven
by wave excitation at Lindblad resonances.
In   \S\ref{linear}  we go on to present linear
calculations for various disc models which confirm
the existence of edge dominated modes for low values of the azimuthal mode number.
In addition we find instabilities for larger values of $m$ for which edge effects
are less dominant but these modes have weaker growth and do not appear in 
nonlinear numerical simulations.

 In \S\ref{hydro} we present results from hydrodynamic simulations 
for a range of disc masses. These are all stable
in the absence of the planet. However, they exhibit low $m$
edge dominated modes once a planetary induced gap is present.
The form and behaviour of these is found to be in accord with linear theory.
However, in the simulations with the lower $Q$ values, the gap is found to 
widen and deepen as the simulation progresses. In addition the global angular momentum
transport through the disc, measured through an effective 
$\alpha$  parameter is increased above the level that would be induced by the planet alone.
In addition the spiral arms associated with the edge instabilities are shown to produce
fluctuating torques acting on the planet. Fast inwards type III migration may occur, but we also 
observed \emph{outwards} migration due to interaction with spiral arms.
This indicates that
migration is unlikely to be a simple monotonic process in a gravitationally active disc.
Finally  in \S\ref{conclusions} we summarise  our results and
conclude.

\section{Basic equations and model}\label{basic_equations}

We describe here the  governing equations and models for the
disc-planet system as used in analytic discussions, linear calculations
and  solved in hydrodynamic simulations.
The system we consider is a gaseous disc of mass $M_d$ orbiting a central star 
of mass $M_*.$  We adopt a cylindrical, non-rotating, co-ordinate
system $(r,\varphi,z)$ centred on the star where $z$ is the vertical
co-ordinate increasing in the direction normal to the disc mid-plane.
It is convenient to adopt 
 a system of equations  in three dimensions to begin the discussion
of angular momentum conservation
in section \ref{barcons} below.  We begin by listing  these  and then go on to explain
how we adapt them to obtain the system governing  a two dimensional razor thin disc that
we solve in hydrodynamic simulations.  They are the  continuity equation 
\begin{align}\label{continuity0}
\frac{\p\rho}{\p t}+\nabla\cdot(\rho \mathbf{u})=0,
\end{align}
and the equation of motion
\begin{align}\label{momentum0}
\rho\left(\frac{\p\bm{u}}{\p t}+\bm{u}\cdot\nabla\bm{u}\right) =
-\nabla p - \rho \nabla\Phi -\rho \nabla\Phi_{ext} + \bm{f}.
\end{align}
Here $\rho$ is the  density, $\bm{u}$ is the velocity field, $\Phi$ is the gravitational
potential due to the disc, $\Phi_{ext}$ is the potential due to external
bodies, 
$p$ is the pressure and $\bm{f}$ is the viscous force. 
The gravitational potential due to the disc satisfies Poisson's equation
\begin{align}\label{Poisson0}
\nabla^2\Phi=4\pi G\rho,
\end{align}
which yields the integral expression
\begin{align}\label{Poissonint}
\Phi= -G\int_{\cal{D}}
\frac{\rho(r^\prime,\varphi^\prime,z')r^\prime dr^\prime d\varphi^\prime dz'}{\sqrt{r^2+r^{\prime 2} -
    2rr^\prime\cos(\varphi-\varphi^\prime)+(z-z')^2 }},
\end{align}
where ${\cal{D}}$ is the domain where $\rho$ is non zero.
This coincides with the disc domain when this is not separated from external material.

For analytic discussion we consider fluids with  a general barotropic equation of state
for which $p=p(\rho)$ and $dp/d\rho =c_s^2,$ with $c_s$ being the local sound speed;
for numerical calculations a locally
isothermal equation of state is adopted, $p=c_s^2\rho$, where
$c_s(r,z)$ is a specified function of $r$ and $z.$

\subsection {Razor thin limit}
To obtain the limiting case of a razor thin disc, we assume
no interior  vertical motions and that the radial and azimuthal velocity
components are independent of $z.$ We then integrate over $z$
so that $\rho$ is replaced by the surface density $\Sigma$ in equation (\ref{continuity0}) and equation 
(\ref{momentum0})
while $p$ becomes a vertically integrated pressure, which in the barotropic case
is assumed to be a function only of $\Sigma$ with
$dp/d\Sigma = c_s^2.$  The $z$ dependence of the potentials is neglected
so that we now have the governing equations
\begin{align}\label{continuity1}
\frac{\p\Sigma}{\p t}+\nabla\cdot(\mathbf{u}\Sigma)=0,
\end{align}
and 
\begin{align}\label{momentum1}
\Sigma \left(\frac{\p\bm{u}}{\p t}+\bm{u}\cdot\nabla\bm{u}\right) =
-\nabla p - \Sigma\nabla\Phi -\Sigma\nabla\Phi_{ext}  + \bm{f}, 
\end{align}
where  $\bm{u} \equiv (u_r,u_\varphi)$ is now a two dimensional  velocity field and
$\bm{f}$ is now the two dimensional viscous force, characterised
by a uniform kinematic viscosity $\nu$ in our models \citep[see][for details]{masset02}.
The midplane disc  potential is  given by 
\begin{align}
\Phi = &-\int_{\cal{D}}
\frac{G\Sigma(r^\prime,\varphi^\prime)}{\sqrt{r^2+r^{\prime 2} -
    2rr^\prime\cos{(\varphi-\varphi^\prime)}+ \epsilon_g^2}}r^\prime dr^\prime d\varphi^\prime,
\label{gpot}\end{align}
where we have introduced a 
 softening  length $\epsilon_g=\epsilon_{g0}H(r^\prime),$ with $\epsilon_{g0}$ being
 a  dimensionless constant
and  a putative
semi-thickness for razor thin discs defined through
$ H(r) \equiv hr = c_s/\Omega_k$ where  $\Omega_k=\sqrt{GM_*/r^3}$
is the Keplerian rotation rate. The dimensionless constant disc aspect ratio is $h.$

The gravitational potential due to external bodies comprising the central star
and an embedded planet   is
\begin{align}
\Phi_{ext} = &-\frac{GM_*}{r} -\frac{GM_p}{\sqrt{r^2+r_p^2 -
    2rr_p\cos{(\varphi-\varphi_p)}+ \epsilon_p^2}}\notag\\
&+ r\int_{\cal{D}}
\frac{G\Sigma(r^\prime,\varphi^\prime)}{r^{\prime2}}\cos{(\varphi-\varphi^\prime)}r^\prime
dr^\prime d\varphi^\prime\notag\\ 
&+\frac{GM_p}{r_p^2}r\cos{(\varphi-\varphi_p)}. 
\end{align}
Here $M_p$ is the fixed mass of the embedded planet with cylindrical coordinates
$(r_p(t),\varphi_p(t))$.
 The associated gravitational  potential is softened 
with   softening  length 
$\epsilon_p=\epsilon_{p0}H(r_p),$ where $\epsilon_{p0}$ is a dimensionless constant.
The last two terms in the expression for
$\Phi_{ext}$ which give  the indirect potential  account for the forces
due to the disc  and embedded planet acting on the central star. They
 occur because we adopt a non-inertial 
frame of reference.  

\subsection{Model setup}
We describe specific disc and planet models used in numerical calculations, which also
motivates the analytical discussion below. We adopt units such that $G=M_*=1$ and the inner boundary radius
of the disc, $r_i,$ is unity.  The Keplerian orbital period at this radius
$r=r_i=1$ is then $P(1)=2\pi$. The disc occupies $r=[r_i,r_o]=[1,10].$  
The disc is taken to have  a locally
isothermal equation of state, $p=c_s^2\Sigma,$ where
$c_s = rh\Omega_k.$ 
We remark that softening is introduced to take account of the 
vertical thickness of the disc. Typically, softening length to semi-thickness
ratios of 0.3---0.6 are used \citep[e.g.][]{masset02,baruteau08}.
Thus fiducial  values for the constants $\epsilon_{p0},$ $\epsilon_{g0}$
and $h,$  of $0.6,$  $0.3$  and $0.05$ are respectively adopted.
The initial  surface density profile is taken to be   
given by 
\begin{align} 
 \Sigma(r) = \Sigma_0 r^{-3/2}\left(1-\sqrt{\frac{r_i}{r+H_i}}\right),
\end{align}\citep[see][]{armitage99}
where $H_i=H(r_i).$
 This form for $\Sigma,$ which  vanishes smoothly a distance $H_i$ inside the disc
inner boundary,  has been  introduced to ensure that both  the
gravitational force   and 
the pressure gradient in hydrostatic equilibrium at the disc inner boundary
 give minor contributions and
 vary smoothly there.
 The constant surface density
scale $\Sigma_0$ is adjusted so that  $ Q(r_o),$ where  
\begin{align}\label{myQ} 
  Q(r) =\frac{c_s\kappa}{\pi G\Sigma} \rightarrow  \frac{hM_*}{\pi r^2\Sigma(r)}
\end{align}
is the Toomre $Q$ parameter,  evaluated  in the limit of  a thin Keplerian disc
for which the epicycle frequency $\kappa =\Omega_k,$
  takes on a specified value $Q_o.$
 Thus
$Q_o$  parametrises the models.

The initial azimuthal velocity is found by assuming  hydrostatic equilibrium
in which the  centrifugal force balances forces due to
stellar gravity and  the disc's self-gravity and pressure gradient. 
Thus the  initial azimuthal velocity is given by
 \begin{align}
 u_{\varphi}^2 = \frac{r}{\Sigma}\frac{d p}{dr} + \frac{GM_*}{r} + r\frac{d\Phi}{dr},
 \end{align}
 For the  local isothermal equation of state we have adopted, the contribution
due to the pressure is given by
\begin{align}
  \frac{r}{\Sigma}\frac{d p}{dr} = c_s^2 
\left\{-\frac{5}{2} + \frac{r\sqrt{r_i
}(r+H_i)^{-3/2}}{2\left[1-\sqrt{r_i/(r+H_i)}\right]}\right\}.
\end{align}
At $r=r_i,$ for $h=0.05,$  this is $\sim 20c_s^2$  which is
approximately $5\%$ of the square of the local Keplerian speed
so that the  contribution  of the pressure force is indeed minor  when compared
to that arising from the gravity of the central star.

The initial radial velocity is set to $u_r = 3\nu/r$, corresponding  to  the 
initial radial velocity velocity of an one-dimensional Keplerian accretion disc with 
$\Sigma\propto r^{-3/2}$ and uniform kinematic viscosity. 
The disc is evolved  for a time  $\sim 280P(r_i)$ before introducing the
planet.  

When a planet of mass $M_p$ is introduced,  it is inserted in a circular
orbit, under the gravity of both the central star and disc,
at a distance  $r_p = r_p(t=0)$ from the central star.
We  quote time in units  of the Keplerian orbital period at the
planet's initial radius, which is given by $P_0=2\pi/\Omega_k(r_p(t=0)).$ 
If the planet
is held on fixed circular orbit then $r_p(t)=r_p(t=0).$  

\section{Numerical Simulations}\label{motivation}

In order to motivate and  facilitate our analytic discussion, linear analysis and
interpretation of the instabilities we find in a self-gravitating disc
with a surface density gap or dip induced by
a massive planet, we here provide a brief demonstration 
of their existence. We show global instabilities associated
with a surface density depression made
self-consistently  by a giant planet in a fixed circular orbit  
by means of numerical simulations.
Details of the  numerical approach are given
in \S\ref{hydro} where additional hydrodynamic simulations 
of this type as well as others,  where the planet is allowed 
to migrate,  will be investigated
more fully.

\subsection{The nature of edge modes at 
the edges of disc surface density  depressions induced 
 by  interaction with giant planets}\label{Firstsim}

Here, we describe results obtained  for
a disc with the initial density profile
specified above in  which there is an                                
embedded planet with mass 
$M_p=3\times10^{-4}M_*.$ 
 This corresponds to a Saturn mass
planet when $M_*=1M_{\odot}.$
The  disc model is $Q_o=1.5$, corresponding to 
total disc  mass of  $M_d= 0.063M_*.$
The  uniform kinematic viscosity was taken to be  $\nu=10^{-5}$ in
code units. The planet was  introduced
at radius $r_p=5$ at $t=25P(r_p)\equiv 25 P_0.$
Its gravitational
potential was then ramped up  over a time interval  $10P_0.$
For this simulation  it was held on a fixed circular
orbit. 
Without a perturber we expect, and find,  the  disc to be gravitationally stable because
we have $Q(r_p) = 2.62$ and near the outer boundary $Q\sim1.5.$  

\begin{figure}
\centering
\includegraphics[width=0.99\linewidth]{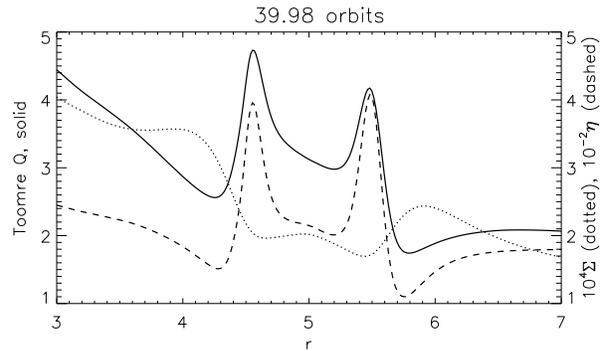}
\caption{Gap profile  produced  by a Saturn-mass planet  embedded 
  in a disc with
  $Q_o=1.5.$ The  azimuthally averaged surface
  density (dotted line), vortensity (dashed line) and  Toomre $Q$ parameter (solid line)
  are plotted. Note that the $Q$ profile shows a maximum and minimum
associated with each gap edge. 
The planet is in a fixed circular orbit
located at $r=5.$ 
\label{Qm1.5_basic}}
\end{figure}

The profile of the gap  opened by the planet is shown in
Fig. \ref{Qm1.5_basic}. The  azimuthally averaged surface
density $\langle \Sigma\rangle_{\varphi}$, vortensity $\langle
  \eta\rangle_{\varphi}$ where  $\eta \equiv \kappa^2/2\Omega\Sigma$,
and Toomre $Q$
value  are plotted. 
The latter is calculated using the 
azimuthally averaged  epicycle frequency.  We remark that
for an axisymmetric disc,
the Toomre parameter is proportional  to  the product of the ratio of the rotation 
frequency $\Omega$ to the epicycle frequency $\kappa$ and the
vortensity $\eta$. It is seen that there is a surface density depression,
referred to a gap, associated with a decrease of surface density
by $\simeq 20\%$ relative to the unperturbed disc.  

 It has been found that 
extrema in the vortensity are associated with instability
\citep{papaloizou89}.
 For 
typical  disc models structured by  disc-planet interactions, such
as those  illustrated in
 Fig. \ref{Qm1.5_basic}, local maxima/minima in $Q$ and $\eta$ 
have been found to approximately coincide. The neighbourhoods of the inner and outer gap edges 
both contain maxima and minima of $Q$ and $\eta.$

In the absence of a planet, $Q$ smoothly decreases outwards. 
In the case  when a planet is present, disc-planet interaction results in  
significant vortensity generation  as material flows through  shocks  \citep{lin10}, 
leading to vortensity maxima. The resulting $Q$ and $\langle\eta\rangle_\varphi$ profiles
are very similar (since $Q=(c_s/\pi G)(2\Omega \eta/ \Sigma)^{1/2}$). 
Fig. \ref{Qm1.5_basic} shows that the planet-modified $Q$ profile may exhibit
 a range of behaviours close to $r_p.$
Vortensity diffusion, e.g., due to significantly 
larger  viscosities than those considered here, 
could render the $Q$ profile to be uniform. In such   cases,
 unstable modes associated with vortensity 
maxima cannot be set up. The unperturbed Toomre-Q profile (prior to
the planet introduction) may also play a role. The Toomre-Q profile
perturbed by the planet is reminiscent of its unperturbed
value, without planet. If the latter happens to be approximately
uniform, the net impact of edge modes on the planet torque could be
negligible.  In our models, the background Toomre $Q$ decreases
with radius, so we expect that the outer gap edge is
more gravitationally unstable than the inner gap edge.


Returning to the present case, Fig. \ref{Qm1.5_basic} shows that
in the   neighbourhood of the outer gap edge, the  maximum  and minimum 
values of $Q$
occur at $r=5.45$  and $r=5.75.$
and  are 
$Q=4.2$ and 
$Q=1.75$ respectively.
 In the region of  the inner gap edge,  the  maximum  and minimum 
values of $Q$
occur at $r=4.55$  and $r=4.25.$
These are 
$Q=4.75$ and 
$Q=2.55$ respectively. 
The extrema are separated by $\simeq 1.4H$ and $1.1H$
in the inner and outer gap edge regions respectively. Thus  the characteristic width
of the gap edge is the local scale-height. On average $Q\sim 2$ for $r>6$.  
Since $Q>1$ everywhere, the disc is stable against local  axisymmetric
perturbations. 

    
\begin{figure}
\centering
\includegraphics[width=0.99\linewidth]{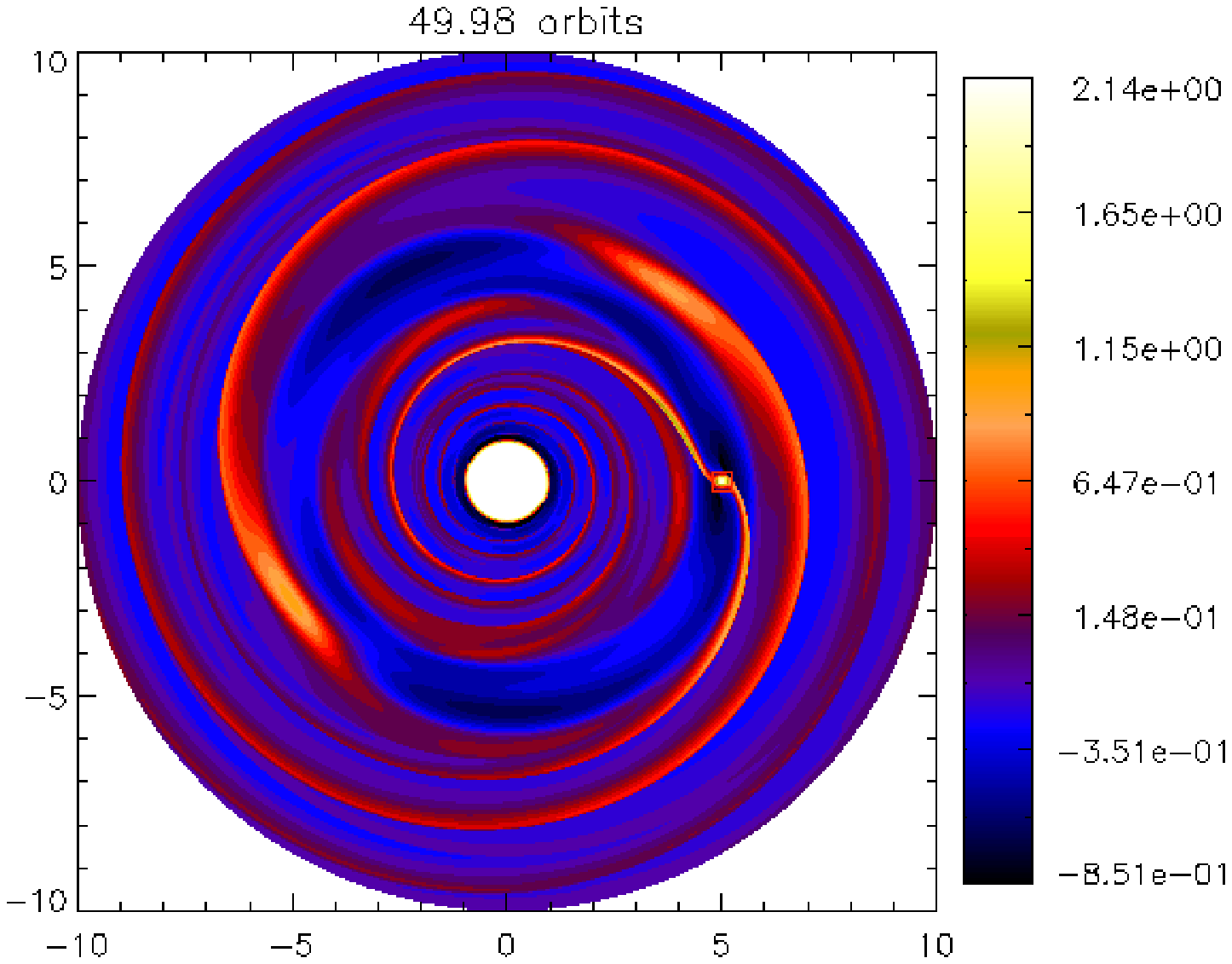}
\includegraphics[width=0.99\linewidth]{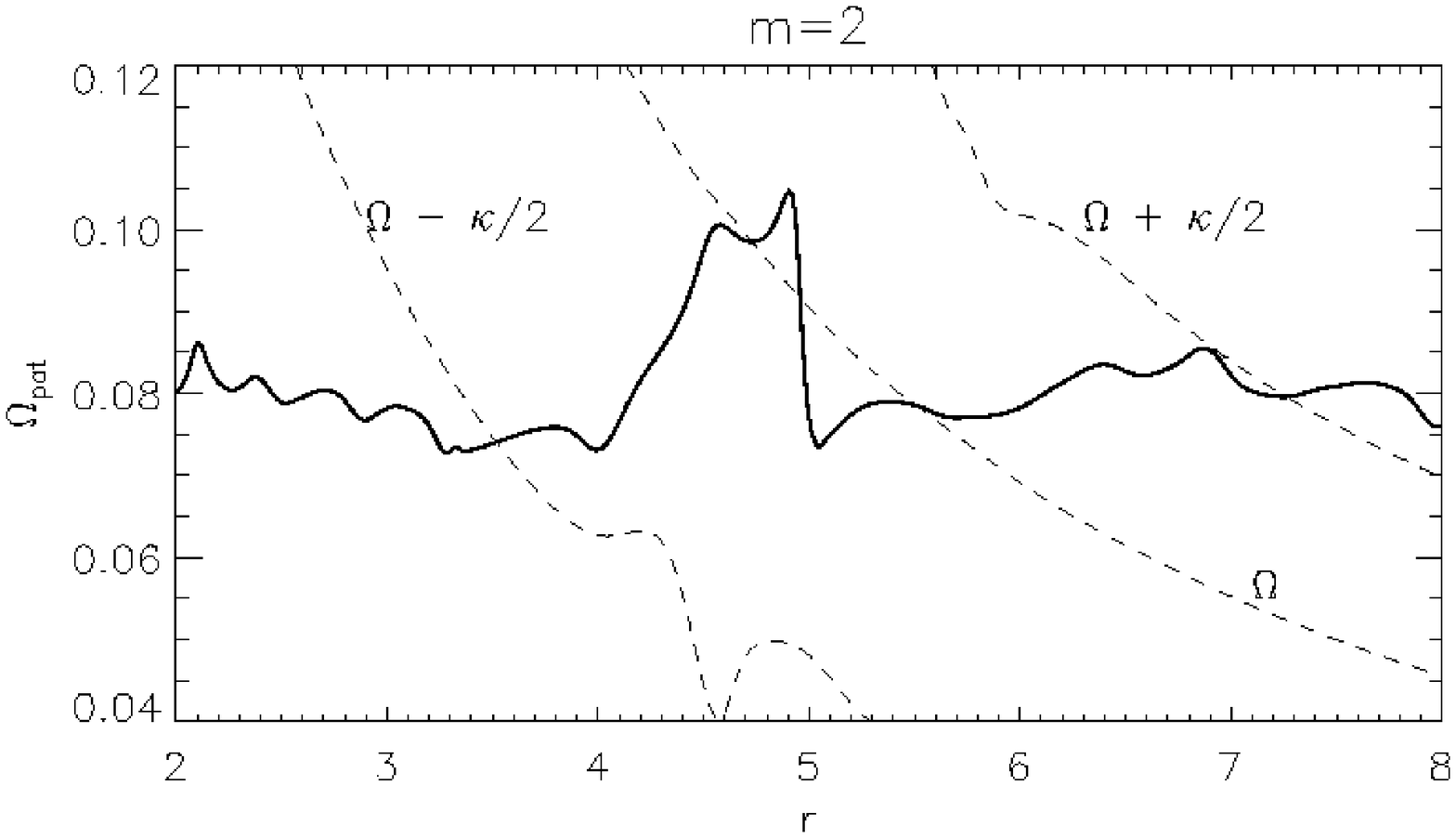}
\caption{Relative surface density perturbation for $Q_o=1.5$ at $t=50P_0$
  (top) and the radial form of the  pattern speed found from  the $m=2$ component
of the Fourier transform of the surface density
(bottom, solid). Note that this mode is associated with the outer disc
and, apart from within the gap,
this mode appears to have a  stable pattern speed of $\Omega_{pat} \sim 0.078$
corresponding to a  corotation 
point at $r=5.5.$
 Plots of  $\Omega$ and $\Omega\pm\kappa/2$ are also given. 
  (bottom, dashed).
\label{Qm1.5_t50}}
\end{figure}

 The simulation shows that both gap edges become unstable. The
 surface density contours  at $t=50P_0$  are   shown in
Fig. \ref{Qm1.5_t50}. We identify a
 mode with $m=3$ and  relative density perturbation
 $\Delta\Sigma/\Sigma\simeq 0.4$ associated with the
inner edge and a  mode with $m=2$ and 
 $\Delta\Sigma/\Sigma\simeq
0.9$ associated with  the outer edge. The modes have become nonlinear, with the development of
shocks, on dynamical time-scales. Spirals do not extend across the gap indicating
effective gap opening by the planet and disc self-gravity is not strong enough to
connect the spiral modes on either side of $r_p.$

Plots of the radial  dependence of  the pattern rotation  speed, $\Omega_\mathrm{pat},$
obtained  from the $m=2$ component of 
the Fourier transform of  the surface density,  together with
the azimuthally
averaged  values of $\Omega$ and $\Omega\pm\kappa/2$   
 are  presented  in Fig. \ref{Qm1.5_t50}.
We  focus  on the mode associated
with  the outer gap edge because  the $m=2$ spiral mode has more
than twice the relative  surface density amplitude compared to  the inner spiral
mode.

By necessity, the Fourier transform includes features due to  the planetary wakes and
inner disc spiral modes as well as the  dominant outer disc mode. 
On average, $\Omega_\mathrm{pat}\sim 0.08$ \footnote{ We have explicitly checked
that this is the pattern speed by measuring the angle through which
the spiral pattern rotates  in a given time interval.}. 
There is a co-rotation point in the outer disc  at $r=5.5$ where
$\Omega=\Omega_\mathrm{pat}\sim 0.078 $, i.e. at  the local vortensity
(surface density) maximum (minimum) of the basic state, which is just
within the gap (see Fig. \ref{Qm1.5_basic}).
This is consistent with pattern speeds obtained from linear calculations presented
later. Fig. \ref{Qm1.5_t50} indicates another co-rotation point at $r\simeq 4.7$ but
this is due to  inclusion of features associated with the planetary wakes. 


We refer to unstable spiral modes identified here as \emph{edge modes}. Their properties  can
be compared to those of \emph{groove modes} in particle discs \citep{sellwood91} or 
fluid discs \citep{meschiari08}. 
\citeauthor{sellwood91} describe groove modes as  gravitational instabilities
associated with  the  coupling of disturbances associated with two edges 
  across the gap between them.  However,
our analytic description of the edge mode instability is a coupling between a
disturbance at a single   gap edge and the disturbance it excites in the adjoining  smooth  disc  away from  co-rotation.
The other gap edge plays no role.
Although both types of mode  are associated with vortensity maxima, the groove modes described above would have
co-rotation at the gap centre midway between the edges,  whereas  in our case co-rotation is at the gap edge.

 This is significant
because in our model there will be differential rotation between the spiral pattern and the planet.
In addition, unstable modes on the inner and outer gap edges need not have the same
$m$ (Fig. \ref{Qm1.5_t50} shows $m=3$ on the inner edge and $m=2$ on the outer edge). 
If there was a groove mode with co-rotation at the gap centre, then the spiral pattern  would  co-rotate  with the planet
and the coupled edges must have disturbances dominated by the same value of $m$.






\subsection{Comparison to vortex instabilities}


We have demonstrated a global instability in self-gravitating
disc-planet systems with a gap. It is interesting to compare
these spiral modes to the well-known localised vortex-forming instabilities that occur
in  non-self-gravitating or weakly self-gravitating discs near gap edges \citep{li09,
lin10}. However, the vortex instability requires low
viscosity. For comparison purposes, here we use a kinematic viscosity of
$\nu = 10^{-6}$ (or an $\alpha$ viscosity parameter $ O(10^{-4})$).
This value of $\nu$  is an order of magnitude smaller than what has been  
typically adopted for protoplanetary disc models. We 
compare the behaviour of disc models with $Q_o=1.5$ and $Q_o=4.0,$ 
the former having strong self-gravity and the latter weak self-gravity.

\begin{figure}
  \centering
\includegraphics[scale=.275,clip=true,clip=1.6cm 0cm 0cm 0cm]{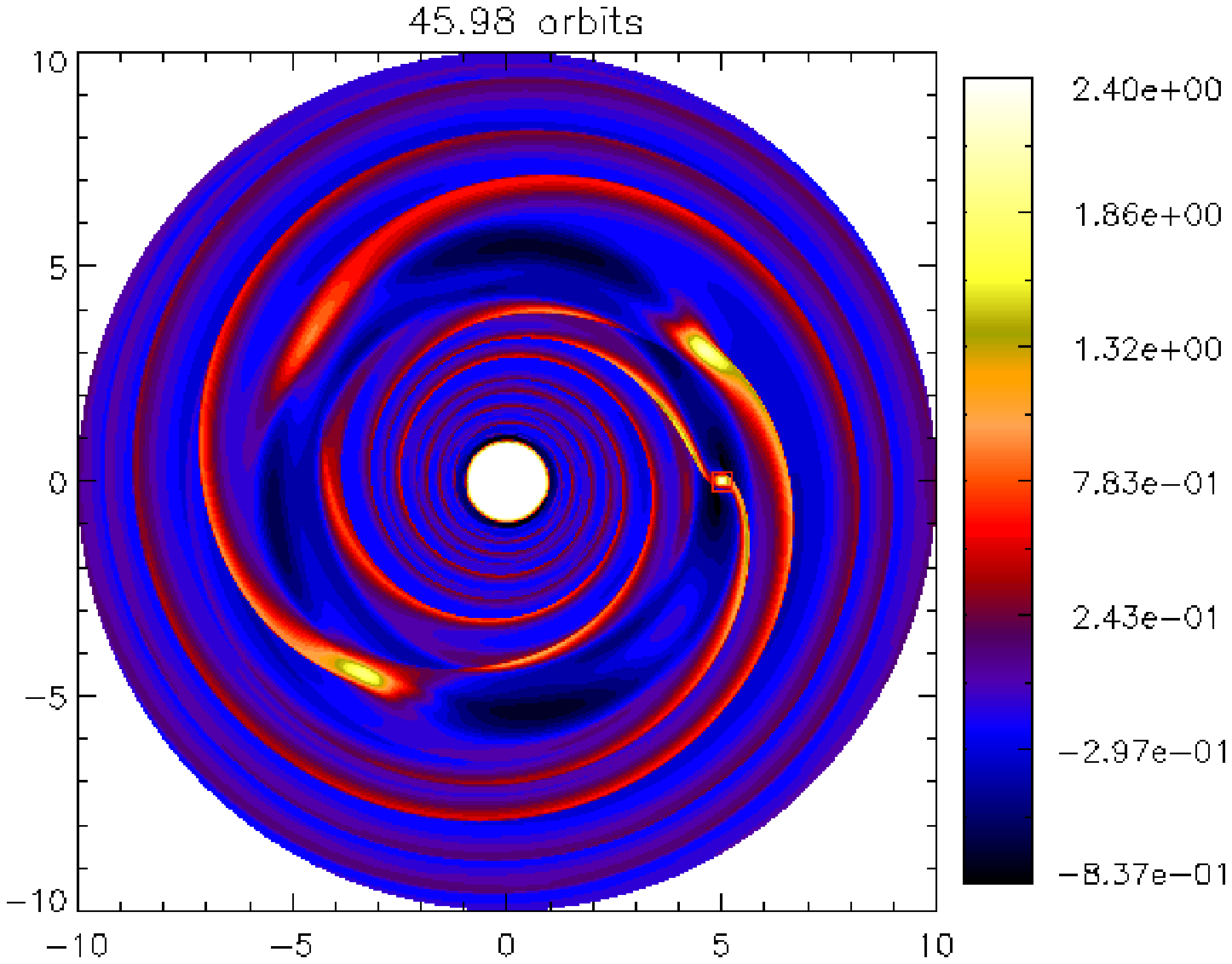}\includegraphics[scale=.275,clip=true,trim=1.6cm 0cm 0cm 0cm]{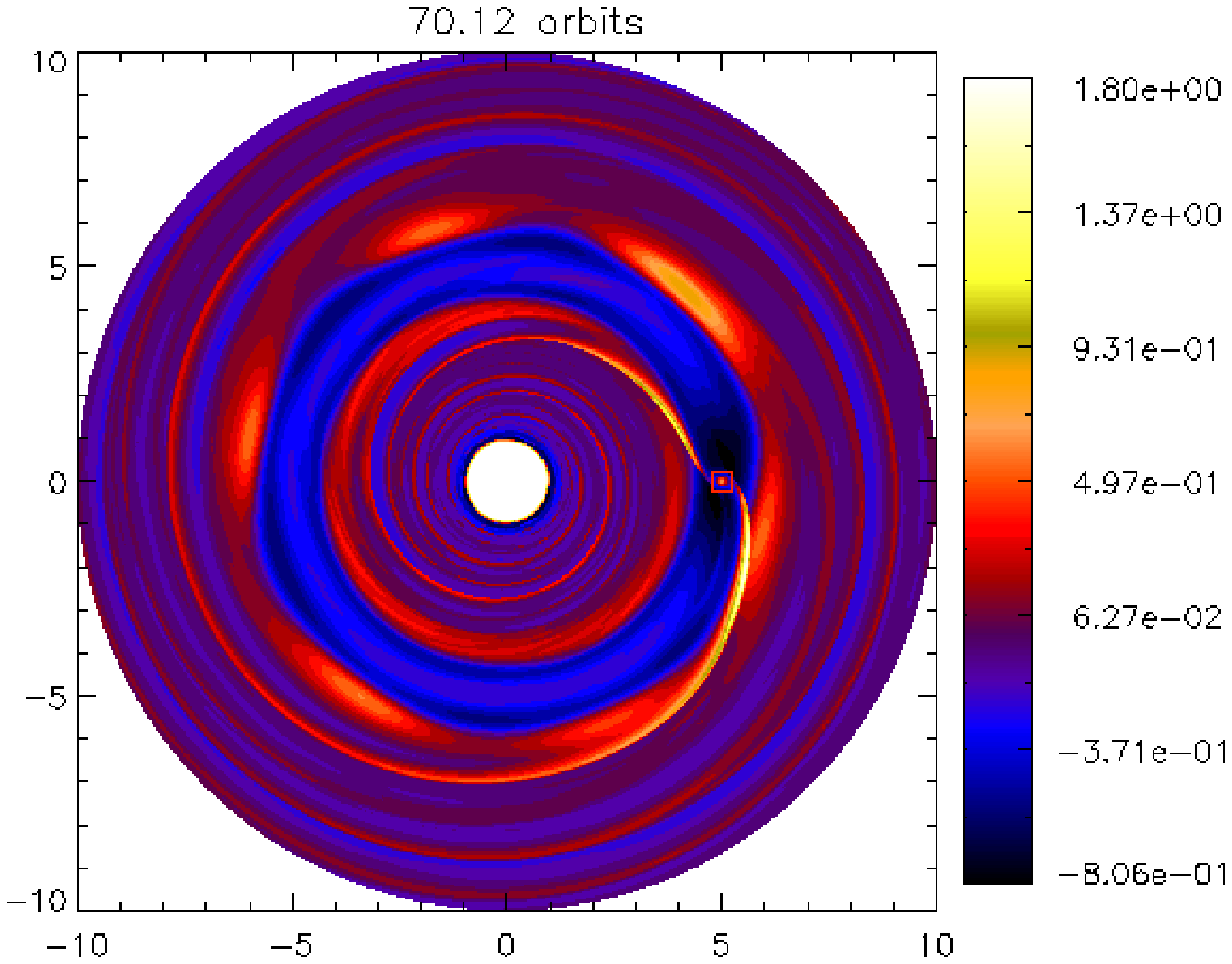} 
 \caption{Relative surface density perturbation for the $Q_o=1.5$
    disc (left) and the $Q_o=4.0$ disc (right) with
    $\nu=10^{-6}$. These plots show that the gap 
    opened by a Saturn-mass planet supports vortex instabilities in low
    mass discs and spiral instabilities  in sufficiently  massive discs. 
    \label{vortex_edge}}
\end{figure}

Fig. \ref{vortex_edge} shows two types of instabilities depending on
disc mass. The lower  mass disc with $Q_o=4$ develops six vortices localised 
in the vicinity of the outer gap edge. 
The more massive disc with $Q_o=1.5$  develops  edge modes of the type identified
in the earlier run with $\nu=10^{-5}.$
Here, the $m=3$ spiral mode is favoured because of the smaller
viscosity coefficient used.  

Vortex modes are said to be local because instability is associated with flow in
the vicinity of co-rotation \citep{lovelace99} and does not require the 
excitation of waves. Edge modes are said to be global because 
instability requires interaction between  an edge disturbance 
and waves launched at Lindblad resonances, away from co-rotation,
in the smooth part of the disc. Vortices perturb the disc even without 
self-gravity \citep{paardekooper10}.  Although the vortex mode in Fig. \ref{vortex_edge} shows
 significant wave perturbations 
in the  outer parts of the disc,  these should be seen as  a consequence 
of unstable vortices  forming around co-rotation, rather than the cause of  a linear instability.

We found the vortex modes have co-rotation close to local vortensity
{\it minima} in the undisturbed gap profile. The 
vortex modes are localised with high $m$ being dominant ($m>5$).
 On the other hand the spiral modes found here are global with low $m<5$. Edge
modes make the gap less identifiable than vortex modes do  because the
former, with corotation closer to $r_p$, 
protrudes the gap edge more.  While the edge mode only takes a few 
$P_0$ to become non-linear with  spiral
shocks attaining comparable amplitudes to the planetary
wakes, the vortex mode takes a significantly longer time 
(the plot for $Q_o=4$ is $30P_0$ after gap formation).

\section{Analytical discussion of edge modes}\label{analytic1}
The fiducial disc simulated above  is massive with
$M_d\sim 0.06M_*$ but it is still stable against gravitational
instabilities in the sense that $Q\geq 1.5$ everywhere
and the disc without an embedded planet, which has
a smoothly varying surface density profile,  does not
exhibit the spontaneous growth of spiral instabilities. When a
planet of Saturn's mass is inserted {\bf ($M_p=3\times10^{-4}M_*$)}, it
is expected to only open a partial
gap  ($\sim 30\%$ deficit in surface density), 
but even this is sufficient to trigger a $m=2$ spiral
instability. 

The fiducial case above thus suggests spiral modes
may develop under conditions that are not as extreme as those
considered by \cite{meschiari08}. They used a
prescribed gap profile with a gap depth of $90\%$ relative to the background,
which is three times deeper than in our models. Their gap corresponds to
$M_p=0.002M_*$, although a planet potential was not explicitly included. 
In both their model and ours,
the gap width is $\simeq 2r_h$ where $r_h=(M_p/3M_*)^{1/3}r_p$ is the Hill radius,
but since they effectively  used a  two Jupiter mass planet, at the same $r_p$ their 
gap is about  1.9 times wider than ours.

 We devote this section and the next
to a theoretical discussion of spiral modes associated with planetary
gap edges. This  work is based on the an analysis of
 the governing equations for linear
perturbations. 
As angular momentum balance is important for 
enabling small perturbations to grow unstably,
 we begin by formulating the conservation of angular
momentum for linear perturbations.

\subsection{The conservation of angular momentum for a  perturbed disc}\label{barcons}
We  derive  a conservation law for the  angular momentum associated with the
perturbations of a disc
that enables the angular momentum density and flux to be identified within the framework of
linear perturbation theory.
The behaviour of these quantities is found to be important for
indicating the nature of the angular momentum balance in a system
with a neutral or weakly growing normal mode and how positive (negative) angular momentum fluxes
associated with wave losses may drive the instability of a disturbance
that decreases (increases)  the local angular momentum density.
\subsection{Barotropic discs}
 We begin   by writing down the linearised equation of motion in three dimensions for the
Lagrangian displacement $\bm{\xi}\equiv (\xi_r,\xi_{\varphi},\xi_z)$
for a differentially rotating  fluid with a  barotropic equation of state  and  with
 self-gravity included 
{\citep[see eg.][]{lbo67,lin93b}} in the form
\begin{align}\label{lagrangian}
\frac{D^2\bm{\xi}}{D t^2} + 2\Omega\hat{\bm{k}}\wedge\frac{D\bm{\xi}}{D
 t}  + r\hat{\bm{r}}\left(\bm{\xi}\cdot\nabla\Omega^2\right) = -\nabla S' 
 - \nabla\Phi'_\mathrm{ext},
\end{align}
where
\begin{align}
  S' = &c_s^2 \frac{\rho'}{\rho}+ \Phi'. 
\end{align}
Here we adopt a cylindrical polar coordinate system $(r,\varphi,z),$ perturbations to quantities are denoted
with a prime while unprimed quantities refer to the background state,  the operator $D/Dt\equiv \p/\p t +\Omega\p/\p\varphi$ 
is the convective derivative following the unperturbed motion and $\hat{\bm{k}}$
is the unit vector in the $z$  direction  which is normal to the disc mid-plane.
For  perturbations that depend on $\varphi$ through  a factor
$\exp({\rm i}m\varphi),$  $m$ being the azimuthal mode number, 
assumed positive, 
that we  consider, the operator $D/Dt$ 
reduces  to  the operator 
 $\left(    \partial/\partial t  + {\rm i}m\Omega\right).$
In addition we recall that for a barotropic equation of state, 
$\Omega =\Omega(r)$
depends only on the radial coordinate.
Then (\ref{lagrangian}) becomes
\begin{align}\label{lagrangianm}
 \frac{\partial^2\bm{\xi}}{\partial t^2} + 
2\Omega\hat{\bm{k}}\wedge\frac{\partial\bm{\xi}}{\partial t}
+2{\rm i}m\Omega\frac{\partial\bm{\xi}}{\partial t}
+2{\rm i}m\Omega^2\hat{\bm{k}}\wedge{\bm{\xi}}\nonumber  \\
+ r\hat{\bm{r}}\left(\bm{\xi}\cdot\nabla\Omega^2\right)
-m^2\Omega^2\bm{\xi} = -\nabla S'
- \nabla\Phi'_\mathrm{ext}.
\end{align}
The perturbation to the density is given by
\begin{align}\label{3Dcont}
\rho'=-\nabla\cdot(\rho \bm{\xi}).
\end{align}
The perturbation to the disc's gravitational potential is given
by linearising Poisson's equation to give
\begin{align}\label{3DPoisson}
\nabla^2\Phi'=4\pi G\rho'.
\end{align}
We have also  added  an external potential perturbation  $\Phi'_\mathrm{ext}$  to assist with 
the
identification of angular momentum flux later on. 

By taking the scalar product of (\ref{lagrangianm}) with ${\bm{\xi}}^*,$
multiplying by the background 
density, $\rho,$ and taking the imaginary part, after making use of  
(\ref{3Dcont}) and (\ref{3DPoisson})  we obtain a conservation law
that expresses the conservation of angular momentum for 
the perturbations in the form.
\begin{align}\label{3DCONS}
 \frac{\p\rho_J}{\p t} + \nabla\cdot(\bm{F}_A + \bm{F}_G +
  \bm{F}_\mathrm{ext}) =  T,
\end{align}
where
\begin{align}
  \rho_J &\equiv -\frac{m\rho}{2} 
  {\rm Im}\left(\bm{\xi}^*\cdot\frac{\p\bm{\xi}}{\p t} +
  \Omega{\hat{\bf k}}\cdot\bm{\xi}\wedge\bm{\xi}^* +
  im\Omega |\bm{\xi}|^2\right)\label{rhoJ} \\
\bm{F}_A &= -\frac{m\rho}{2}{\rm Im}\left(\bm{\xi}^*S'\right)\label{FA} \\
\bm{F}_G &= -\frac{m}{2}{\rm Im}\left(\frac{1}{4\pi
      G}\Phi'\nabla \Phi'^*\right)\label{FG} \\
\bm{F}_\mathrm{ext} &=
-\frac{m\rho}{2}{\rm Im}\left(\bm{\xi}^*\Phi'_\mathrm{ext}\right)\label{Fext}
\end{align}
and
\begin{align}\label{extTorque}
T=\frac{m}{2}{\rm Im}\left(\Phi'_{ext}\rho'^{*}\right).
\end{align}
Here $\rho_J$ is the angular momentum density and the angular momentum 
flux is split into three contributions, $\bm{F}_A$ being 
proportional to the Lagrangian displacement, is the advective
angular momentum flux, $\bm{F}_G$ is the flux associated with
the perturbed gravitational stresses and $\bm{F}_{ext}$
is the flux associated with the external potential
which is inactive for free perturbations.

The quantity $T$ is a torque density associated with the external
potential as can be seen from that fact that
the real torque integrated over azimuth and divided by $2\pi$ is
\begin{align}\label{extTorque1}
-\frac{1}{2\pi}\int^{2\pi}_0{\rm Re}(\rho'^{*}){\rm Re}\left(\frac{\p\Phi'_{ext}}{\p \varphi}\right)d\varphi
 =T=\frac{m}{2}{\rm Im}\left(\Phi'_{ext}\rho'^{*}\right).
\end{align}
This justifies the scalings used for the angular momentum density
and fluxes.

\subsection{The conservation of angular momentum in the two dimensional  razor thin disc limit}
The form of the conservation law for a  razor thin disc
is obtained by integrating equation({\ref{3DCONS}) over the vertical coordinate.
For terms $\propto \rho$ the integrand is non zero only within the
disc. $\bm{\xi}$ and  the gravitational potentials
may be assumed to depend only
on $r$ and $\varphi.$ The effect of the integration is simply to replace
$\rho$ and $\rho' $ by the surface density $\Sigma$  and its perturbation 
 $\Sigma'$ respectively. The fluxes  become vertically integrated fluxes.
For the term associated with the gravitational stresses, assuming the
potential perturbation vanishes at large distances, we have
\begin{align}\label{vintFG}
\int^{\infty}_{-\infty}\nabla\cdot{\bf F}_Gdz
= -\frac{m}{2}{\rm Im}\left(\Phi'(r ,\varphi,0)\Sigma'^{*}\right)
=\frac{1}{r}\frac{\p\left( r F_{Gr}\right)}{\p r} 
\end{align} 
where
\begin{align}\label{FGrdef}
 F_{Gr}
=-\frac{m}{8\pi G}{\rm Im}\int^{\infty}_{-\infty} {\Phi'}\frac{\p \Phi'^*}{\p r}dz .
\end{align}
Thus for a two dimensional  razor thin disc the conservation law is of the
same form as (\ref{3DCONS}) but with $\rho_J$,
${\bf F}_A,$ and ${\bf F}_{ext}$
given by (\ref{rhoJ}), (\ref{FA}) and (\ref{Fext}) with $\rho$ replaced by
$\Sigma$ respectively.  As the vertical direction has been integrated
over only the radial components of the fluxes contribute.
From the above analysis the vertically integrated
 angular momentum flux due to gravitational stresses
becomes
\begin{align}\label{FG2D}
{\bf F}_G \to F_{Gr}{\hat {\bf r}},
\end{align}
with $F_{Gr}$ given by (\ref{FGrdef}).
Accordingly this term, unlike the others involves an integration
over the vertical direction.
We shall  obtain  explicit expressions for the fluxes in terms of the
perturbation $S'=\Sigma'c_s^2/\Sigma +\Phi'$ appropriate to the razor thin disc  below.

\subsection{The conservation of angular momentum  for a disc with a
 locally isothermal equation of state}
It is possible to repeat the analysis of section \ref{barcons} when the disc has a locally isothermal equation of state.
In this case $P=\rho c_s^2,$ where $c_s$ is a prescribed function of position and
the linearised equation of motion (\ref{lagrangian}) is modified to read
\begin{align}\label{lagrangianli}
\frac{D^2\bm{\xi}}{D t^2} + 2\Omega\hat{\bm{k}}\wedge\frac{D\bm{\xi}}{D
 t}  + r\hat{\bm{r}}\left(\bm{\xi}\cdot\nabla\Omega^2\right) = -\nabla S' +\frac{\rho'}{\rho}\nabla(c_s^2)
 - \nabla\Phi'_\mathrm{ext}.
\end{align}

An expression of the form (\ref{3DCONS}) may be derived
but now the torque density $T$ takes the form
\begin{align}\label{extTorque2}
T=\frac{m}{2}\left[{\rm Im}\left(\Phi'_{ext}\rho'^{*}\right) - 
  {\rm Im}\left(\rho'\bm{\xi}^*\cdot\nabla(c_s^2)\right) \right].
\end{align}
When $c_s$ is constant corresponding to a strictly isothermal equation of state, 
the fluid is again barotropic and previous expressions recovered. Equation \ref{extTorque2}
contains terms in addition to  external contributions. In general
these imply the possibility of an exchange of angular momentum between perturbations
and the background.
A very similar discussion applies to the razor thin limit.
We now go on to apply the above analysis to determining
the angular momentum balance for the  linear perturbations
of razor thin discs  in more detail
and thus determine  properties of and conditions for  unstable normal  modes.
 
\subsection{Properties of the linear modes of a razor thin disc}
 We assume linear perturbations for which the $\varphi$ and $t$
dependence  is through a factor  of the form 
 $e^{i(\sigma t + m\varphi)}$ where $\sigma$ is the
complex eigenfrequency and $m$ is the azimuthal mode-number. 
From now on this factor is taken as read.
Linearising the hydrodynamic equations (\ref{continuity1}) and (\ref{momentum1})
for the inviscid case and ignoring external potentials,
we obtain
\begin{align}
  \ {\rm i}\sbar\Sigma' & = -\frac{1}{r}\frac{d(\Sigma r u_r')}{dr}
-\frac{{\rm i}m \Sigma u_{\varphi}'}{r}\label{cl}\\
  \ {\rm i}{\bar{\sigma}}u_r' -2\Omega u_{\varphi}' &= -\frac{dS'}{dr}
 +\frac{\Sigma'}{\Sigma}\frac{d c_s^2}{dr}\label{mrl}\\
 \ {\rm i}{\bar{\sigma}}u_{\varphi}' + \frac{\kappa^2}{2\Omega}u_r' &=-\frac{{\rm i}m S'}{r},
\label{mphil}\end{align}
where $\bar{\sigma} = \sigma + m\Omega$ is the
shifted mode frequency.
 Using equations (\ref{mrl}) and (\ref{mphil}) to eliminate $u_r'$ and $u_{\varphi}'$
in equation (\ref{cl}) we obtain
\citep[see eg.][]{papaloizou91}

\begin{align}\label{barotropic2}
 r\Sigma' = & \frac{d}{dr}\left[\frac{r\Sigma}{D}
\left(\frac{dS'}{dr} +
    \frac{2m\Omega\bar{\sigma}S'}{r\kappa^2}\right)\right]\nonumber \\
&-\frac{2m\Omega\bar{\sigma}\Sigma}{\kappa^2D}\left(\frac{dS'}{dr} +
  \frac{2m\Omega\bar{\sigma}S'}{r\kappa^2}\right)
+ \frac{mS'}{\bar{\sigma}}\frac{d}{dr}\left(\frac{1}{\eta}\right)\nonumber \\ 
&-\frac{4m^2\Omega^2\Sigma S'}{r\kappa^4}
 -\frac{d}{dr}\left[\frac{r\Sigma'}{D}\frac{dc_s^2}{dr}\right]
+\frac{2m\Omega\Sigma'}{Dr\bar{\sigma}}\frac{dc_s^2}{dr},
\end{align}
where $D\equiv \kappa^2 - \bar{\sigma}^2,$ and recall
 $\eta = \kappa^2/2\Omega\Sigma$ is the vortensity. This form
shows that a corotation singularity where  $\bar{\sigma}=0$
can be avoided if corotation corresponds to a vortensity extremum. 

The above expressions may be applied to a disc, either  with a locally
isothermal equation of state, or with a barotropic 
 equation
of state, where the integrated pressure
$P=P(\Sigma)$ and the soundspeed $c_s^2\equiv dP/d\Sigma$. 
To obtain the latter case , the quantity  $dc_s^2/dr$
is simply replaced by zero.
We remark that inclusion of additional terms dependent on this 
has been found numerically to produce only a slight modification of the discussion
that applies when they are neglected.
This is because $c^2_s$
 varies slowly compared to either the linear perturbations
or the surface density in the vicinity of the edge, thus we shall neglect
$(r/c_s^2)dc_s^2/dr$ from now on. We also recall that $ S' = \Sigma'c_s^2/\Sigma+\Phi' $ 
and  the gravitational potential is given by the vertically integrated Poisson integral as
\begin{align}
\Phi' & = S'-\frac{\Sigma'c_s^2}{\Sigma} = - G\int_{\cal D} K_m(r,r')\Sigma'(r')r'dr' \label{2DPoisson},
\end{align}
where
\begin{align}
  K_m(r,r') &= \int^{2\pi}_0  \frac{\cos(m\varphi)d\varphi}{\sqrt{r^2+r'^2 -
      2rr'\cos{(\varphi) + \epsilon_g^2}}}.\label{VVVVV}
\end{align}
Here the domain of integration ${\cal D}$ is that of the disc provided
there is no external material. However, in a situation where density  waves propagate beyond the disc
boundaries ${\cal D}$ either should be formally extended or the
form of $K_m$ changed to properly reflect the boundary conditions.
We remark that (\ref{2DPoisson}) enables $\Sigma'$ to be determined 
in terms of $S'$ and if this is used in (\ref{barotropic2}) a single eigenvalue
equation for $S'$ results \citep[eg.][]{papaloizou91}.


We shall now consider the angular momentum flux balance for normal
modes. For simplicity we shall consider the barotropic case.
The analytic discussion concerns a disc model with one boundary
being a sharp edge with  arbitrary  propagation of density waves 
directed away from this boundary being allowed.
When applied to a gap opened by a planet in a global disc, 
the analytic model corresponds the section of the disc from the inner disc boundary to the
inner gap edge or the section from the outer gap edge to the outer disc boundary. 
These two sections are assumed to be decoupled 
in the analytic discussion of stability but not in the numerical one.
In addition, although it produces the
background profile,
the planet is assumed to have no effect on the linear stability analysis
discussed here, but it is fully incorporated in nonlinear simulations.
The correspondence between the various approaches adopted indicates
the above approximations are reasonable.
The discussion  of the outer and inner disc sections is very similar,
accordingly  these  are considered below  together.

\subsection{Angular momentum flux balance for normal modes}\label{Fluxbal}
The vertically integrated 
angular momentum flux associated with perturbations can be related
to the background vortensity profile. For this discussion we use
the governing equation in the form given by equation(\ref{barotropic2})
\citep{papaloizou91}. We assume a barotropic disc model and neglect terms
involving $dc_s/dr$. Multiplying this equation by $S'^*$ and
integrating, we get
\begin{align}\label{angmom_balance}
&{\cal {F}}(S',\sigma) = -\int_{r_i}^{r_o} r\Sigma' S'^* dr 
\pm\left. \left[ \frac{r\Sigma}{D}\left(\frac{d S'}{d r}  +
 \frac{2m\Omega\bar{\sigma}}{r\kappa^2}S'\right)
 S'^*\right]\right|_{r_b}\nonumber \\
&- \int_{r_i}^{r_o} 
 \frac{r\Sigma}{D}\left(\frac{d S'}{d r}  +
 \frac{2m\Omega\bar{\sigma}}{r\kappa^2}S'\right)
 \left(\frac{d S'^*}{d r}  +
 \frac{2m\Omega\bar{\sigma}}{r\kappa^2}S'^*\right)dr
 \nonumber\\ 
& -\int_{r_i}^{r_o}
 \frac{4m^2\Omega^2\Sigma|S'|^2}{r\kappa^4}dr + \int_{r_i}^{r_o}
 \frac{m|S'|^2}{\bar{\sigma}}\frac{d}{dr}\left(\frac{1}{\eta}\right) dr 
=0,\end{align}
where, here and below,  the positive sign alternative applies to the case where the disc domain
has an inner sharp edge where $r=r_i$ and then $r_b=r_o,$ the outer boundary radius.
The negative sign alternative applies to the case where the disc domain
has an outer sharp edge where $r=r_o$ and then  $r_b=r_i,$ the inner boundary radius.
In both cases waves may propagate through $r=r_b.$
Note also  that there is no contribution from edge boundary terms
as the surface density is assumed to be negligible  there.

We may now express the terms in the above integral relation,
which for convenience we have taken to define a functional of $S'$
that depends on  $\sigma,$
in terms of quantities related to the transport of angular momentum
by taking its imaginary part.
We shall begin by assuming that $\sigma$ is real or, more precisely,
that we are in the limit that marginal stability is approached.

 By making use of equation (\ref{vintFG}) we find that
\begin{align}
&{\rm Im}\left[\int_{r_i}^{r_o} r\Sigma' S'^* dr\right]=
{\rm Im}\left[\int_{r_i}^{r_o} r\Sigma'\left( \frac{\Sigma'^*c_s^2}
{\Sigma} +\Phi'^*\right)dr\right]\nonumber\\
&={\rm Im}\left[\int_{r_i}^{r_o} r\Sigma'\Phi'^*dr\right]=\pm\frac{2}{m}
\left. [rF_{Gr}]\right|_{r_b}.
\end{align}
We recall that $F_{Gr}$ is the vertically integrated 
 angular momentum flux transported by gravitational stresses. 
Note that $F_{Gr}$ at the sharp edge is ignored. Equation \ref{vintFG},
together with the requirement that $F_{Gr}$ is regular for $r\to0$ and vanishes
more rapidly than $1/r$ for $r\to\infty$, implies that $F_{Gr}$ is zero at the sharp
edge separating the disc domain and the exterior (assumed) vacuum or very low density region. 

In addition we remark that from equations (\ref{mrl} ) and (\ref{mphil}), 
the radial Lagrangian displacement is given by
\begin{align}
\xi_r &= \frac{u'_r}{{\rm i} \sbar}= -\frac{1}{D}\left(
\frac{d S'}{d r} +
    \frac{2mS'\Omega}{r\bar{\sigma}}\right).
\end{align}
From this it follows that the  imaginary part of the   second term on the right hand side of
equation (\ref{angmom_balance}) is
\noindent  $-~\pm~{\rm Im}~[r\Sigma\xi_rS'^*]|_{r_b} = -\pm(2/m)[rF_{Ar}]|_{r_b}.$
Using the above, taking the imaginary part of (\ref{angmom_balance}) yields the remarkably simple 
expression
\begin{align}\label{angmom_balance2}
\pm\left[r(F_{Gr}+F_{Ar})\right]|_{r_b}  
= \frac{m}{2}{\rm Im}\left(\int_{r_i}^{r_o} \frac{m|S'|^2}{\bar{\sigma}}
\frac{d}{dr}\left(\frac{1}{\eta}\right) 
 dr\right).
\end{align}

Note that we have been assuming that $\sigma$ is real
 and  that the mode is at marginal stability.
Then the right hand side of (\ref{angmom_balance2})  is apparently real.
However, the integrand is potentially singular at corotation  where
$\sbar=0.$ Thus the approach to marginal stability
has to be taken with care.

Setting $\sigma=\sigma_R-{\rm i}\gamma,$ where
$\sigma_R,$ being  the real part of $\sigma,$
defines the corotation radius, $r_c,$  through $\Omega(r_c)=-\sigma_R/m$
and the growth rate as  $\gamma,$ with $-\gamma$ being the imaginary part of $\sigma$. 
Marginal stability can be approached by  
assuming that  $\gamma$ has a vanishingly small magnitude
but  is positive.
Then we  can use the Landau prescription and substitute $1/\sbar$ by $ 
{\cal{P}}(1/\sbar) + i\pi\dd(\sbar),$
where  ${\cal{P}}$ indicates that the principal value
of the integral is to be taken and $\dd$ denotes Dirac's delta function.
Adopting this (\ref{angmom_balance2}) becomes
\begin{align}\label{angmom_balance3}
\left[r(F_{Gr}+F_{Ar})\right]|_{r_b}
= \pm\frac{\pi |m|}{2}\left.\left[\frac{|S'|^2}{|d\Omega/dr|}
\frac{d}{dr}\left(\frac{1}{\eta}\right)\right]\right|_{r_c}.
\end{align}
The above relation can be viewed as stating that at marginal stability,
either corotation is at a vortensity extremum and both terms in equation
(\ref{angmom_balance3}) vanish, or
angular momentum losses as a result of waves
passing through $r=r_b,$ are balanced by torques
exerted at corotation. The latter torque is proportional to the gradient
of the vortensity $,\eta $ \citep{goldreich79} 
and recall that $\eta^{-1}$ scales with $\Sigma$.


The propagating waves  quite generally
carry negative angular momentum   when located in an inner disc
and positive angular momentum  when located in  an outer disc \citep{goldreich79}.
Thus  a balance   may be possible between  angular momentum losses resulting from
the propagation of these waves out of the system  and
angular momentum gained or lost by  an  edge disturbance
as expressed by equation (\ref{angmom_balance3}).
For an edge disturbance
in a region of increasing (decreasing) surface density, such as an inner (outer) 
edge to an outer (inner) disc, the edge disturbance loses (gains) angular momentum. 

In practice, an inner (outer) edge disturbance
may be associated with a positive (negative) surface density
slope  near a vortensity maximum  or near a vortensity minimum
occurring as a result of the variation of both  $\kappa^2$ and $\Sigma.$
The former case is associated with discs for which self-gravity is important
and spiral density waves are readily excited. The latter is associated with weakly or non
self-gravitating  discs and is associated with vortex formation at gap edges
as has already been discussed by several authors 
(see  eg. \cite{lin10} and references therein).

So far we have not discussed the  boundary condition at $r=r_b$ (the non-sharp boundary). 
One possibility is that this corresponds to stipulating outgoing waves or a radiation condition,
so that  the fluxes in (\ref{angmom_balance3}) are non zero.
Another possibility is to have a surface density taper to zero 
which would mean wave reflection and removal of the  boundary flux terms.
In fact we expect that the issue of stability is not sensitive to this.  We remark that
in the case of non self-gravitating discs and the low $m$ modes of
interest,  the  regions away from the edge  are evanescent and amplitudes
are small there, the disturbance being localised in the vicinity of the  edge.
For self-gravitating discs, as we indicate below,
 instability can appear because of an unstable interaction between
{\it outwardly propagating positive} ({\it inwardly propagating negative}) angular momentum density waves
 and a {\it negative} ({\it positive})  angular momentum edge disturbance localised 
near an {\it inner} ({\it outer})
disc gap edge. As long as these waves are excited it should
not matter whether they are reflected or transmitted at large distance,
as long as their angular momentum density remains in the wider disc
and is not fed back to the exciting  edge disturbance.

For these reasons and also simplicity we shall assume that at the boundary
where $r=r_b,$  either
the  surface density  has tapered
 to zero, or there is reflection such  that 
the  boundary terms in expressions like (\ref{angmom_balance}) may  be dropped.
Equation(\ref{angmom_balance3}) then simply states that marginal stability
occurs when corotation is at a vortensity extremum.
We find that the vortensity {\it maximum}  case
is associated with an instability in self-gravitating discs that is
associated with strong spiral waves. On the other hand
the   vortensity {\it minimum}  case is associated with 
the vortex-forming instability in  non-self gravitating discs
that has been previously studied (see \cite{lin10} and references therein)
and will not be discussed further in this paper ( but see \cite{lin11} for
a discussion of the effect of weak self-gravity on vortex modes).

\section{Description of the edge disturbance associated with
a vortensity maximum in a self-gravitating disc}\label{analytic2}
We now focus on the description of the 
instability in a  self-gravitating disc as being
due to a disturbance associated with the 
edge causing the excitation of spiral waves  that propagate away from it 
 resulting  in destabilisation. This occurs because the emitted waves
carry away  angular momentum which has the opposite
sign to that associated with the edge disturbance 
 ( see section \ref{Fluxbal} and equation (\ref{angmom_balance3})).
Accordingly the excitation process is expected to lead to the growth
of this disturbance.

We thus  consider the disturbance to be localised in the vicinity of the
edge and the potential perturbation it produces to excite spiral density
waves in the  bulk of the disc.
Our numerical calculations show that edge dominated modes of this type
occur for low $m$ and are dominant in the nonlinear regime.
We  assume the mode is weakly growing corresponding
to the back reaction of the waves on the edge disturbance being weak.
Thus  in the first instance we calculate a neutral edge
disturbance and then calculate the wave emission as a perturbation.
  We emphasise again that an 
important feature is that corotation is at a vortensity {\it maximum}
(in contrast to the non self-gravitating case for which corotation
is located at a vortensity {\it minimum}).
We show that such modes require self-gravity.
In an average sense they  require a sufficiently small $Q$ value and so  
cannot occur in the non self-gravitating limit ($Q\to\infty$).

\subsection{Neutral edge disturbances with corotation at a vortensity maximum}\label{Neutedge}
We start from
equation (\ref{barotropic2}) for $\Sigma'.$
 We assume that only the third term on the right hand side need be retained.
This is because this term should dominate near corotation in the presence
of large vortensity gradients that are presumed to occur near the edge
and the disturbance is assumed localised there 
(see also the discussion of groove modes in collisionless
particle discs of \cite{sellwood91}, where similar assumptions are made).
Thus we have
\begin{align}\label{sigma_bigS}
  \Sigma' = \frac{S'}{r{\bar{\omega}}}\frac{d}{dr}\left(\frac{1}{\eta}\right),
\end{align}
where $\bar{\omega} = \sbar/m$. The associated gravitational potential is given by
\begin{align}\label{PPOT}
\Phi' =-
G\int_{r_i}^{r_o} 
K_m(r,r')\frac{S'(r')}{{\bar{\omega}}(r')}\frac{d}{dr'}\left[\frac{1}{\eta(r')}\right]
dr'.
\end{align}
From the relation $S =\Sigma'c_s^2/\Sigma+\Phi'$
we thus obtain the integral equation
\begin{align}
&S'(r)\left[ 1-
  \frac{c_s^2}{r\Sigma{\bar{\omega}}}\frac{d}{dr
}\left(\frac{1}{\eta}\right)\right] =\nonumber\\
&- \int_{r_i}^{r_o}GK_m(r,r')\frac{S'(r')}{{\bar{\omega}}(r')}\frac{d}{dr'}\left[\frac{1}{\eta(r')}\right]
dr'.\label{Inteq}
\end{align}
For a disturbance dominated by self-gravity, pressure should be negligible. 
Under such  an approximation we have $S'\sim\Phi'.$ 
Equation (\ref{sigma_bigS}) then implies that to obtain a negative (positive) potential perturbation
for a  positive (negative) surface density perturbation, we need
$\mathrm{sgn}{(\bar{\omega})}=\mathrm{sgn}(d\eta/dr)$. At co-rotation where $\bar{\omega}=0$
and there is a vortensity extremum, 
this requirement becomes $\mathrm{sgn}{(d\bar{\omega}/dr)}=\mathrm{sgn}(d\Omega/dr)
=\mathrm{sgn}(d^2\eta/dr^2)$. Since typical rotation profiles have $d\Omega/dr\equiv\Omega' 
< 0$, the physical 
requirement that potential and surface density perturbations have opposite signs implies that
$d^2\eta/dr^2<0$ at co-rotation, i.e. the vortensity is a maximum. Thus our discussion applies to
vortensity maxima only. 
For the  class of vortensity  profiles for which 
$\mathrm{sgn}(d\eta/dr) = \mathrm{sgn}(\bar{\omega}),$
we demonstrate below that 
the integral equation (\ref{Inteq}) may be transformed into a Fredholm  integral equation
with symmetric kernel.

Noting that corotation is located at a vortensity maximum,
the requirement above imply the vortensity  
increases (decreases) interior (exterior) to corotation
( we comment that the discussion may also be extended to the case when that is true 
with $d^2\eta/dr^2$ vanishing at corotation).
As we expect the disturbance to be localised around corotation,
it is reasonable to assume this holds and that we may contract
the integration domain $(r_i,r_o)$ to exclude regions which do not conform
without significantly affecting the problem.

Proceeding in this way we introduce the function $\mathcal{H}$ such that
\begin{align}  S'(r) =
  \mathcal{Z}(r) 
\mathcal{H}(r),
\end{align}
where
\begin{align}
\mathcal{Z}(r)=
\left[-\frac{d}{dr}\left(\frac{1}{\eta}\right)\frac{1}{\bar{\omega}}\right]
^{-1/2} 
\left[  1-\frac{c_s^2}{r\Sigma\bar{\omega}}
\frac{d}{dr}\left(\frac{1}{\eta}\right)\right]^{-1/2}.
\end{align}
Defining the new symmetric kernel $\mathcal{R}(r,r')$ as
\begin{align}
\mathcal{R}(r,r') \equiv  &  GK_m(r,r')\mathcal{Y}(r)\mathcal{Y}(r')
\end{align}
where
\begin{align}
\mathcal{Y}(r)=
\left[-\frac{d}{dr}\left(\frac{1}{\eta}\right)\frac{1}{\bar{\omega}}\right]
^{1/2}
\left[  1-\frac{c_s^2}{r\Sigma\bar{\omega}}
\frac{d}{dr}\left(\frac{1}{\eta}\right)\right]^{-1/2}
\end{align}
we obtain the integral equation
\begin{align}
  \mathcal{H} = \int_{r_i}^{r_o}\mathcal{R}(r,r')\mathcal{H}(r')dr'.
\end{align}
In other words, $\mathcal{H}$ is required to be
the solution to
\begin{align}\label{fredholm}
  \lambda \mathcal{H} = \int_{r_i}^{r_o}\mathcal{R}(r,r')\mathcal{H}(r')dr',
\end{align}
with $\lambda=1$. However, Eq. \ref{fredholm} is just a standard
homogeneous Fredholm integral  equation of the second kind. Thus the
existence of a nontrivial solution ($\mathcal{H}\neq 0$) implies  $\mathcal{H}$ is an
eigenfunction of the kernel $\mathcal{R}$ with unit eigenvalue.

The kernel
 $\mathcal{R}$ is  positive and  symmetric.  Accordingly the Fredholm
equation above has the property that the maximum eigenvalue  $\lambda > 0$
 and is the maximum of the quantity $\Lambda$ given by 
\begin{align}\label{CauSCh}
  \Lambda = \frac{\int_{r_i}^{r_o}\int_{r_i}^{r_o}\mathcal{R}(r,r')
    \mathcal{H}(r')\mathcal{H}^*(r)drdr'}{\int_{r_i}^{r_o}|\mathcal{H}(r)|^2 dr}
\end{align}
over  all $\mathcal{H}$. Thus, if it is possible
to show that we must always have  $\mathrm{max}(\Lambda) < 1,$ then it
is not possible to have a non-trivial solution corresponding
to a neutral  mode. Upon application of the
Cauchy-Schwarz inequality (twice), one can deduce
\begin{align}
\Lambda^2 \leq
\int_{r_i}^{r_o}\int_{r_i}^{r_o}|\mathcal{R}(r,r')|^2drdr'
 = \Lambda_{est}^2 \ge \mathrm{max}(\Lambda^2) .
\end{align}
Thus the existence  of a neutral mode
of this type is not possible if $\Lambda_{est}^2$  is less than unity.

To relate the necessary condition for mode existence to physical
quantities, we can estimate $\Lambda_{est}^2$.
The Poisson kernel $K_m(r,r')$ is largest at $r = r'$. Other factors in
the numerator of  $\mathcal{R}$ involve the factor $1/\bar{\omega}$. We assume
these factors have their largest contribution at  corotation where $r=r_c$ and
$\bar{\omega} = 0.$ 
 Hence, $ \Lambda_{est}^2 \sim
\int_{r_i}^{r_o}\int_{r_i}^{r_o}|\mathcal{R}(r_c,r_c)|^2drdr'
$. Assuming the edge region has width of order $L_c,$ taken to
be much less than the local radius but not less than  the local scale-height,
we estimate
\begin{align}\label{lambda_est}
\Lambda_{est} \sim &|\mathcal{R}(r_c, r_c)|L_c\notag\\
& =
GK_m(r_c,r_c)\left|\frac{1}{\Omega^\prime}\left(\frac{1}{\eta}\right)^{\prime\prime}\right|L_c
\left/ \left| 1 - \frac{c_s^2}{r_c \Sigma
      \Omega^\prime}\left(\frac{1}{\eta}\right)^{\prime\prime}\right|\right..
\end{align}
All quantities are evaluated at co-rotation  and the  double prime denotes $d^2/dr^2$.
Because the edge is thin
by assumption, we can further approximate the Poisson kernel by
\begin{align}
K_m(r,r') \sim \frac{2}{\sqrt{rr'}}K_{0}\left(\frac{m}{\sqrt{rr'}}\sqrt{|r
    - r'|^2 + \epsilon_g^2} \right),
\end{align}
where $K_{0}$ is the  modified Bessel function of the second kind of order
 zero  and recall that
$\epsilon_g$ is the softening length. If pressure 
effects are negligible ($c_s^2$ being small), we obtain
\begin{align}\label{maxL}
  \Lambda_\mathrm{est} \sim
  \frac{2GK_{0}(m\epsilon_g/r_c)}{r_c}
  \left|\frac{1}{\Omega^\prime}
\left(\frac{1}{\eta}\right)^{\prime\prime}\right|_{r_c}L_c.
\end{align}
Since $\eta = \kappa^2/2\Omega\Sigma$, the requirement that $\Lambda^2_\mathrm{est}$
exceeds unity, leads  us to  expect that modes of the type we have been 
discussing
 can exist  only for sufficiently large surface density scales. We have approximately
\begin{align}\label{maxL1}  \Lambda_\mathrm{est} \sim
\frac{4G\Sigma K_{0}(m\epsilon_g/r_c)}{\Omega^2 L_c}.
\end{align}
Thus taking $L_c\sim H$ and $\epsilon_g$ being a fraction $\epsilon_{g0}$ of the local
scale height, we obtain $ \Lambda_\mathrm{est} \sim 4\ln(1/(m\epsilon_{g0}h))/(\pi Q).$
On account of the logarithmic factor,
the condition $\Lambda_\mathrm{est}>1$  suggests
that edge disturbances which lead to instabilities can be present
for $Q$ significantly larger than  unity, as is confirmed numerically.
However, a surface density threshold must be exceeded.
Thus such modes associated with vortensity
{\it maxima} do not occur in a non self-gravitating disc.
Finally we remark that although we applied a model assumption to 
obtain an integral equation with symmetric kernel in the above analysis,
which enables the existence of solutions to be shown,
the demonstration of a surface density threshold  does not depend on this. 
The application of the Cauchy-Schwartz inequality to (\ref{CauSCh})
may be carried out in a similar manner for the integral equation (\ref{Inteq})
leading to similar conclusions.
The fundamental quantity is the vortensity profile,
(\ref{maxL}) indicates that if it is not
sufficiently peaked,  there can be no mode.
Similarly, mode existence becomes less favoured  for increasing
softening and/or increasing $m$. 

At this point we remark that an analysis of the type discussed
above does not work for vortensity minima. An equation
 similar to (\ref{fredholm})
could be derived but in this case the  corresponding kernel $\mathcal{R}$ 
would be negative, implying inconsistent negative eigenvalues $\lambda.$ 
This situation results from the surface density perturbation
leading to a gravitational potential perturbation with the
wrong sign to satisfy the condition (\ref{sigma_bigS}).
Thus we do not expect the type of edge disturbance considered
in this section to be associated 
with vortensity minima.
The fact that edge modes associated with vortensity maxima
require a threshold value of $Q^{-1}$ to be exceeded separates
them from vortex forming modes which are {\bf  modes}  associated with vortensity
minima and require $Q^{-1}$ not to be above a threshold in order
to be effective. For most 
disc models considered in this paper, with minimum $Q\le 2$, vortex modes do not occur.
Discs with minimum $Q\ga 2$ are considered in \cite{lin11}.

\subsubsection{Implications for disc-planet systems}
For the locally isothermal disc models adopted in this paper, with a small aspect-ratio
$h=0.05$, edge modes could appear more easily as the disc is cold. 
This is because if  the disc thickness sets the length  scale 
of the edge,  equations  (\ref{lambda_est}), (\ref{maxL}) and (\ref{maxL1})
indicate that, for fixed $Q$, lowering sound-speed and hence the disc aspect ratio
increases 
$\Lambda_{est}$. 

We can apply these equations to gaps opened by a Saturn mass
planet, which is considered in linear and nonlinear numerical calculations later on.   
Without instabilities,
these gaps  deepen with time.  There is also vortensity mixing in the co-orbital region 
as fluid elements pass through shocks and  repeatedly execute
horseshoe turns close to the planet.  In a fixed orbit this reduces  the gap surface density 
and the magnitude of the edge  vortensity peaks. In this way  the conditions for edge modes 
become less favourable with time. In this case  edge modes 
are expected to  develop early on during gap formation, if at all. 
However, this effect may be less pronounced if the orbit evolves  because the planet migrates.

For larger planetary masses such as Jupiter, a deeper gap will be opened
but stronger shocks are also induced, which may lead to stronger vortensity peaks. 
Thus, in view of potentially competing effects, the conditions for edge modes 
as a function of planet migration and planetary mass must be investigated 
numerically and  this will be undertaken in future studies.   

\subsection{Launching of spiral density waves}\label{GTexcite}
Although localised at the  gap  edge, the edge disturbance
perturbs the bulk of the  disc through its gravitational potential
exciting density waves. This is expected to be through torques
exerted at Lindblad resonances (Goldreich \& Tremaine 1979).
When the disc is exterior (interior) to the edge
the wave excitation can be viewed  as occurring at the
the outer (inner) Lindblad resonance respectively. These resonances
 occur  where $\sigma/m=-\Omega\mp\kappa/m.$
Here the negative (positive) sign alternative applies to the outer (inner) 
Lindblad resonance respectively.
The perturbing potential is given by (\ref{PPOT}) and from now
on this potential is given the symbol $\Phi_{edge}.$

The total conserved  angular momentum  flux   associated 
with the launched waves, when they are assumed to propagate out 
of the system is given by \cite{goldreich79} as 
\begin{align}
F_T = \frac{\pi^2 mr\Sigma }{\beta}\left|\frac{d\Phi_{edge}}{dr} + \frac{2m\Omega\sbar
    \Phi_{edge}}{r\kappa^2}\right|^2.
\end{align}
where $\beta = 2\kappa(\mp \kappa^\prime -  m\Omega^\prime).$
These waves carry positive angular momentum outwards when excited by an inner edge,
or equivalently negative angular momentum inwards when excited by an outer edge. 
Thus they  will destabilise
 negative angular momentum disturbances at an inner disc edge
or   positive angular momentum disturbances  at an outer disc edge that cause their emission.
They will lead to an angular momentum flux at the boundary where $r=r_b$  given by
$(2\pi)\left[r(F_{Gr}+F_{Ar})\right]|_{r_b}=F_T$

\subsection{ Spiral density waves and the growth of edge modes}
We now investigate the effects of  wave losses at the non-sharp boundary as a perturbation.
To do this we relax the assumption that the surface density tapers to zero at
the  boundary where $r=r_b.$  Instead we  adopt a small value there such that 
self-gravity is not important for the density waves.
Thus we retain the domain ${\cal D}$ for evaluating the gravitational
potential to be $(r_i,r_o).$
Then, for real frequencies,  all the terms in equation (\ref{angmom_balance})
apart from the term inversely proportional 
to $\bar{\sigma}$ and the  boundary  term  associated with the advective angular flux
at $r=r_b$  are real. 
The imaginary part of the latter
term was shown to be proportional to the wave angular momentum flux.
We  assume that the mode is close to marginal stability (subscript `ms' below) with corotation
at a vortensity maximum where $\sigma=\sigma_r$ 
 and  write
equation  (\ref{angmom_balance})
in the form 
${\cal {F}}(S,\sigma )={\cal {F}}(S,\sigma_r +\delta\sigma)=0.$
Assuming $\delta\sigma$ is small, we then  expand to first order in $\delta\sigma,$  obtaining
\begin{align}\label{FINM}
\left.\frac{\partial {\cal {F}}}{\partial \sigma }\right|_{ms}\delta \sigma
 + {\cal {F}}(S,\sigma_r)\equiv (D_r+ {\rm i}D_i)\delta\sigma +{\cal {F}}(S,\sigma_r) =0.
 \end{align}
Differentiating 
 (\ref{angmom_balance})  we  find
\begin{align}\label{margnx4a}
D_r=& \left.\left(
 {\cal P}\int_{r_i}^{r_o}
\frac{m|S|^2}{\eta^2\bar{\sigma^2}}\frac{d\eta}{dr} dr\right)
\right|_{\sigma=\sigma_r}\nonumber\\
& -\left.{\frac{\partial}{\partial\sigma}\left(\int_{r_i}^{r_o}
\frac{r\Sigma}{D}\left|\frac{d S}{d r}  +
\frac{2m\Omega\bar{\sigma}}{r\kappa^2}S\right|^2
 dr\right)}\right|_{\sigma=\sigma_r}
\end{align}
and
\begin{align}\label{margnx5a}
D_i=\left.-
\frac{\pi |S|^2}{m\Omega'|\Omega'|}
\frac{d^2}{dr^2}\left(\frac{1}{\eta}\right)\right|_{ms}.
\end{align}
Setting $\delta\sigma = \delta\sigma_r-i\gamma,$ where $\gamma$ is the growth rate
and taking the imaginary part of (\ref{FINM}), noting that
the imaginary part of 
the first two terms on the right hand side
 of equation (\ref{angmom_balance})
contribute to the imaginary part of $\mathcal{F}$  giving this as 
$\mathrm{Im}(\mathcal{F})= -\pm (2/m)\left[r(F_{Gr}+F_{Ar})\right]|_{r_b}= \mp (1/(m\pi))F_T$
with $F_T$ given above, we obtain
\begin{align}\label{margnx6}
\gamma=-\pm\frac{F_T}{m\pi D_r}+\delta\sigma_r \frac{D_i}{D_r}.
\end{align}
Similarly the real part of (\ref{FINM}) gives
\begin{align}\label{margnx6real}
 D_r\delta\sigma_r = - \gamma D_i -\mathrm{Re}(\mathcal{F}),\end{align}
where $\mathrm{Re}(\mathcal{F})$ is the real part of $\mathcal{F}$.
We suppose that 
the waves emitted by the edge disturbance result in $\gamma\neq0,$
but that the back reaction   does not change the location of the
co-rotation point from that of  the marginally
 stable mode (as indicated by numerical results). Thus 
co-rotation remains at the original vortensity maximum, implying   that
conditions  adjust  so that  $\delta\sigma_r=0.$
 We then have: 
\begin{align}\label{margnx7}
\gamma=\mp\frac{F_T}{m\pi D_r},
\end{align}
where the upper (lower) sign applies to an inner (outer) edge.
In this case, instability occurs when  wave emission
causes negative (positive) angular momentum to be  transferred to a negative (positive)
angular momentum edge disturbance located at an inner (outer) sharp edge. 

According to equation (\ref{margnx7}), to 
obtain instability $(\gamma>0)$ for an inner edge disturbance we need 
$D_r<0$ (since $F_T>0$). For a disturbance concentrated at corotation near  an inner sharp edge,
with the disc lying beyond, the first term on the right hand side 
of (\ref{margnx4a}) dominates. As corotation is at a  vortensity
maximum, we expect the contribution to the first integral of $D_r$ from
the region just beyond corotation,  
where $d\eta/dr < 0$, to be negative
and the contribution from the region just interior to corotation,  
where $d\eta/dr>0$, to be positive. 
Because the region exterior to an inner edge has higher 
surface density than that interior to it, and the integrand for the first term 
in $D_r$ is proportional to $\Sigma$, we may expect the region exterior to co-rotation
dominates the contribution to $D_r$, making it negative and therefore unstable. 
Indeed, we find $D_r<0$ for the numerical fiducial case presented in section \ref{num_fiducial}. 
In this case there is instability due to the reaction of 
the emitted outwardly propagating positive angular  momentum wave on
the negative angular momentum inner gap edge  disturbance.
A corresponding discussion applies to a disc with an outer sharp edge.

We comment that the above considerations depend on the excitation
of waves at a Lindblad resonance that were transported with
a conserved action or angular momentum flux towards a boundary
where they were lost. However, our linear calculations and simulations
described below indicate lack of sensitivity to  such a boundary condition.
This can be understood if it is emphasised that the significant issue
is that after  emission, the wave action density should not return
to the edge disturbance responsible for it. This is possible for example if the
waves become  trapped in a cavity in the wider disc in which case a radiative
boundary would not be needed.

\section{Linear calculations}\label{linear}

In this section, we present numerical solutions to the linear normal mode
problem where the background self-gravitating disc contains  a gap,
presumed to have been  opened by
a planet. The basic state is obtained  from hydrodynamic 
simulations by azimuthally averaging the surface density and azimuthal 
 velocity component 
fields to obtain one dimensional profiles. The basic state is thus
axisymmetric.  The   radial velocity is assumed to be zero.  

We adopt a local
isothermal equation of state  $P=c_s^2(r)\Sigma$
with  sound-speed $c_s = h\sqrt{GM_*/r}$ and
$h=0.05$ for consistency with nonlinear simulations.
The softening prescription used is that presented in
\S\ref{basic_equations}.   
As for the analytic discussion,  the gravitational potential 
due to the  planet   
and viscosity are neglected.
 The linearised  equations follow from equations 
(\ref{cl})~-~(\ref{barotropic2}) together with  
equations (\ref{2DPoisson}) and (\ref{VVVVV}).
However, rather than solve
in terms of the quantity $S'=\Sigma'c_s^2/\Sigma +\Phi',$
which was convenient for analytic discussion, we find it more convenient
here to work in terms of the relative surface density perturbation,
$W=\Sigma'/\Sigma$ for which a single governing equation may be
written down explicitly.
In terms of this, the  velocity perturbations $u_r^\prime,\,u_\varphi^\prime$
can be written in the form
\begin{align}
u_r^\prime &= -\frac{1}{D}
\left[i\bar{\sigma} \left(c_s^2\frac{dW}{dr} + \frac{d
      \Phi^\prime}{dr}\right) + \frac{2im\Omega}{r}\left( c_s^2W +
    \Phi^\prime\right)\right]\label{linnum}\\  
u_\varphi^\prime &= \frac{1}{D}
\left[\frac{\kappa^2}{2\Omega}\left(c_s^2\frac{dW}{dr} +
    \frac{d\Phi^\prime}{dr}\right) + \frac{m\bar{\sigma}}{r}\left( c_s^2 W
  + \Phi^\prime\right)\right].\label{linnum1})
\end{align}
The gravitational potential perturbation $\Phi'$
can be  expressed in terms of $W$ through equations 
(\ref{2DPoisson}) and (\ref{VVVVV}) and
we  note that  $\kappa^2 \equiv r^{-3}d(r^4\Omega^2)/dr.$
The derivatives
 $d \Phi'/dr$ and $d^2 \Phi'/dr^2$ can be computed by 
replacing $K_m(r,r')$ with $\partial K_m/\partial r$ and $\partial^2
K_m/\partial r^2$, respectively. Inserting  (\ref{linnum}) and (\ref{linnum1})
into the linearised continuity  equation  
yields the governing equation for $W$ 
\begin{align}\label{localiso}
&\frac{d}{dr}\left[\frac{r \Sigma}{D}\left(c_s^2\frac{dW}{dr} +
    \frac{d\Phi^\prime}{dr}\right)\right] 
+ \left[ \frac{2m}{\bar{\sigma}}
 \left(\frac{\Sigma\Omega}{D}\right)
 \frac{dc_s^2}{dr} - r\Sigma \right]W \notag\\
& + \left[\frac{2m}{\bar{\sigma}}\frac{d}{dr}
 \left(\frac{\Sigma\Omega}{D}\right)
 -\frac{m^2\Sigma}{rD}\right]\left(c_s^2 W +
 \Phi^\prime\right )\equiv {\mathcal L}(W) = 0.
\end{align}
Note that the the term in $dc_s^2/dr$ has been kept for consistency
with simulations used to setup the basic state. We checked that this term
has negligible effect on the results obtained below, by solving the linear problem
without this term. This is because $c_s$ varies on a global scale, whereas the edge
disturbance is associated with strong local gradients.

To  solve equation  (\ref{localiso})  we 
discretised it on  an equally spaced grid applied to  the domain
$r=[1,10].$ In practice this employed   $N_r=1025$ grid  points. 
Equation (\ref{localiso}) together with the applied boundary conditions (see below)
is then converted  into a
matrix equation of the form $\sum_{j=1}^{N_r}\mathcal{L}_{ij}(\sigma)W_j = 0,$ 
where the matrix  ${\large(}\mathcal{L}_{ij}(\sigma){\large)}$  gives
the discretised linear operator, which is a function of the
frequency $\sigma,$  and $W_j$ is an approximation to
$W$ at the $j$th grid point.
This problem cannot be solved  for any value of  $\sigma.$ 
This has to be consistent with the condition that the determinant
of the system of linear equations be zero. Thus $\sigma$ is an
eigenvalue although it is not an eigenvalue
of ${\large (}\mathcal{L}_{ij}{\large )}$ in the conventional sense.

We  proceed by  taking
${\large (} \mathcal{L}_{ij} {\large )} $
to be a  function of $\sigma.$ For a specified  value
of  $\sigma$ we  solve the usual eigenvalue problem
$\mathcal{L}_{ij}(\sigma)W_j = \mu(\sigma) W_i$ for an eigenvalue,
$\mu,$ which may  also be considered to be a function of $\sigma.$
The  Newton-Raphson method
is then  used to solve the
equation  $\mu(\sigma) = 0.$
The values of $\sigma$ so obtained are the required  eigenvalues
associated with the physical normal modes of the system.

Because simulations suggest the important
disturbances are associated with the outer gap edge, it is reasonable
to search for values of  $\sigma$
 such that co-rotation lies near  the outer gap
edge. We checked that the reciprocal of
the final matrix condition number is small (at the level of  machine precision) in
order for results to be accepted. For simplicity,  the boundary 
 condition $dW/dr= 0$
 was applied at $r=r_i=1$ and $r=r_o=10.$
As discussed in section \ref{Fluxbal},
  modes are  found to be 
driven by the back reaction of emitted
density waves on  a disturbance located at an outer gap edge,
a process  expected to be insensitive to boundary  conditions.
Indeed, we find that our results are insensitive as  to whether
the above or other boundary conditions are used (see section \ref{edge_mode_bc}). 



\subsection{A fiducial case}\label{num_fiducial}
In order to provide a fiducial case, we study the 
disc model with a gap, for which  $Q_o=1.5,$  that is adopted 
in \S\ref{hydro}.  
However, the simulations described in
\S\ref{hydro} from which  the model was extracted had 
  $\nu=10^{-5},$ whereas our linear calculations
described below are for inviscid discs.
 
The basic state gap profile is illustrated
in  Fig. \ref{linear_basic} where 
the surface density $\Sigma,$ the relative deviation
of the  angular velocity from the Keplerian value  $\Omega/\Omega_k -1$
and the vortensity $\eta$ are plotted.
 As expected from our analytic discussion
of modes near to marginal stability,  local extrema in
$\eta$ in the vicinity of gap edges are closely
 associated with  instability. This is manifest through
mode corotation points being very close to them.  

\begin{figure}
  \centering
  \includegraphics[width=0.45\textwidth]{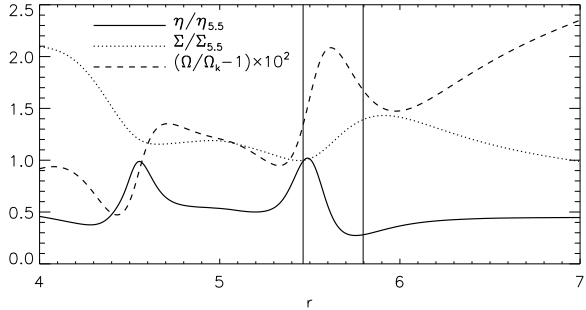}
  \caption{ Gap profile produced  by a Saturn-mass planet in the $Q_o=1.5$
    disc with viscosity $\nu=10^{-5}$. The surface
    density $\Sigma$ and vortensity $\eta$ are scaled by
    their values at $r=5.5.$ The relative  deviation of the angular velocity
     from the Keplerian value is also plotted. Vertical lines indicate
     co-rotation radii $r_c$ (where
    ${\mathrm{Re}}(\bar{\sigma})=0$).  For the $m=2$ mode $r_c = 5.5$,
    close to a vortensity maximum. When self-gravity is neglected $r_c = 5.8$,
    close to a vortensity minimum.
\label{linear_basic}}
\end{figure}

The magnitude of relative surface density perturbation, $|W|$, for
self-gravitating (SG) and non-self-gravitating (NSG) responses are
shown in Fig. \ref{Qm1.5_linear}. For NSG cases we set  $\Phi^\prime = 0.$
 The eigenfrequencies
for SG (NSG) modes are given by  $-\sigma = 0.1587 +
i0.4515\times10^{-2}$ ($-\sigma = 0.1458 + i0.2041\times10^{-2}$).
 These correspond to co-rotation points at  $r_c=5.4626,\,5.7959$ for the
SG and NSG modes, respectively. The SG growth rate corresponds to $\sim
3$ times the local orbital period. Thus although $\gamma/\sigma_R \sim
0.03,$ is relatively small, the  instability grows  on a
dynamical timescale.

The co-rotation points of SG and NSG modes are very close to  local
maximum and minimum  of  $\eta$, respectively
(Fig. \ref{linear_basic}). 
The SG mode grows twice as fast as the NSG mode, consistent with
the observation made when comparing edge modes and vortex
modes in \S\ref{motivation}, where the former became non-linear
sooner. 

The SG and NSG eigenfunctions are similar around co-rotation ($r\in [5,6]$), although
the SG mode has a larger width and is shifted slightly to the left 
(Fig.\ref{Qm1.5_linear}).
 As expected from the discussion in section \ref{GTexcite}, 
the SG mode has significant wave-like region
interior to the inner Lindblad resonance ($r\leq 3.4$) and exterior to
the outer Lindblad resonance at ($r\geq7.2$). By contrast, the  NSG mode
 has negligible amplitude outside $[5,7]$ compared to that at
co-rotation, whereas the
SG amplitude for  $r\in [8,10]$ can be up to $\simeq 56\%$ of
the peak amplitude near co-rotation. 
Increasing $m$ in the NSG calculation increases the amplitude in the wave regions,
but even for $m=6$, we found the  waves in $[8,10]$ for the NSG case 
have an amplitude of about 33\% of that
at co-rotation, a smaller value than that pertaining to
the  SG case with  $m=2.$

The  behaviour  in the wider disc, away from co-rotation, shows that the
for low $m,$ the NSG mode is a  localised vortex mode whereas the  SG mode
corresponds to  an edge mode  with global spirals. Noting that
maxima in the vortensity and Toomre $Q$ nearly  coincide,
it makes sense that the SG mode
is global because away from co-rotation, the background Toomre $Q$ is
decreasing, which makes it easier to excite density waves, because
the evanescent zones between corotation and the Lindblad resonances
that are expected from WKBJ theory,  narrow accordingly. 
\begin{figure}
  \centering
  \includegraphics[width=0.45\textwidth]{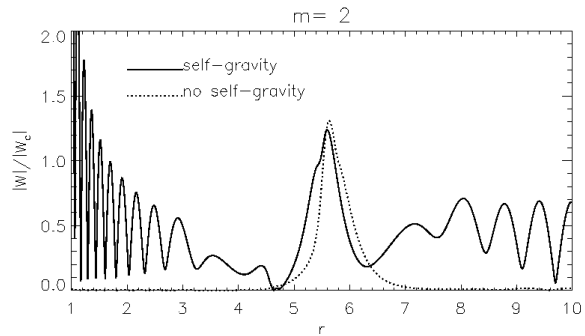}
    \caption{$m=2$ unstable modes found from linear stability analysis 
	for the fiducial model, with
      self-gravity (solid) and without self-gravity (dotted). $|W|$ has been
      scaled by its value at corotation.  
      \label{Qm1.5_linear}}
\end{figure}

Comparing the SG and NSG modes shown in 
 Fig. \ref{Qm1.5_linear}, we see that 
including self-gravity in the linear response enables  additional waves
in the disc at low $m$. 
Fig. \ref{Qm1.5_linear_pot} shows the gravitational 
potential perturbation for the
SG mode. A comparison with $|W|$ in Fig. \ref{Qm1.5_linear} shows that
the surface density perturbation around $r_c$ has an
associated potential perturbation that varies on a more global scale.
The peak in $|W|$  about $r_c$ is confined to $r\in [4.6,6.2]$ whereas
that for $|\Phi'|$ is confined to $r \in [3.2,7.5],$ 
overlapping
Lindblad resonances (see Fig. \ref{Qm1.5_t50}}) in the latter case.
 Thus $\Phi'$ is less localised  around co-rotation
than $|W|$.

 Rapid oscillations seen in $|W|$  for $r>7$ are  not observed for
$|\Phi'|$ in the same region. Furthermore, $\Phi'(r>7)$ is at most
$\simeq 20\%$ of the co-rotation amplitude, which is a smaller ratio
than that for $|W|$. This leads to the notion that the disturbance at
the outer gap edge is driving the disturbance in the outer disc as in
our analytical discussion given  above. In effect, the disturbance at 
 co-rotation  acts like an external perturber (e.g. a planet) to drive
density waves in the outer disc, through its gravitational field. 
Clearly, this is only possible in a self-gravitating disc.   


\begin{figure}
  \centering
  \includegraphics[width=0.99\linewidth]{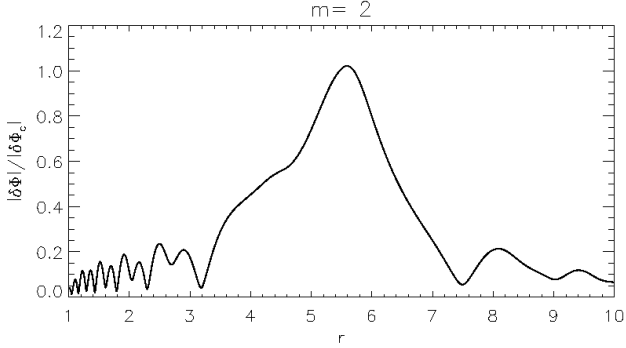}
  \caption{Gravitational potential perturbation $\Phi'$ ($\equiv\delta\Phi$ in the plot), 
      scaled by its value
    at $r=5.5$ for the $m=2$ mode with  self-gravity. 
\label{Qm1.5_linear_pot}}
\end{figure}

\subsection{Energy balance of edge and wavelike disturbances} \label{energy_balance}
The analysis in \S\ref{analytic2} describes the edge mode as
as being associated with a 
 coupling between disturbances associated with  vortensity
extrema  near an edge 
 and the smooth regions interior or exterior to the edge. We
apply this idea to the reference case (with self-gravity) by
considering the region $r>r_p$. Specifically, we focus on $r\in [5,10]$
because hydrodynamic simulations indicate the spiral arms are more
prominent in the outer disc (\S\ref{motivation} and \S\ref{hydro}). 


Multiplying the governing equation ( \ref{localiso})  by
$S'^*$, integrating over $[r_1,r_2]$ and then taking the  real part,
we obtain  

\begin{align}\label{balance}
{\mathrm{Re}}\int_{r_1}^{r_2}{\varrho}dr = 
{\mathrm{Re}} \int_{r_1}^{r_2}( \varrho_\mathrm{corot} + \varrho_\mathrm{wave})dr,
\end{align}
where
\begin{align}
  &\varrho \equiv r \Sigma W S'^* = r(c_s^2
|\Sigma'|^2/\Sigma + \Sigma'\Phi'^*) \label{totalenerg},\\
  &\varrho_\mathrm{corot} \equiv \frac{2m}{\bar{\sigma}}\frac{d}{dr}
  \left(\frac{\Sigma\Omega}{D}\right)|S'|^2,\\
  &\varrho_\mathrm{wave} \equiv S'^*\frac{d}{dr}\left[\frac{r \Sigma}{D}\left(c_s^2\frac{dW}{dr} +
    \frac{d\Phi'}{dr}\right)\right] \notag \\ 
&\phantom{\varrho_\mathrm{wave} \equiv}+\frac{2m}{\bar{\sigma}} 
  \left(\frac{\Sigma\Omega}{D}\right)
  \frac{dc_s^2}{dr}WS'^*   -\frac{m^2\Sigma}{rD}|S'|^2.\label{rho_wave}
\end{align}
$\varrho$ has dimensions of
energy per unit length and, for a normal mode,
 its real part is  four times the energy   per unit length associated with
pressure perturbations (the term  $\propto c_s^2$ ) and gravitational potential
perturbations (the term  $\propto \Phi'^{*}$).  As the factor of four 
is immaterial to the discussion we simply call the integral of this
quantity over the region concerned the thermal-gravitational energy (TGE). 
We remark that when self-gravity dominates, the TGE is negative 
and when pressure dominates, it is positive.

When the integral is performed, one sees that the TGE is balanced by various terms on the RHS of
 equation (\ref{balance}).  For simplicity we   split the terms on the RHS into just
two parts that are integrals 
of  $\varrho_\mathrm{corot}$ and $\varrho_\mathrm{wave}$ over the region of interest.  
This is of course not  a unique procedure. The
vortensity term $\varrho_\mathrm{corot}$ has been isolated because it
contains the potential  co-rotation  singularity  which can be amplified by 
large vortensity gradients at the gap edge. 
The rest  of the RHS
is collected into $\varrho_\mathrm{wave}$. 
Note that the term in $dc_s^2/dr$ is included in 
$\varrho_\mathrm{wave}$, despite being proportional to $1/\sbar$. 
This is motivated by trying to keep $\varrho_\mathrm{corot}$ as close to
the vortensity term identified in the analytical formalism 
as possible. However, we have considered attributing the $dc_s^2/dr$ term to 
$\varrho_\mathrm{corot}$ or neglecting it altogether. In both cases it made 
negligible difference compared to the splitting adopted above. Again, this is because
$c_s^2$ varies on a much larger scale than vortensity gradients. 

It is important to note that while the real part of the
\emph{combination} $ \varrho_\mathrm{corot} + \varrho_\mathrm{wave}$ gives
rise to the TGE, we can not interpret ${\mathrm{Re}}(\varrho_\mathrm{corot})$ as
an energy density of the co-rotation region. It
contains a term contributing to the TGE
that is proportional to the vortensity gradient (see below) 
and potentially associated with a corotation singularity. 
Similar arguments apply to
$\varrho_\mathrm{wave}$ which, in the strictly isothermal case, can
be seen to be associated with density waves
(see section \ref{Fluxbal} above).  We use this splitting  to show that
for the modes of interest, the vortensity term contributes  most to
the TGE.  



\begin{figure}
  \centering
\includegraphics[width=0.99\linewidth]{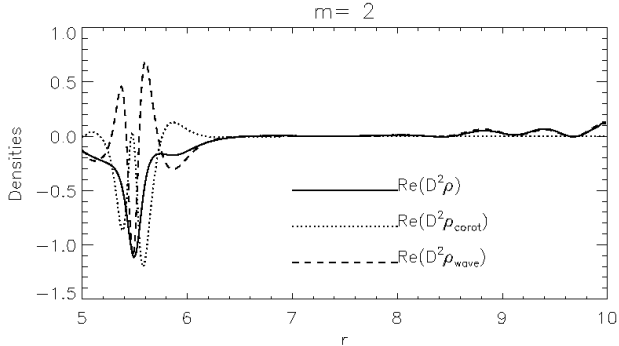}
  \caption{One-dimensional modified energy densities computed from the
    eigenfunctions $W,\, \Phi'$ for the fiducial case $Q_o=1.5$. 
\label{energy_den}}
\end{figure}

Numerically however, quantities defined  above 
are inconvenient because  of the vanishing of  $D$
for neutral modes at Lindblad resonances.  
To circumvent  this we work with  $D^2\varrho,\,
D^2\varrho_\mathrm{corot},$ and $D^2\varrho_\mathrm{wave}.$
We call these modified energy densities.
This change does not disrupt our purpose because
the most important balance  turns out to be between  
the terms involving ${\mathrm{Re}}(D^2\varrho)$ and
${\mathrm{Re}}(D^2\rho_\mathrm{corot})$, both focused near co-rotation, where $D\sim\kappa^2$. 
In the region near Lindblad resonances and beyond, 
${\mathrm{Re}}(\varrho)$ and ${\mathrm{Re}}(\varrho_\mathrm{corot})$ are  small.  
Thus the incorporation of the $D^2$ factor does not influence 
conclusions about the TGE balance.
The  modified energy densities are
plotted in Fig. \ref{energy_den} for the $m=2$ mode in the fiducial case.
The curves share essential features 
with  those obtained  using the original definitions without the additional
factor of $D^2$ (these are
presented and discussed in  Appendix \ref{energy}). 

Fig. \ref{energy_den} shows that ${\mathrm{Re}}(D^2\varrho)$ is negative around co-rotation 
which is located near the
outer gap edge, $r\in[5,6]$). Accordingly  ${\mathrm{Re}}(\varrho)<0$ and  must be dominated by the
gravitational energy contribution since the pressure contribution is positive
definite.  This supports the interpretation of the disturbance as an
edge mode where self-gravity is important. If it were the 
vortex mode then  ${\mathrm{Re}}(D^2\varrho)>0$ near corotation,
because in that case, self-gravity is unimportant compared to
pressure.

For $r\geq 8.4$, ${\mathrm{Re}}(D^2\rho)$ becomes positive due to pressure effects  and
oscillates towards the outer boundary as the perturbation becomes
wave-like. Similarly, the pressure perturbation dominates towards the
inner boundary (not shown). This signifies that self-gravity  
becomes unimportant relative to pressure within these regions. This is
consistent with the behaviour of the gravitational potential perturbations, which are 
largest near co-rotation. We checked explicitly that the gravitational
energy   contribution to the TGE is largely focused around the gap edge
($r\in [5,6]$), even more so than the gravitational potential perturbation.

Fig. \ref{energy_den} shows  that ${\mathrm{Re}}(D^2\rho_\mathrm{corot})$ and 
${\mathrm{Re}}(D^2\rho_\mathrm{wave})$ have their largest amplitudes for
$r\in[5,6]$.
Integrating over $r\in [5,10]$, we find $\tilde{U}<0$, 
and using a normalisation such that
\begin{align*}
  &\tilde{U}\equiv {\mathrm{Re}} \int_5^{10} D^2\varrho dr = -|\tilde{U}|,
\end{align*}
we find
\begin{align*}
  &\tilde{U}_\mathrm{corot}\equiv{\mathrm{Re}} \int_5^{10}
  D^2\varrho_\mathrm{corot} dr \simeq -0.822|\tilde{U}|,\\ 
  &\tilde{U}_\mathrm{wave}\equiv{\mathrm{Re}} \int_5^{10} D^2\varrho_\mathrm{wave} 
  dr \simeq -0.179|\tilde{U}|.  
\end{align*}
This implies the TGE is negative, and hence a
gravitationally-dominated disturbance. We have
$|\tilde{U}_\mathrm{corot}/\tilde{U}|\sim 0.8$, suggesting the TGE is 
predominantly   balanced by the vortensity  term which from
Fig. \ref{energy_den}, is localised in the gap edge
region, balances the gravitational energy of the mode. Because
vortensity gradients are largest near the gap edge, gravitational
energy is most negative here  overtaking  pressure, resulting in a
negative TGE.

\subsubsection{ Further analysis of the vortensity  term}\label{furtherdecomp}
The vortensity term  $\rho_\mathrm{corot}$ that we  defined above differs from  that
naturally identified in the alternative form of the governing equation
in \S\ref{analytic1}. However, this difference is not significant.
The vortensity term  can be further decomposed as
\begin{align*}
  &\varrho_\mathrm{corot} = \varrho_\mathrm{corot,1} +
  \varrho_\mathrm{corot,2}\\
  &\varrho_\mathrm{corot,1}  =
  \frac{m}{\bar{\sigma}(1-\bar{\nu}^2)}\frac{d}{dr}\left(\frac{1}{\eta}\right)|S'|^2,\\ 
  &\varrho_\mathrm{corot,2}  =
  \frac{2m}{\kappa\eta(1-\bar{\nu}^2)^2}\frac{d\bar{\nu}}{dr}|S'|^2,
\end{align*}
where $\bar{\nu}=\bar{\sigma}/\kappa$. Let us now temporarily consider
$\varrho_\mathrm{corot,1}$ as the new vortensity term, so that
$\varrho_\mathrm{corot} \to \varrho_\mathrm{corot,1}$ and attribute
$\varrho_\mathrm{corot,2}$ to the wave term so that 
$\varrho_\mathrm{wave}\to \varrho_\mathrm{wave} + \varrho_\mathrm{corot,2}$. 
The new vortensity
term is proportional  to the vortensity gradient explicitly. 
With the new definitions we  find
\begin{align*}
|\tilde{U}_\mathrm{corot}/\tilde{U}|\sim 0.690.
\end{align*} 
  Thus  the new co-rotation
term alone accounts for $\sim 70\%$ of the (modified) TGE and
giving  almost the same result as the previous one. In addition as
most of the contribution comes from  around co-rotation, where $|\bar{\nu}|\ll 1$. Replacing the
factor $(1-\bar{\nu}^2)$ by unity has a negligible effect on 
the results. However, making this replacement gives 
\begin{align*}
  \rho_\mathrm{corot,1}\to
  \frac{m|S'|^2}{\sbar}\frac{d}{dr}\left(\frac{1}{\eta} \right),
\end{align*}
which  corresponds to the vortensity term used in the analysis given in
\S\ref{analytic1}. 
Whether this term contributes positively or negatively to the TGE depends on 
the sign of $\int_{r_1}^{r_2}\varrho_\mathrm{corot,1}dr$. We expect most of the contribution
to this integral comes from co-rotation where $\sbar\simeq0$. 
Then $\mathrm{sgn}\left(\int_{r_1}^{r_2}\varrho_\mathrm{corot,1}dr\right)
=\mathrm{sgn}(\left.\varrho_\mathrm{corot,1}\right|_{r_c})$.  
Evaluating this for a neutral mode with co-rotation at a
vortensity extremum, 
we find  $\left.\varrho_\mathrm{corot,1}\right|_{r_c} =
\left.-\eta^{-2}|S'|^2(d^2\eta/dr^2)/\Omega^\prime\right|_{r_c}$.
Since quite generally,  $\Omega^\prime < 0,$
we conclude that if co-rotation occurs at a maximum value of $\eta$
then the vortensity term contributes negatively to the TGE.
As stated above, when self-gravity dominates pressure, the TGE is negative,
so it
can then be mostly accounted for  by the vortensity term. 
This is precisely the type of 
 balance assumed for the neutral edge mode modelled 
in section {\ref{Neutedge}.

However, a consequence of the above discussion is that  if co-rotation occurs at
a minimum value of $\eta,$ then the vortensity
term contributes positively to the TGE.
This can only give the dominant contribution when the TGE is
positive, which can only be the case when the pressure contribution
dominates over that from self-gravity. 
In the limit of weak self-gravity,
 the TGE will  be $>0$ and it can be balanced by
a localised disturbance with co-rotation at a  vortensity minimum 
as happens  for the vortex modes.

\subsection{Dependence on $m$ and disc mass} \label{varm_varQ}
We have solved the linear eigenmode  problem for discs with $1.2\leq Q_o \leq
1.6$. The basic state is set up from hydrodynamic simulations as
described in \S\ref{motivation}, with $\nu=10^{-5}$. 
The local $\mathrm{max}(Q)$ near the outer gap edge ranges 
between $Q=3.3$ and $Q=4.4$. Self-gravity is included in the
response ($\Phi'\ne 0$) .

Fig \ref{varm_and_Q}(a) compares eigenfunctions $,|W|,$ for different $m$ for the
$Q_o=1.2$ disc. Increasing $m$ increases the amplitude in the
wave region relative to that around co-rotation ($r_c \simeq
5.5$), resulting in $m=6$ being more global than $m=3$. 
High $m$ modes do
not fit our description of edge modes in the region of interest ($r>5$), 
because the disturbance at co-rotation becomes comparable to  
or even smaller than that in  the wave region beyond the outer Lindblad resonance.
It is then questionable to interpret the waves
as a secondary phenomenon induced by the
edge disturbance, even though the physics of the modes,
as being entities driven by an unstable interaction between edge and wave
disturbances,  may be
more or less  the same.  
Hence, we typically find  what we describe as
edge-dominated modes with $m\la 3$
in simulations where the surface density perturbation is maximal  near
the gap edge.  For fixed $m=2$, hence in that  regime,
Fig. \ref{varm_and_Q}(b) shows  that lowering $Q_o$ makes the co-rotation
amplitude larger relative to that in the wave region. This is expected because
increasing the level of self-gravity means the necessary condition for 
edge disturbance is more easily satisfied (section \ref{Neutedge}). 

\begin{figure}
  \includegraphics[width=0.45\textwidth,clip=true,trim=0cm 0.5cm 0cm 0cm]{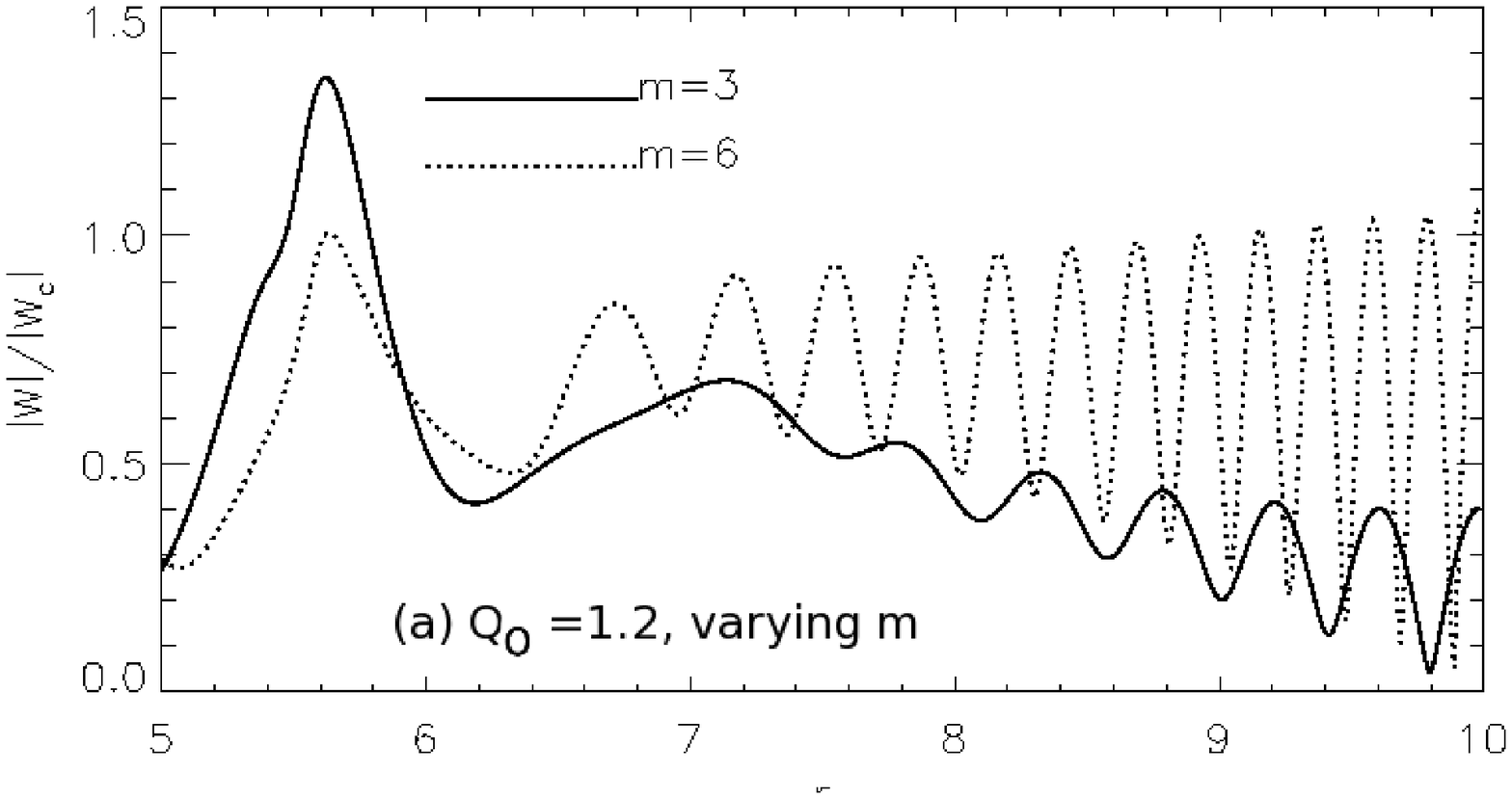}
  \includegraphics[width=0.45\textwidth,clip=true,trim=0cm 0cm 0cm
  1cm,keepaspectratio=false]{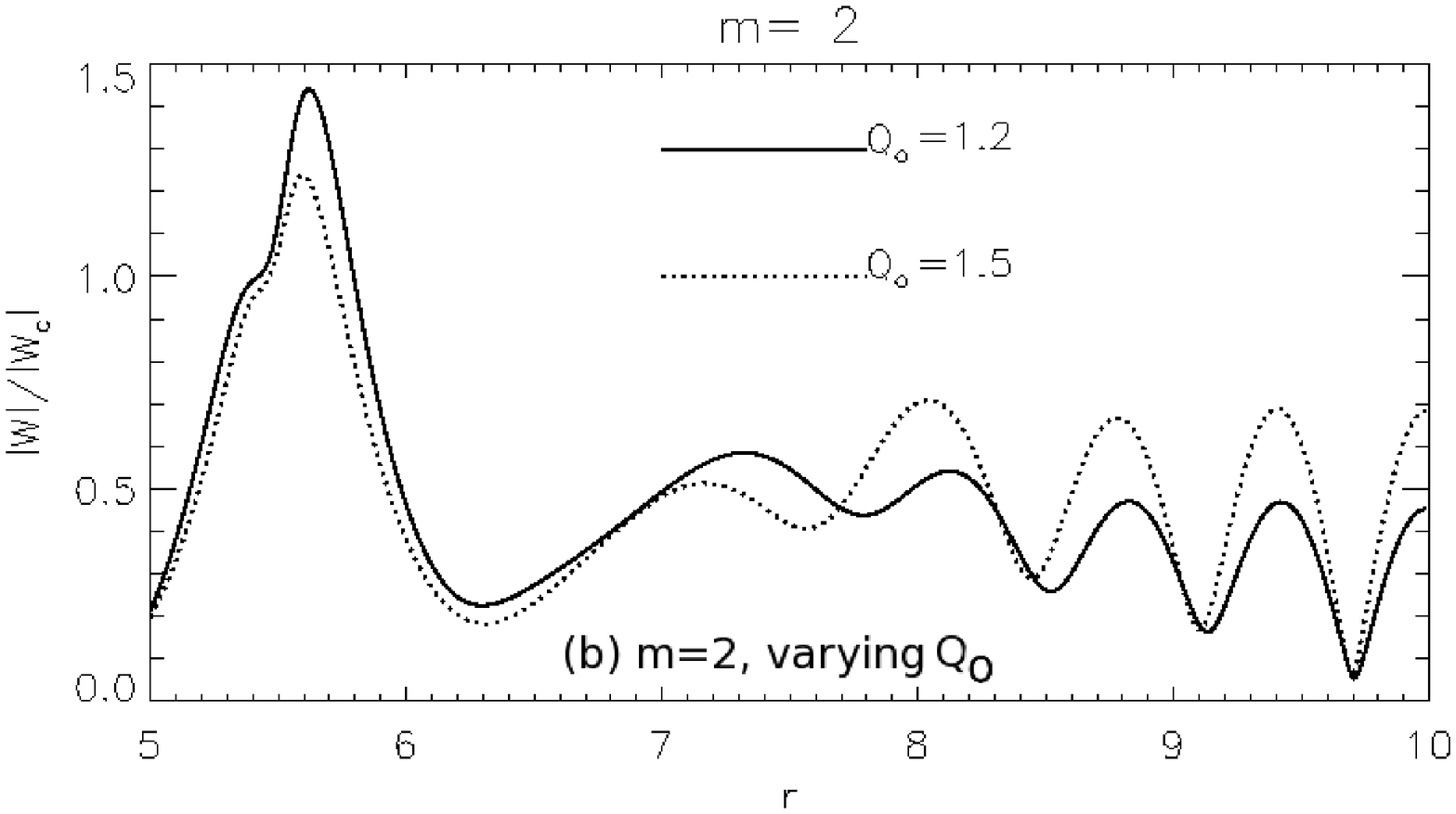} 
  \caption{The effect of azimuthal wave-number $m$ (a) and the effect
    of disc mass, parametrised by $Q_o$ (b) on linear edge modes. \label{varm_and_Q}}
\end{figure}

In
Fig. \ref{compare_gamma_and_corot} we plot the locations
of the corotation points and the growth rates
for the unstable modes we found as a function of the azimuthal mode number, $m$
and a range of disc masses (parametrised by $Q_o$). Note that changing $Q_o$ affects
both the background state and  the linear response,
while $m$  acts as  a parameter of the linear response only. 
The plots in   Fig. \ref{compare_gamma_and_corot}(a) show
that the co-rotation radius tends to move outwards with
increasing $m$ and/or $Q_o.$ 
However, corotation is always located in the edge region where the vortensity
is either decreasing or maximal.
The largest growth rates are found for low $m$ ($\la 3$) edge-dominated modes.

Referring to the basic state (Fig. \ref{linear_basic}), we see that
increasing $m$ and/or $Q_o$, which weakens self-gravity,
shifts co-rotation towards $\mathrm{max}(\eta^{-1})$ (the vortensity minimum),
which disfavours edge dominated modes.  For sufficiently low mass or
non-self-gravitating discs, we expect co-rotation associated with
vortensity minima  and therefore vortex modes to dominate
(e.g. $Q_o=4$ in \S\ref{motivation}).  Thus  we only expect edge
dominated  modes for
low $m$ and sufficiently large  disc mass.   
An  exception to the general trend above is that  the $Q_o=1.6,\,m=6$ mode,
has a smaller co-rotation radius  than the corresponding
mode  with $Q_o < 1.6$. This
may be a boundary effect because for  this value of  $m$ and $Q_o$
the modes are distributed through the outer disc  and
require the implementation of  accurate  radiative
boundary conditions, rather than the simplified  boundary conditions 
actually applied.  
However, such high $m$ modes typically have growth rates
smaller than  those for lower $m$, accordingly
we  do not expect them to dominate,
nor are they observed in the  nonlinear simulations discussed later.

Fig.\ref{compare_gamma_and_corot}(b) shows that
growth rates increase as $Q_o$ is
lowered (increasing surface density scale). 
We found that decreasing $Q_o$ results in deeper gaps, because
disc material trapped in the vicinity of the planet adds to the
planet potential, so that the effective planet mass is larger for
smaller $Q_o$. Steeper gaps are expected to be more
unstable, therefore there is a contribution to the increased growth
rate  from changes to the background profile as the disc mass is increased. 

We also expect the effect of lowering $Q_o$, through the linear response, 
to de-stabilise edge dominated modes as they rely on self-gravity. 
However, the effect of self-gravity through the response is weaker for
larger $m$ values because the size of the Poisson kernel decreases. 
Hence, the increase of the growth rates with disc mass 
for $m=4$---6  is  likely more attributable to the effect of self-gravity on the basic state.    
Growth rates for  the most unstable  mode approximately doubles as the  disc mass is increased by
25\% ($Q_o=1.5\to Q_o=1.2$). 
Notice for fixed $Q_o\le1.5$, the $m=3$---5 growth rates are similar but $m>3$ are not edge modes.

For $Q_o=1.2,\,1.3$ the most
unstable mode has  $m=3,$ whereas for $Q_o=1.5,$ the $m=2$ mode has
the largest growth rate. As discussed in \S\ref{analytic2}, edge dominated 
modes require sufficient disc mass 
and/or low $m$. For fixed $m$, they are stabilised with increasing $Q_o$. 
Hence, for edge modes to exist with decreasing disc mass, they must shift to 
smaller $m$. The higher $m$ modes are less affected by self-gravity. 
This may explain the double peak in growth rate plotted as a function of $m$ 
for $Q_o=1.6$ as we move from $Q_o=1.5$ (we remark that for $Q_o=1.5$ the $m=3$ growth rate
is also slightly smaller than $ m= 2,\,4$).

\begin{figure}
  \centering
  \includegraphics[width=0.45\textwidth,clip=true,trim=0cm 0.5cm 0cm 0cm]{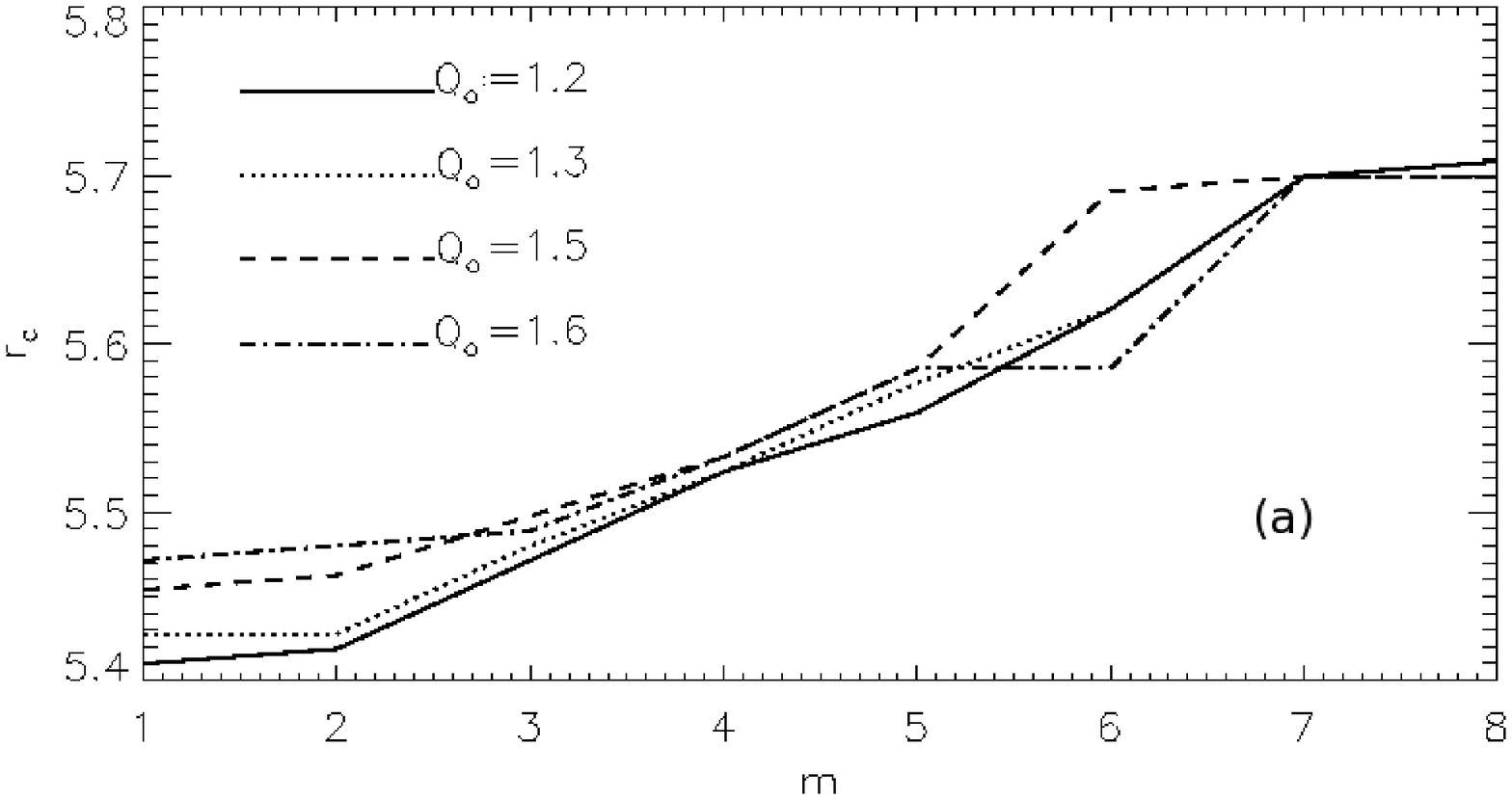} 
  \includegraphics[width=0.45\textwidth,clip=true,trim=0cm 0cm 0cm 1cm]{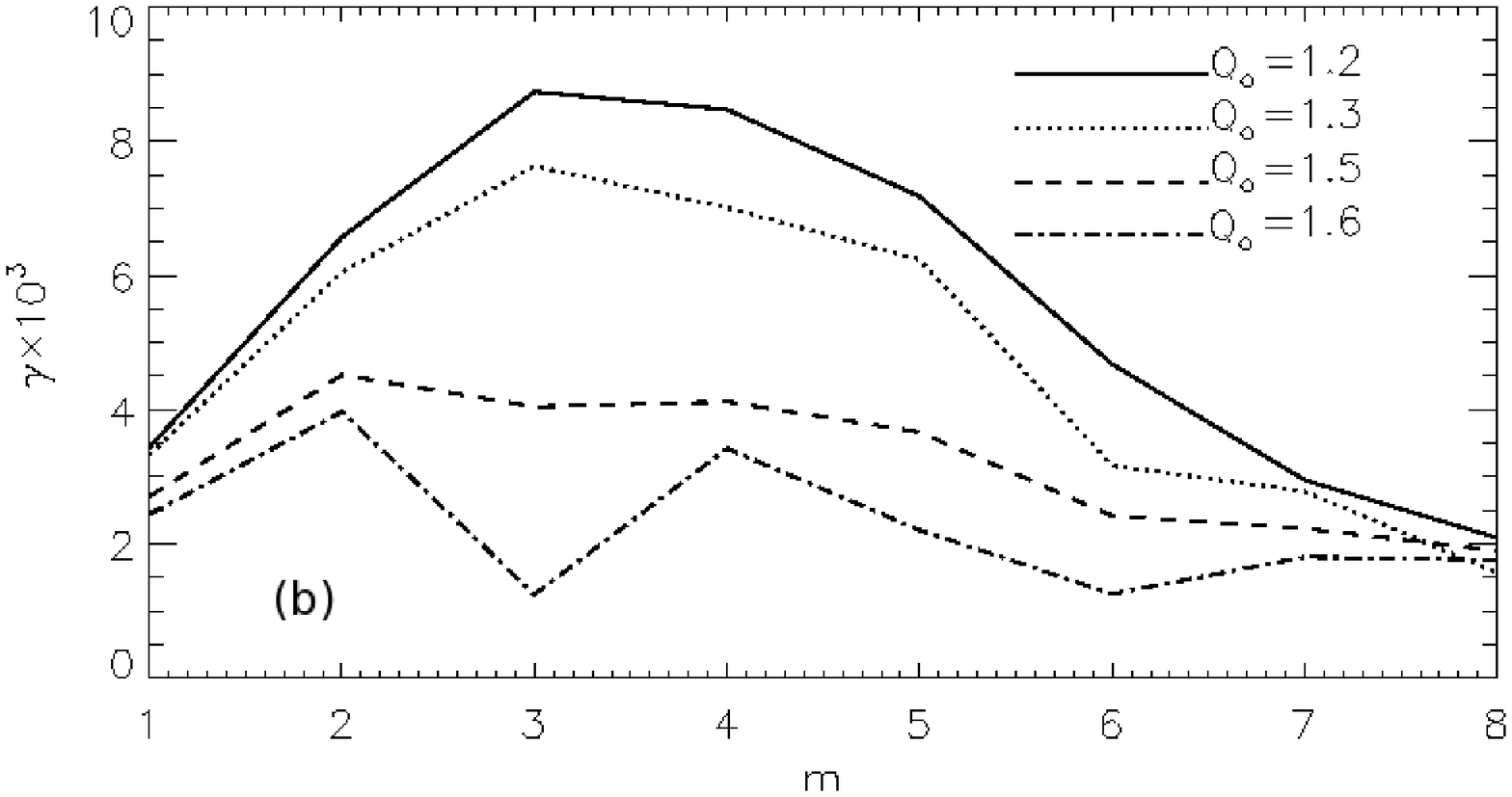}
  \caption{Co-rotation radii (a) and growth rates (b)
    as a function of azimuthal wave-number $m$, for a range of $Q_o$
    values. 
    \label{compare_gamma_and_corot}}
\end{figure}


  The integrals of the modified energy densities  for each
azimuthal wave-number are given for the
fiducial case with $Q_o=1.5$ in Table \ref{Qm1.5_varm}.
 Only the  modes with $m\leq2$ are edge dominated modes
 because  for these sum of the integrals contributing to the (modified) TGE 
 is negative, corresponding to a gravitationally dominated disturbance
and over 50\% of the energy is accounted by the vortensity edge term. 
The mode with $m=2$ had the largest growth rate 
and corotation radius at $r=5.46.$ This is fully consistent
with the nonlinear simulation of the fiducial case performed 
in section \ref{Firstsim}. There the dominant mode in the outer disc had $m=2$
and corotation radius at $r\sim 5.5.$

For $m>2$, the total (modified) energy  becomes positive, which
must be due to the pressure term ($c_s^2|\Sigma'|/\Sigma$) 
becoming dominant. The vortensity term alone cannot
balance this  because it contributes negatively. 
For $m\geq 4$, the vortensity term contribution is small consistent with high 
$m$ solutions being predominantly wave-like(Fig. \ref{varm_and_Q}). 
However, note that inspite of this the vortensity term may still play some role
in driving the modes through corotation torques.
These trends have also been found for other models
so they appear to be general. However, high $m$ global modes are unimportant
for the problems we consider, 
as they are not seen to develop in nonlinear simulations. 










\begin{table}
  \centering
  \caption{Eigenfrequencies expressed in units of $\Omega_k(r=1),$
       corotation radii and 
     modified energy integrals  for
    different  azimuthal mode numbers, $m$, for the fiducial  disc with
    $Q_o=1.5$. The modified energy integrals are taken over  the outer disc 
    $r\in[5,10]$.\label{Qm1.5_varm}}   

  













  \begin{tabular}{ccccll}
    \hline\hline
    $m$ & $-\sigma_R$ & $\gamma\times10^{2}$ & $r_c$ &
    $-\tilde{U}_\mathrm{corot}/|\tilde{U}|$ & $-\tilde{U}_\mathrm{wave}/|\tilde{U}|$\\ 
    \hline
    1 & 0.079 & 0.270 & 5.454  &
    0.65 & 0.35 \\
    2 & 0.159 & 0.452 & 5.463  &
    0.82 & 0.18 \\
    3 & 0.237 & 0.404 & 5.498  & 
    0.34 & -1.24 \\
    4 & 0.313 & 0.412 & 5.533  &
    0.026 & -1.03 \\
    5 & 0.386 & 0.366 & 5.585  & 
    0.0048 & -1.004 \\
    6 & 0.462 & 0.242 & 5.603  &
    0.0015 & -0.998 \\
    7 & 0.524 & 0.223 & 5.699  & 
    0.0010& -0.997 \\
    8 & 0.599 & 0.189 & 5.699  & 
    0.00099 & -0.995 \\
   \hline
  \end{tabular}
\end{table}

\subsection{Softening length}
Gravitational softening  has to be used in a
two-dimensional calculation.  It prevents a singularity at $r=r'$ in
the Poisson kernel and approximately accounts for the disc's vertical
dimension but is associated with some uncertainty.  
Apart from the linear response, softening also has an effect through 
the gap profile set up by our nonlinear simulations. A fully self-consistent 
treatment requires a new simulation with each new softening considered.
For reasons of numerical tractability though,
we  only performed experiments using two values
of softening parameter $\epsilon_{g0,\mathrm{e}}$ in simulations
to setup the gap profile
and range of values $\epsilon_{g0,\mathrm{l}}$ 
used in the linear response. 

Note that 
in nonlinear simulations a single disc potential softening parameter $\epsilon_{g0}$ is used.
Our results are summarised in Table \ref{soft_expts}. The integrated total
modified energy values indicate we have found dominated  edge modes 
since the major contribution is  due to the co-rotation/vortensity term. 

Comparing growth rates for fully self-consistent cases
$\epsilon_{g0,\mathrm{e}}=\epsilon_{g0,\mathrm{l}} =0.3,\,0.6$
 shows  that
increasing the softening length stabilises edge modes, which is expected
since they are driven by self-gravity. In addition  all growth rates for
$\epsilon_{g0,\mathrm{e}}=0.6$ are smaller than those for
$\epsilon_{g0,\mathrm{e}}=0.3$
independently of $\epsilon_{g0,\mathrm{l}}$.
This indicates  stabilisation of  edge
modes via the basic state when softening is increased. 
For fixed
$\epsilon_{g0,\mathrm{e}}$, $\gamma$ decreases as
$\epsilon_{g0,\mathrm{l}}$ increases.  
We could not find edge modes for $\epsilon_{g0,\mathrm{l}} \geq
0.6$ with $\epsilon_{g0,\mathrm{e}}= 0.3$, nor for
$\epsilon_{g0,\mathrm{l}} \ga 0.83$ with $\epsilon_{g0,\mathrm{e}}=
0.6$.
 An upper limited is expected from analytical considerations
(see \S\ref{Neutedge}) because if self-gravity in the linear response is
too weak, the vortensity/co-rotation disturbance cannot be maintained
so edge dominated modes are suppressed. This is intuitive because the edge mode
requires gravitational interaction between gap edge and the smooth disc. 

\subsubsection{Convergence issues}\label{linear_convergence}
For fixed $\epsilon_{g0,\mathrm{e}}=0.3$, the energy ratios in
Table \ref{soft_expts} indicate convergence as
$\epsilon_{g0,\mathrm{l}}\to0$. Perhaps surprisingly, 
$|\tilde{U}_\mathrm{corot}/\tilde{U}|$ decreases with softening. 
We found this is because $\varrho_\mathrm{wave}$, as defined  by equation
(\ref{rho_wave}), includes a term proportional to $d^2\Phi'/dr^2$ which is the dominant 
contribution to $|\tilde{U}_\mathrm{wave}|$, and its contribution mostly comes from 
the co-rotation region (as the potential perturbation is largest there).  
As the strength of self-gravity is increased by decreasing softening, 
the edge disturbance is modified, but the subsequent effect of the
$d^2\Phi'/dr^2$ term on the energies cannot be anticipated a priori. 
 
For $\epsilon_{g0,\mathrm{l}}=0.03$, the vortensity term does not
dominate the modified total energy as much. 
However, this case is in fact not self-consistent 
because the softening used to setup the gap profile is an order of 
magnitude larger than that used in linear calculations. 
The limit of zero softening is also physically irrelevant because 
the disc has finite thickness. For the self-consistent cases with 
reasonable softening values ($0.3,\,0.6$), the vortensity
term dominates the energy balance, so the analytic description developed 
above works reasonably well. 

\begin{table}
  \centering
  \caption{As for table \ref{Qm1.5_varm}
 but for linear calculations with different softening
    parameters, for the $Q_o=1.5$  fiducial case and azimuthal mode number
    $m=2$. Subscripts `l' denote the softening parameter used in the linear
    response, and `e' denotes the  softening  parameter used in setting up the basic
    state. Note that although growth rates vary
by about a factor of two, the location of the corotation
radius does not vary  significantly. \label{soft_expts}}   


    \begin{tabular}{cccccl}
    \hline\hline
    $\epsilon_{g0,l}$ & $\epsilon_{g0,e}$ &
    $-\sigma_R$ & $\gamma\times10^{2}$  &
    $-\tilde{U}_\mathrm{corot}/|\tilde{U}|$ &  $-\tilde{U}_\mathrm{wave}/|\tilde{U}|$\\ 
    \hline
   0.03& 0.3 & 0.159 & 0.560 & 0.66 &    
    0.33\\    
    0.05& 0.3 & 0.159 & 0.544 & 0.67 &
    0.32\\ 
    0.1 & 0.3 & 0.159 & 0.508 & 0.69 &
    0.30\\ 
    0.2 & 0.3 & 0.159 & 0.461 & 0.74 &
    0.23 \\ 
    0.3 & 0.3 & 0.159 & 0.452 & 0.82 & 
    0.18 \\
    0.5 & 0.3 & 0.158 & 0.435 & 0.87 &
    0.14 \\
    0.3 & 0.6 & 0.159 & 0.420 & 0.91 &
    0.09 \\
    0.5 & 0.6 & 0.158 & 0.410 & 0.95 &
    0.064 \\
    0.6 & 0.6 & 0.158 & 0.375  & 0.99 &
    0.0017 \\
    0.8 & 0.6 & 0.158 & 0.276  & 1.15 &
    -0.14 \\
   \hline
  \end{tabular}
\end{table}

\subsection{Edge mode boundary issues}\label{edge_mode_bc}
The boundary condition $W^\prime=0$ was applied for
simplicity. Vortex modes in low mass or non-self-gravitating discs are
localised and insensitive to boundary conditions
\citep{valborro07}. The edge dominated modes rely on self-gravity and are
therefore intrinsically global, boundary effects cannot be assumed
unimportant a priori. 

We performed additional experiments with varying boundary
conditions for the fiducial  case ($Q_o=1.5$). These include 
re-locating the inner boundary to $r=1.1,\,2.5$ or the outer boundary to $r=9.8$ and
approximating/extrapolating boundary derivatives using interior grid points
\citep{adams89}. In the last case, the linear ODE is applied at endpoints of the grid
and therefore no boundary conditions imposed. 
%
%
For self-gravitating disc calculations giving edge dominated modes,
these various conditions gave co-rotation radii  varying by about 0.2\% and growth
rates varying by at most 10\%. Overall the eigenfunctions  $W$  are
similar, showing  the essential features of the edge mode, including
relative amplitude of outer disc wave region and the co-rotation
region. We conclude that the existence of edge dominated  modes is not too
sensitive to boundary effects. Physically this is because
edge modes are  mainly driven by the local vortensity maximum or edge, which
is an internal feature away from boundaries.

\section{Nonlinear hydrodynamic simulations}\label{hydro}

We present nonlinear 
hydrodynamic simulations of disc-planet models with $1.2 \leq
Q_o \leq 2$, corresponding to disc masses $0.079M_* \geq M_d \geq
0.047M_*$. Most simulations use a viscosity of $\nu=10^{-5}$
or equivalently $\alpha = O(10^{-3})$, which has been
 typically adopted for protoplanetary discs. 
This value has  also been found to
 suppress vortex instabilities \citep{valborro07}.
We adopt gravitational softening  parameter 
 $\epsilon_{g0}=0.3$. The planet is
set on a fixed circular 
orbit at $r_p=5$ in order  to focus on gap stability.
We briefly explore the effects of varying 
viscosity and softening lengths later in this section and
planetary migration in the next section.

\subsection{Numerical method}
The hydrodynamic equations are evolved with the FARGO code
\citep{masset00}. FARGO is an explicit finite-difference code similar 
to ZEUS \citep{stone92} but customised for disc-planet interactions. 
It circumvents the time step limit imposed by the rotational velocity
at the inner boundary by splitting the azimuthal velocity into mean
and perturbed parts,  azimuthal transport being   performed on each 
separately. Self-gravity for FARGO was implemented and tested by
\cite{baruteau08}. Two-dimensional self-gravity can be calculated
using Fast Fourier Transform \citep{binney87}. This requires the
radial domain to be logarithmically spaced and doubled in extent. The
planet potential is introduced at $25P_0$ and its gravitational
potential ramped up  over $10P_0$, where $P_0$ is the Keplerian period
at $r=5$. 

The disc is divided into $N_r\times N_\varphi = 768\times2304$  grid
points in radius 
and azimuth giving a resolution of $\Delta r/H = 16.7$. The grid cells
are nearly square, with $\Delta  r/r\Delta\phi = 1.1$. We impose 
open boundaries at $r_i$ and non-reflecting boundaries at $r_o$
\citep{godon96}. The latter has also been used in the  self-gravitating
disc-planet simulations of \cite{zhang08}. As argued above, 
because edge modes are
physically driven by an internal edge, we do not expect boundary
conditions to significantly affect whether or not they exist. Indeed, 
simulations with open outer boundaries, or damping boundaries \citep{valborro06}
all show development of edge modes.  
%
%
\subsection{Overview}
We first present an overview of the effect of edge  dominated modes on
disc-planet systems. Fig. \ref{qm_density} shows the relative surface
density perturbation ($\Delta\Sigma/\Sigma$) for $Q_o=1.2,\,1.5$ and
2.0. The profile formed in these cases has local   
$\mathrm{max}(Q)= 3.3, \, 4.2,\,5.3$ near the outer gap edge and the average $Q$ for  $r\in [6,r_o]$ is
$1.6,\,2.0,\,2.6$. Although  the edge mode is associated with the  local
vortensity maximum (located close to  $\mathrm{max}(Q)$ for gap profiles),
it requires coupling to the external smooth disc via
self-gravity. Hence, edge modes only develop if $Q$ in these smooth regions
is sufficiently small, otherwise global disturbances cannot be
constructed.

The case $Q_o=2.0$ gives a standard result for gap-opening planets,  
typically requiring planetary masses comparable or above that of Saturn. 
For that mass a partial gap of depth $\sim
50\%$ is formed,  a  steady state attained and remains stable to the
end of the simulation. This serves as a stable case for comparison
with more massive discs. This simulation was repeated with
$\nu=10^{-6}$, in which case we identified  both vortex and edge
modes. The standard viscosity $\nu=10^{-5}$ thus suppress both types
of instabilities for $Q_o=2$.  Although we adopted  $\nu=10^{-5}$ to avoid 
complications from vortex modes, we note that they are stabilised 
for sufficiently massive  discs anyway (e.g. $Q_o=1.5,\, \nu=10^{-6}$ do not show
vortices, see section \ref{motivation}).

Gap edges become unstable for $Q_m\leq 1.5$ when $\nu= 10^{-5}.$
As implied by linear analysis, the $m=2$ edge mode
dominates when  $Q_o=1.5$ and it  saturates by $t=100P_0.$
It is more prominent in the outer disc, and over-densities deform the gap edge
into an eccentric ring.  The under-density in the gap also becomes  more
non-axisymmetric, being deeper where the gap is wider.  
The inner disc remains fairly circular, though non-axisymmetric disturbance ($m=3$) 
can be identified. 
Note the global nature of the edge modes in the outer
disc: spiral arms can be traced back to disturbances at the gap edge
rather than the planet. Consider the spiral disturbance at the outer
gap edge just upstream of the planet where $\Delta\Sigma/\Sigma$ is
locally maximum. Moving along this spiral outwards,
$\Delta\Sigma/\Sigma$ reaches a local minimum at $(x,y)=(6,-4)$ before
increasing again. This is suggestive of the outer disc spirals being
induced by the edge disturbance.

When the effect of  disc self-gravity is increased to the 
$Q_o=1.2$ case,  there is significant
disruption to the outer gap edge, it is no longer clearly
identifiable. The $m=1$ edge mode  becomes  dominant in the outer
disc, although overall it  still appears to carry  an $m=2$
disturbance because of the perturbation due to the planet.
The inner disc becomes
visibly eccentric with an $m=2$ edge mode disturbance. The shocks due
to edge modes can comparable strength to those induced by the planet. 

\begin{figure*}
\centering
\includegraphics[width=0.32\textwidth,clip=true,trim=1.5cm 0cm 0cm 0cm]{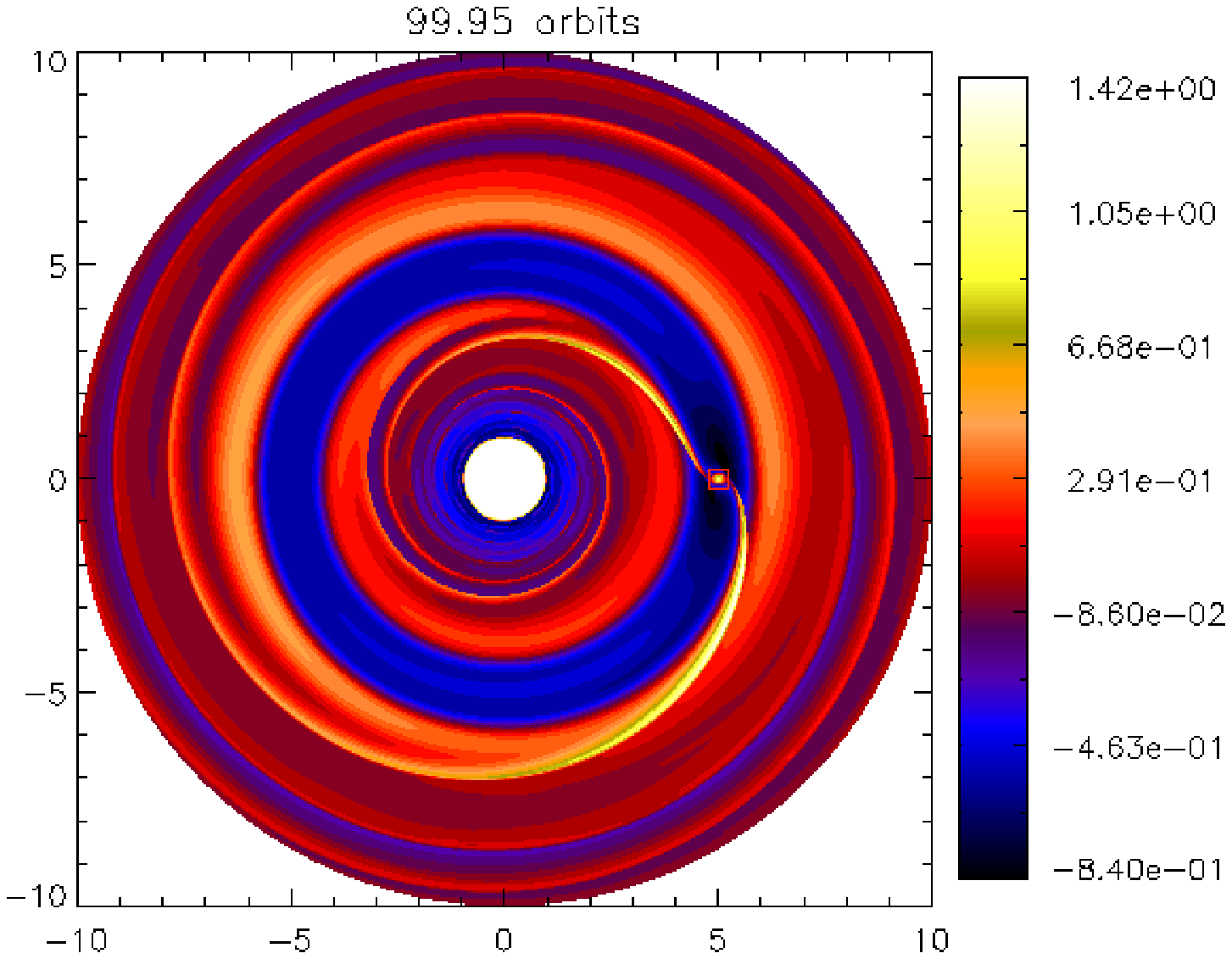}
\includegraphics[width=0.32\textwidth,clip=true,trim=1.5cm 0cm 0cm 0cm]{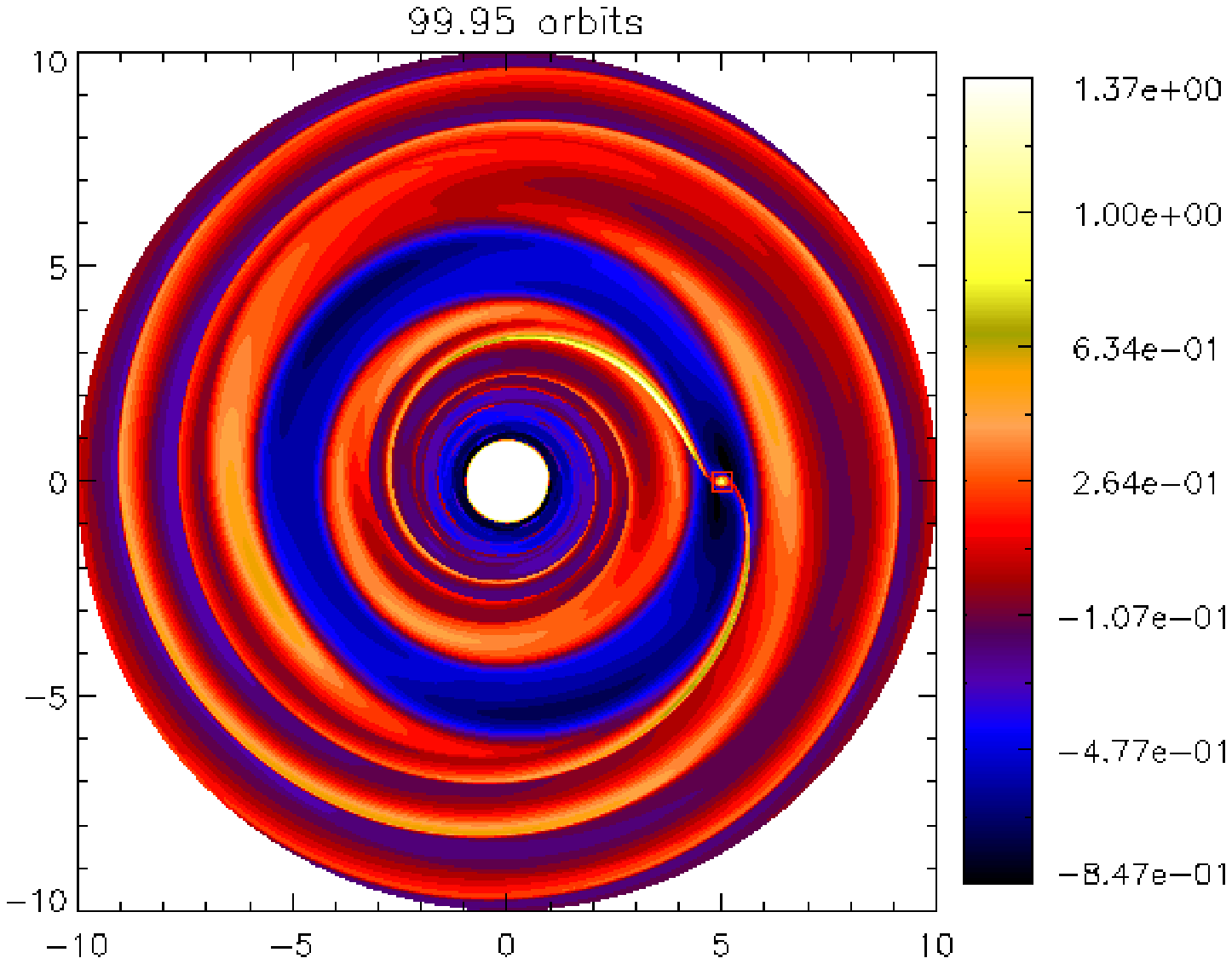}
\includegraphics[width=0.32\textwidth,clip=true,trim=1.5cm 0cm 0cm 0cm]{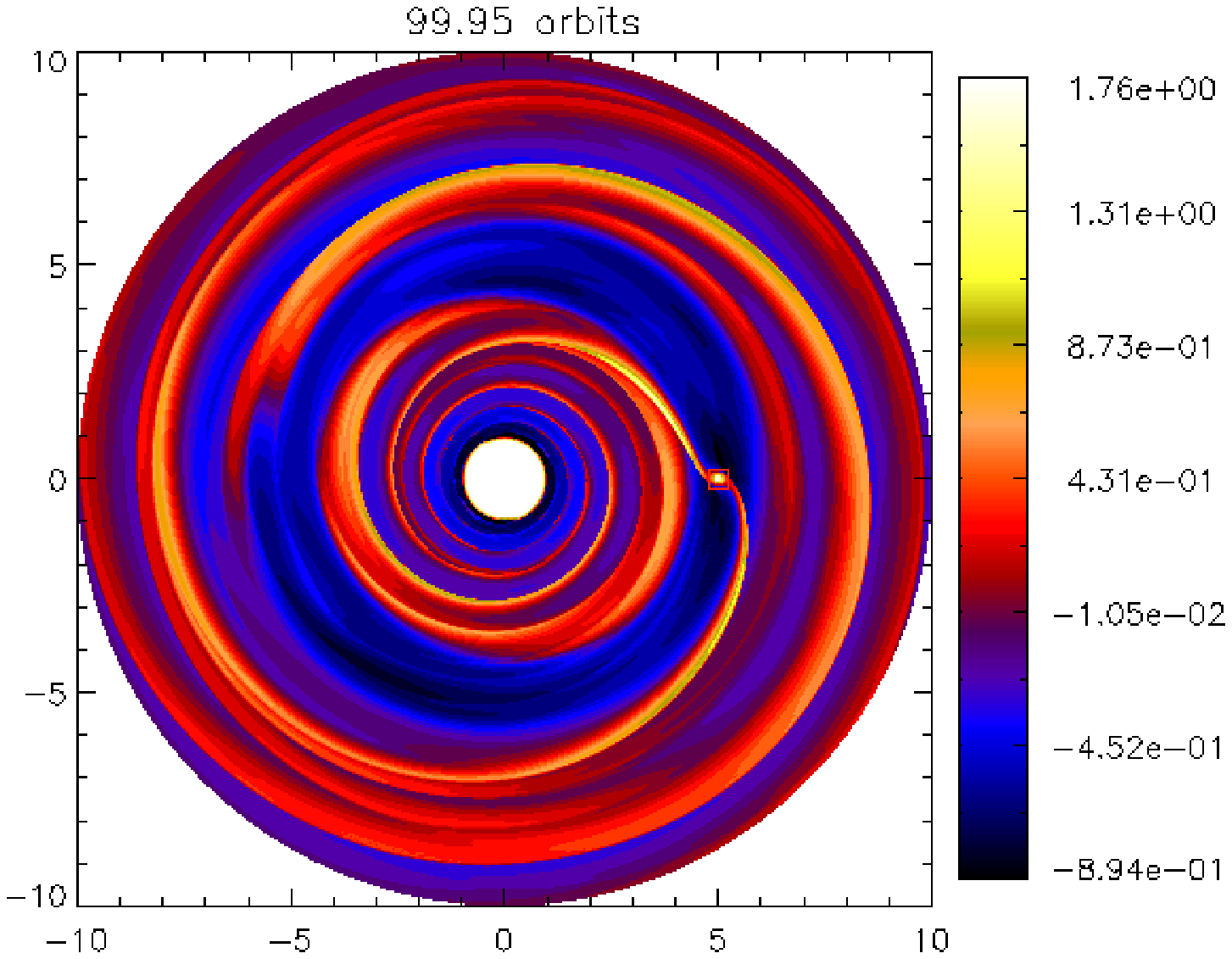}
\caption{Surface density perturbations (relative to $t=0$) at
  $t\sim 100P_0$ as a function the minimum value of Toomre $Q$ in the
  disc, $Q_o$. From left to right: $Q_o=2.0,\,1.5,\,1.2$. 
\label{qm_density}}
\end{figure*}


\subsection{Development of edge modes for  $Q_o=1.5$}

We examine the  fiducial case  with $Q_o=1.5.$
 Fig. \ref{Qm1.5_nonlin1} shows the development of the
edge instability.
 The final dominance of $m=2$ is consistent with linear
calculations.
 The planet opens a well-defined gap by $t\sim 40P_0$, or 
$\sim 5P_0$ after the planet potential  $,\Phi_p,$ has been  fully 
ramped up. At  $t\sim 46P_0,$ two over-dense blobs, associated with the 
edge mode  can be  identified at the outer gap
edge. At this time the gap has $\Delta\Sigma/\Sigma \simeq -0.3$, implying
edge modes can develop \emph{during} gap
formation, since the steady state stable gap in $Q_o=2$ reaches
$\Delta\Sigma/\Sigma \simeq -0.6.$ 
Edge modes are global  and are associated with
long trailing spiral waves  with density perturbation  amplitudes
that are  not small compared to
that at the gap edge.
This is clearly seen at  $t\sim 50P_0$ when spiral
shocks have already developed.
We estimate that the edge modes have a growth rate 
consistent with predictions from linear theory and   become non-linear
within the time frame  $46P_0 < t < 50P_0.$

The edge perturbations  penetrate the 
outer gap edge and trail  an angle  similar to the planetary wake.
Notice the $m=3$ disturbance near $r=4$ in 
the snapshots taken at $t=50P_0$ and $t=56P_0.$  Two of the three
over-densities, seen as a function of azimuth,
correlate to the $m=2$ mode in the outer disc, while the third  over-density
adjacent to the planet correlates with the outer planetary wake. 
These features result from self-gravity and are unrelated to vortex modes.  

The presence of both planetary wakes (with pattern speed $\simeq
\Omega_k(5)$) and edge modes (with pattern speed $\simeq 
\Omega_k(5.5)$) of comparable amplitude enable direct interaction between them.
Passage of edge mode
spirals through the planetary wake may disrupt the former, explaining some of the 
apparently split-spirals in the plots. 
At 
$t\sim 64P_0$  Fig. \ref{Qm1.5_nonlin1} indicates  that a 
spiral density wave feature coincides with the
(outer) planetary wake. 
The enhanced outer planetary wake implies an  increased negative torque 
exerted on the planet at this time. 
This effect occurs during the  overlap of positive
density perturbations. Assume that  at a fixed radius the
edge mode spiral has azimuthal thickness $nh$ where $n$ is a
dimensionless number. The time taken for this spiral to cross the
planetary wake is $  \Delta T = nh/|\Omega_k(5.5) -
  \Omega_k(5)| \simeq 0.06nP_0$. This is $\ll P_0$ for $n=O(1)$. We
expect such crossing spiral waves  to induce  associated  oscillations in the 
disc-planet torque.

The non-linear evolution of edge modes in disc-planet systems can give
complicated surface density fields. The gap can  become  highly 
deformed.  Fig. \ref{Qm1.5_nonlin1} shows that  large voids develop to
compensate for the over-densities  in spiral 
arms, leading to  azimuthal gaps ($t=64P_0$). 
At $t=72P_0$, the gap width ahead of the planet is narrowed by the
spiral disturbance at the outer edge trying to connect  to that at the 
inner edge. 
However, this gap-closing effect is opposed by the planetary  torques
which tend to open it. 

A quasi-steady state is reached by $t\geq 76P_0$  showing an eccentric
gap. For $t=99.5P_0$ additional contour lines are
plotted to indicate  two local surface density maxima along the edge mode
spiral adjacent to the planet. This spiral is disjointed around
$r\sim7.5$ where on traversing an arm,  there is a minimum in $\Delta\Sigma/\Sigma$.
 Taking co-rotation of the spiral at $r_c=5.5$ gives the
outer Lindblad resonance of an  $m=2$ disturbance at $r=7.2,$ assuming  a
Keplerian disc. This is within the disjointed region of the spiral
arm, which supports our interpretation that the disturbance at the
edge is driving activity in the outer disc by perturbing it
gravitationally and launching waves at outer Lindblad resonances. 

\begin{figure*}
  \centering
  \includegraphics[scale=0.47          ,clip=true,trim=0cm
  1.83cm 1.645cm 0cm]{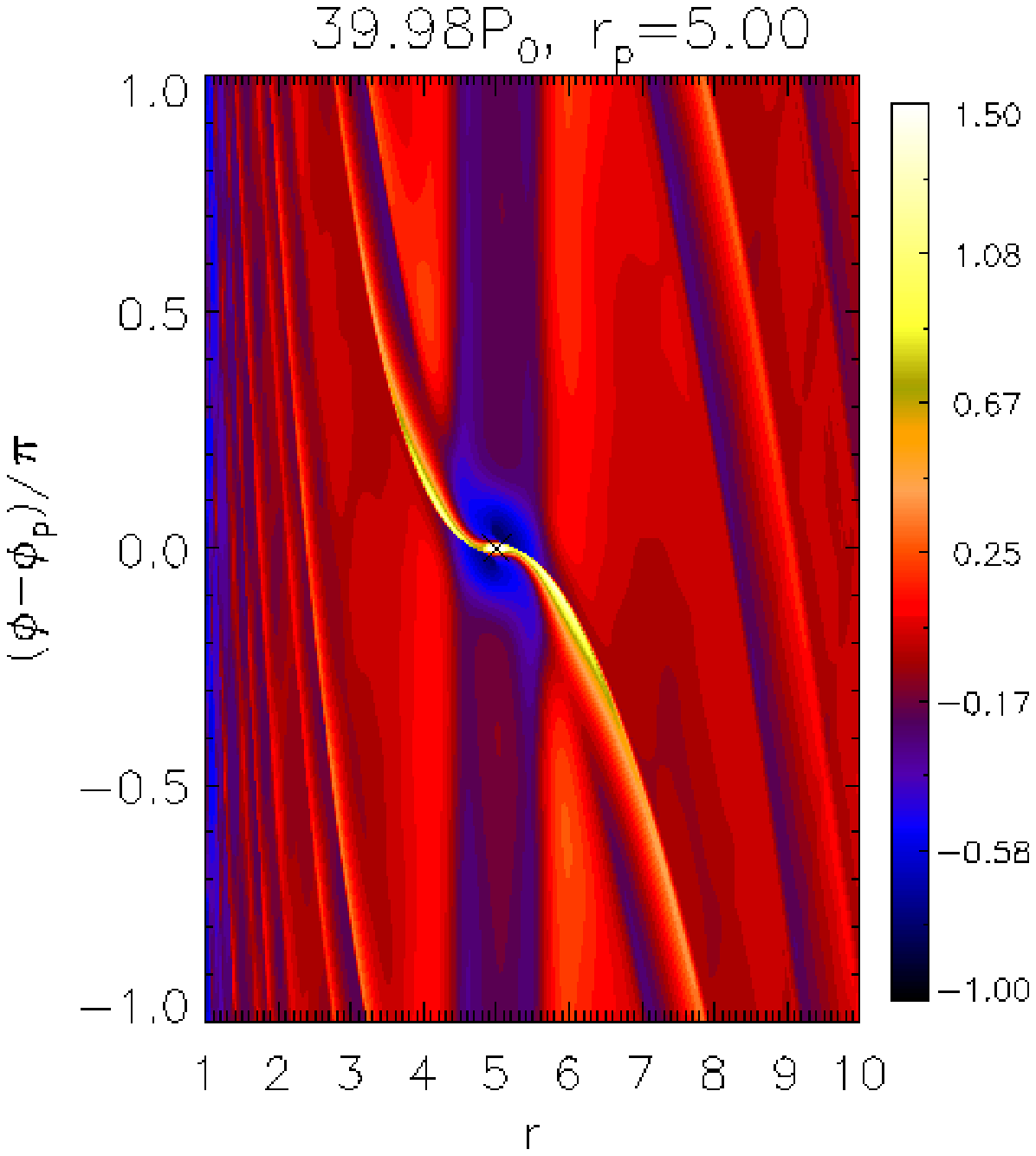}
  \includegraphics[scale=0.47                 ,clip=true,trim=2.34cm
  1.83cm 1.645cm 0cm]{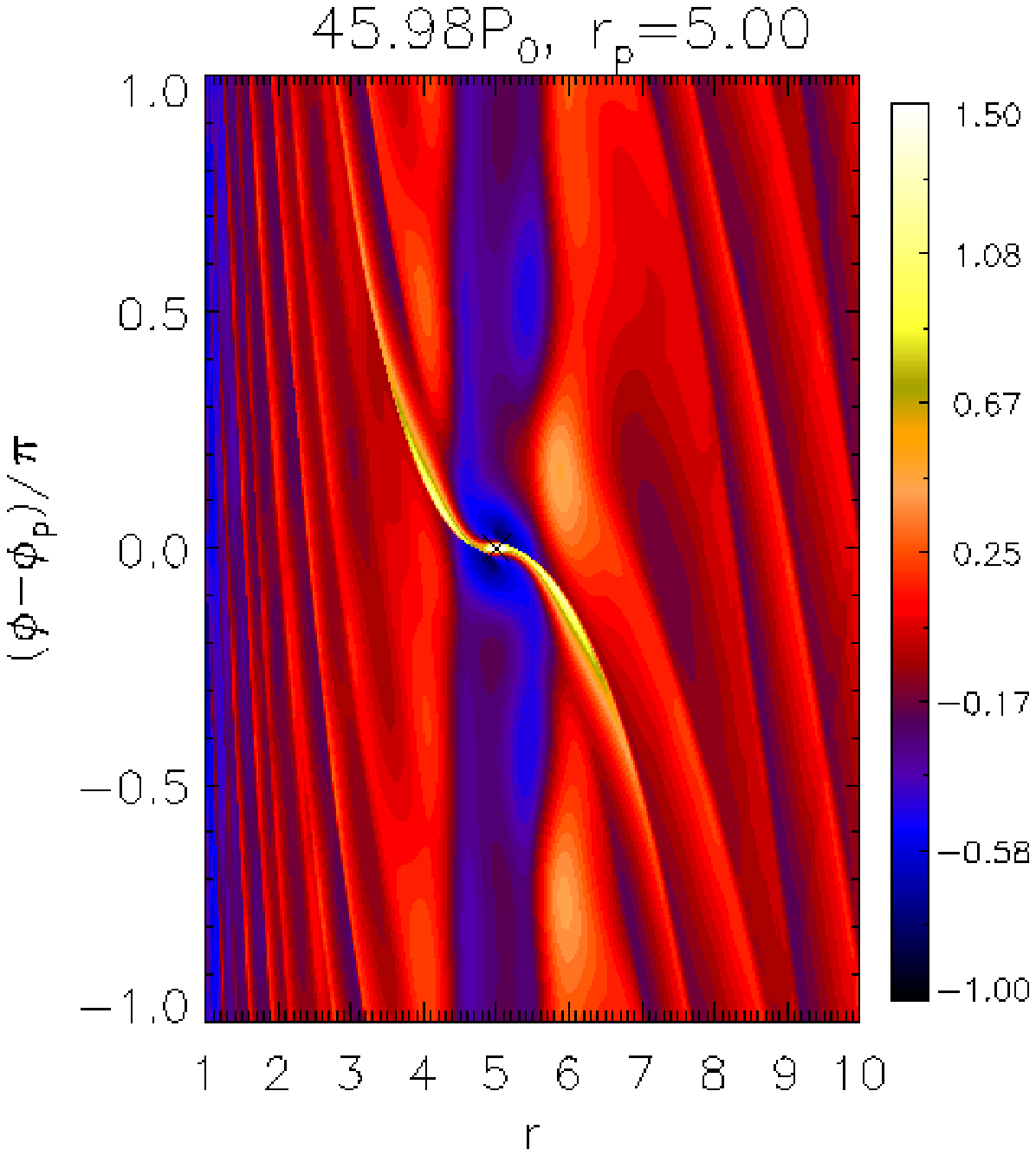}
  \includegraphics[ scale=0.47                  ,clip=true,trim=2.34cm
  1.83cm 1.645cm 0cm]{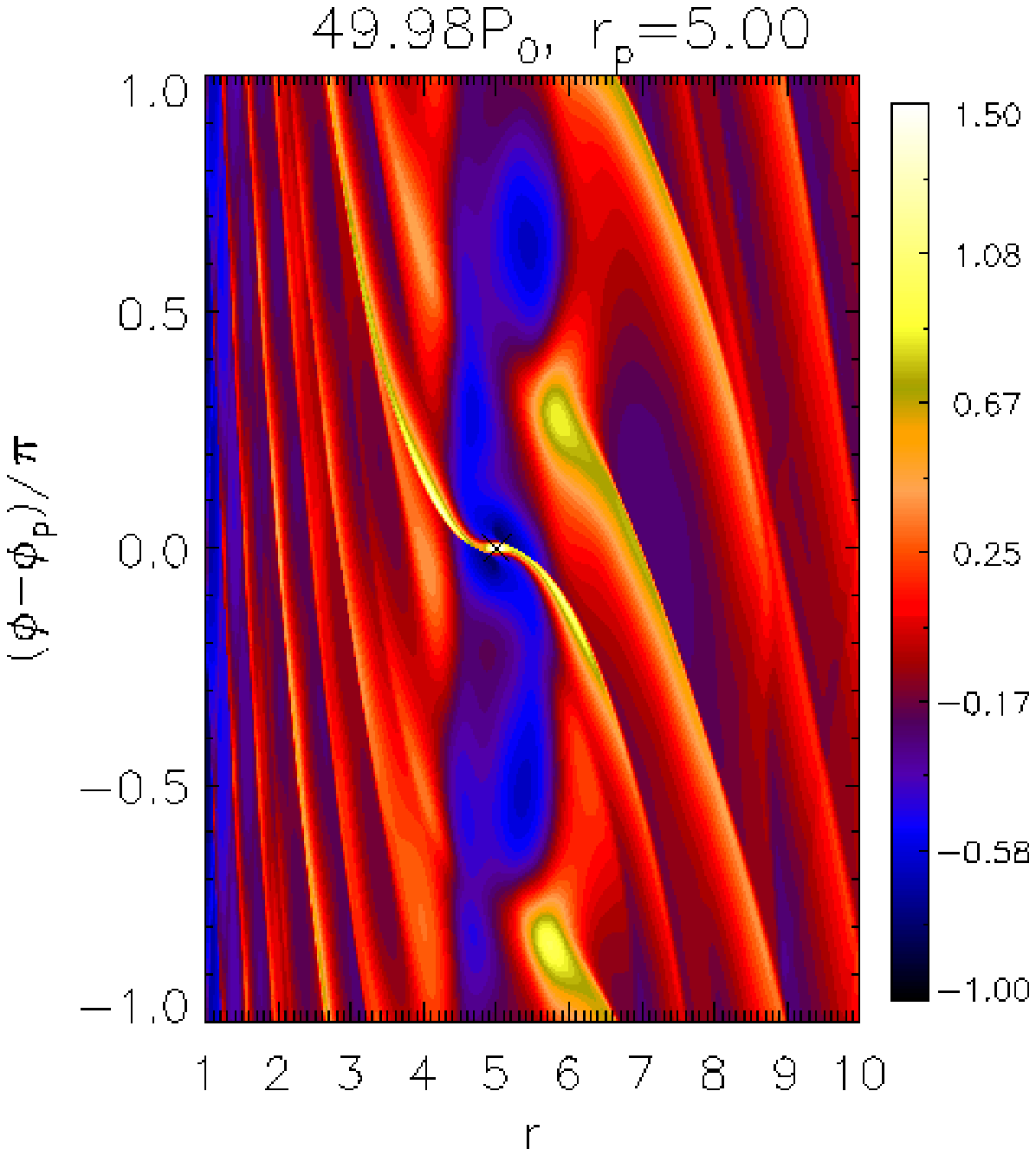}
  \includegraphics[ scale=0.47                  ,clip=true,trim=2.34cm
  1.83cm 0cm 0cm]{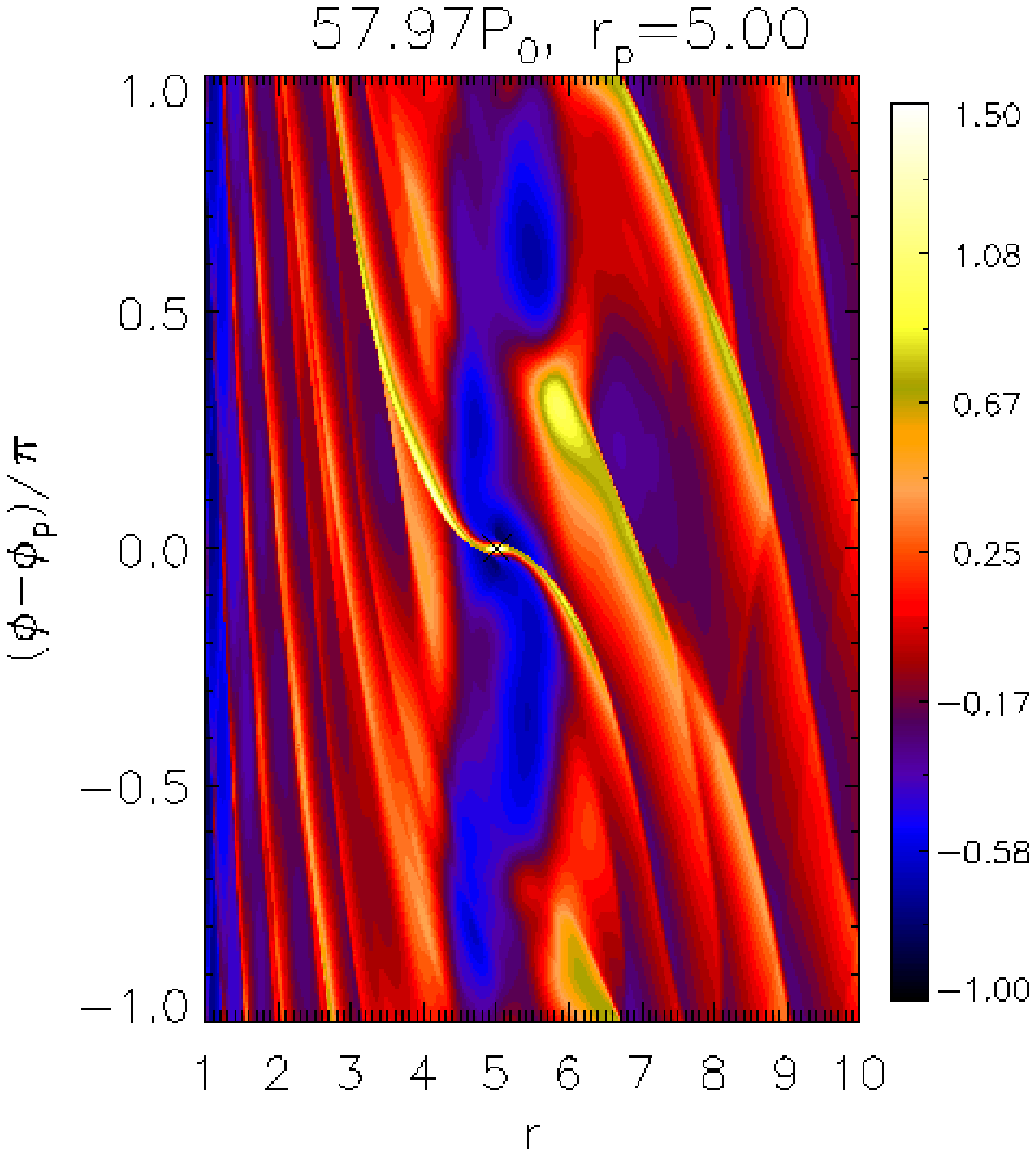}
  \includegraphics[scale=0.47       ,clip=true,trim=0cm
  0cm 1.645cm 0cm]{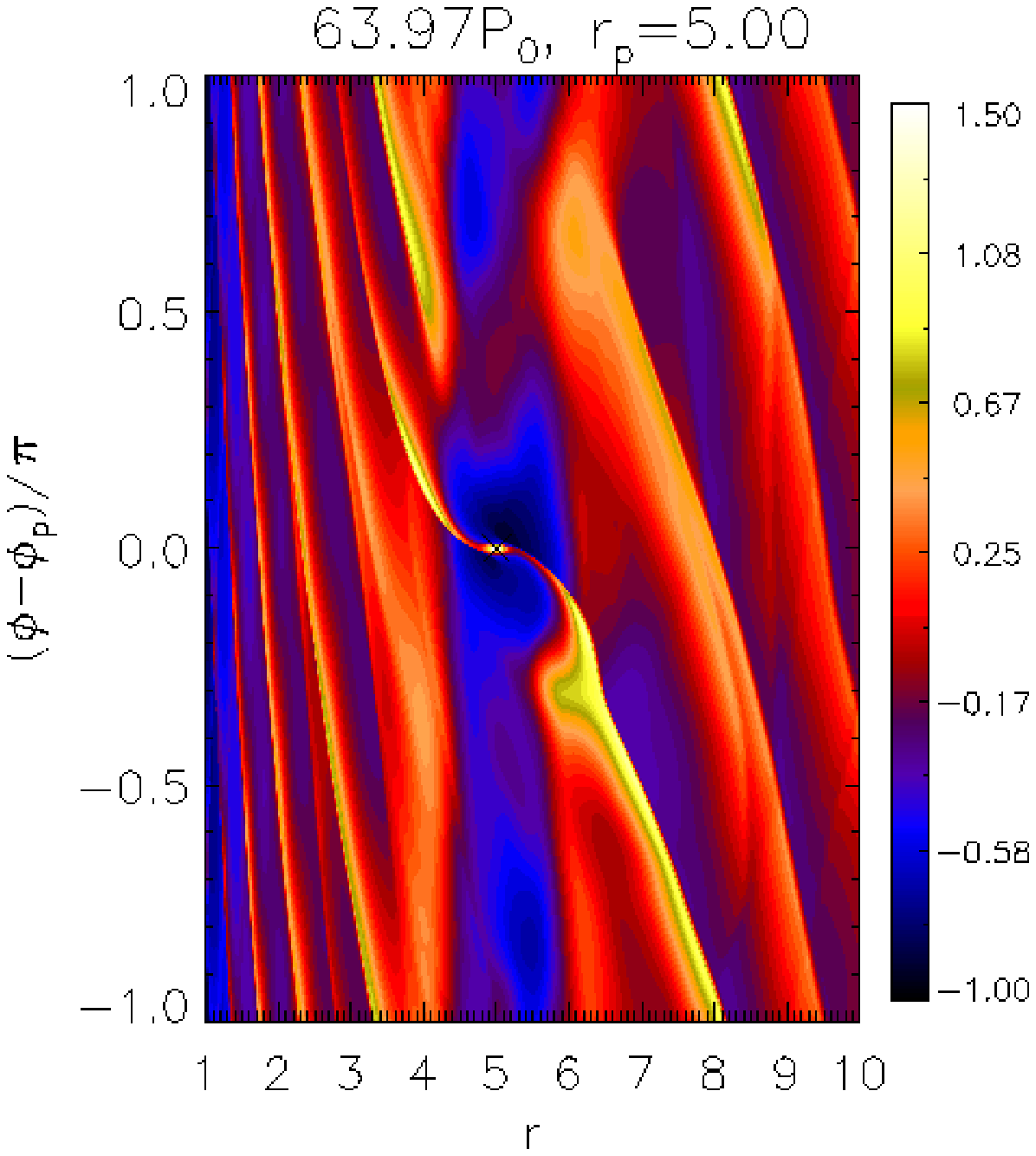}
  \includegraphics[scale=0.47                 ,clip=true,trim=2.34cm
  0cm 1.645cm 0cm]{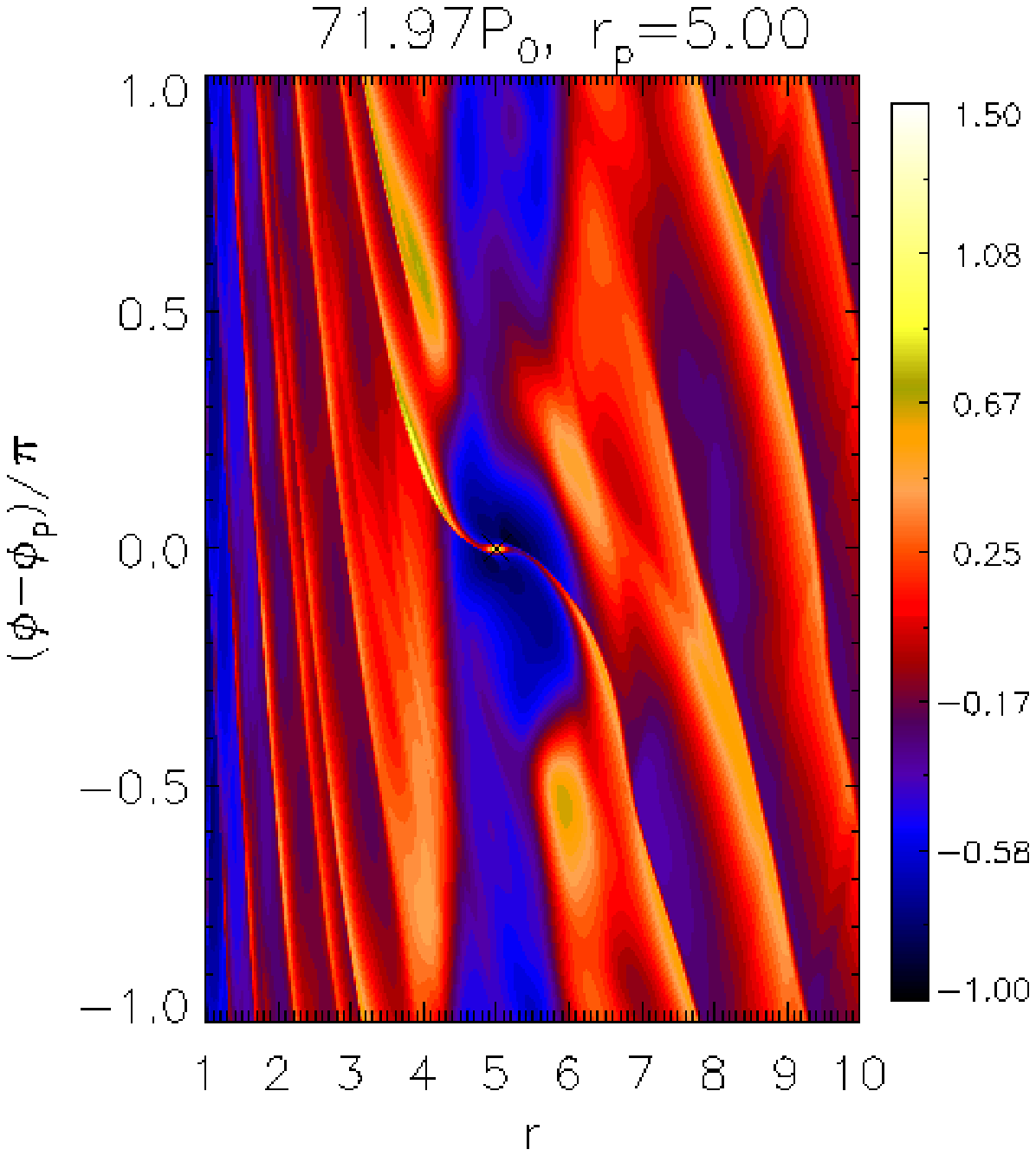}
  \includegraphics[ scale=0.47                  ,clip=true,trim=2.34cm
  0cm 1.645cm 0cm]{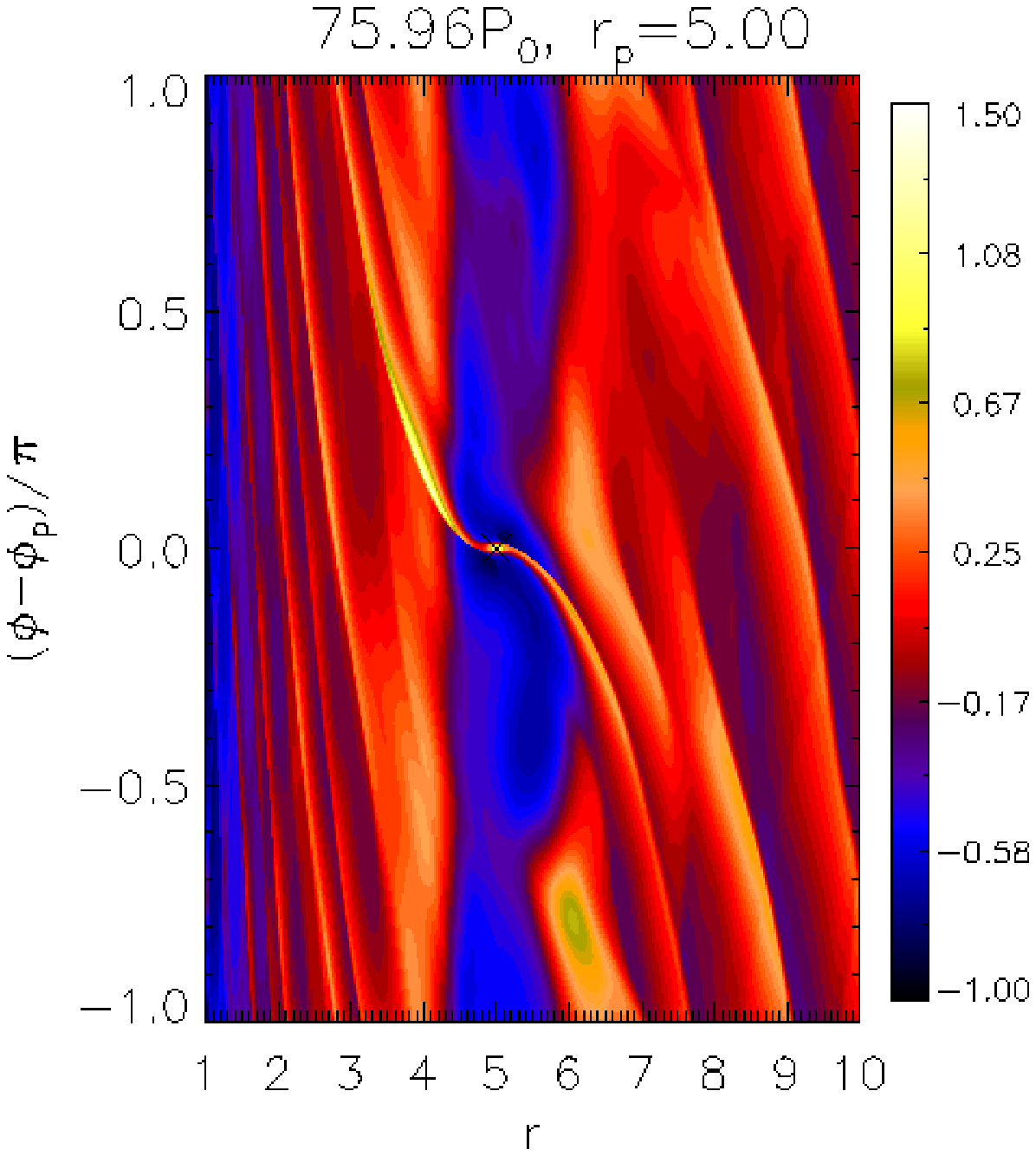}
  \includegraphics[ scale=0.47                  ,clip=true,trim=2.34cm
  0cm 0cm 0cm]{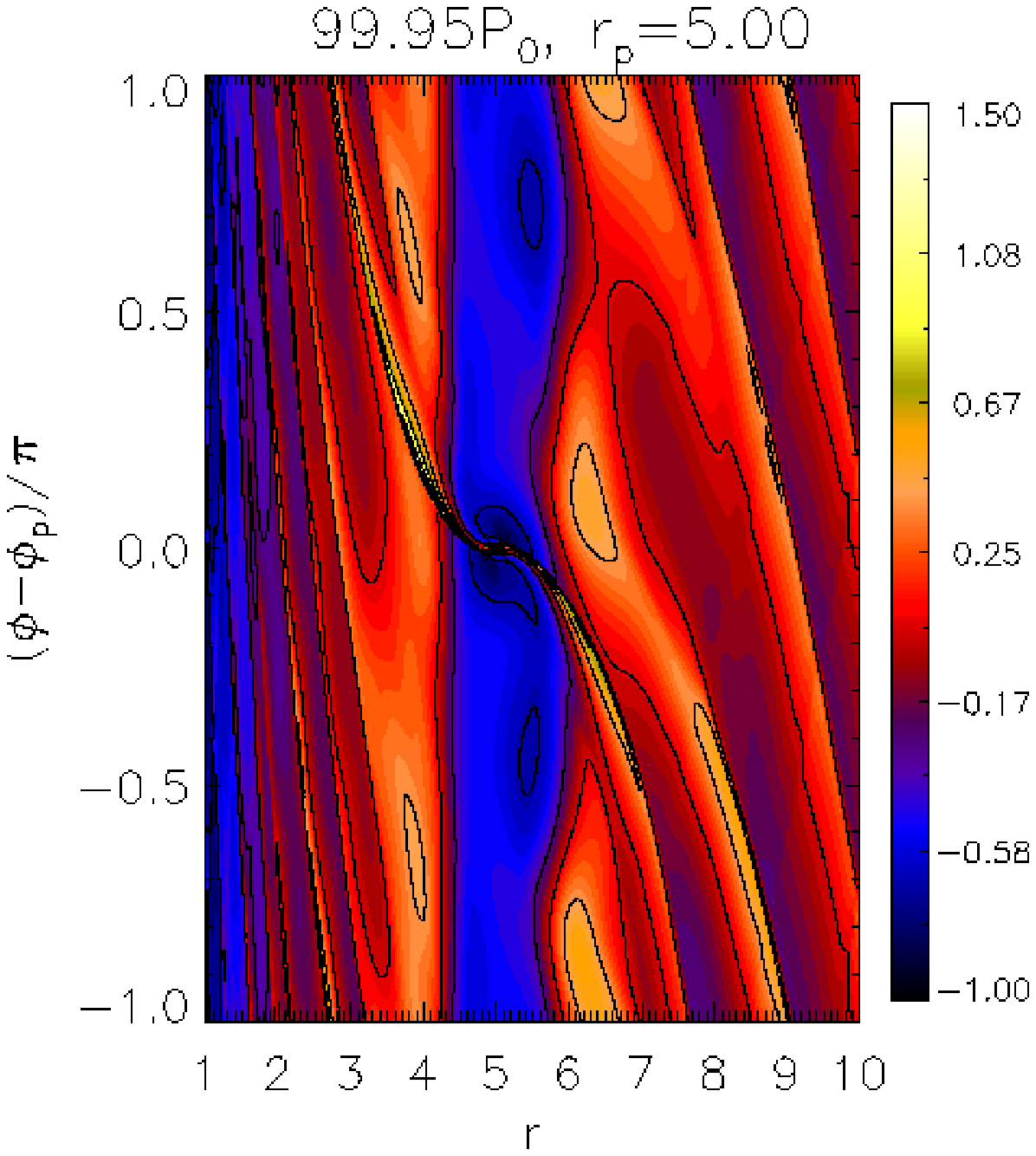}
  \caption{Evolution of  the surface density perturbation 
(relative to $t=0$ ) for the $Q_o=1.5$
    case from $t\sim 40P_0$ to $t\sim 100P_0$.   
    \label{Qm1.5_nonlin1}}
\end{figure*} 




\subsubsection{Eccentric gaps}
Several snapshots of the  $Q_o=1.5$ run  
taken during  the second half of the simulation are
shown in Fig. \ref{Qm1.5_ecc}, which  display  the precession of an
eccentric outer gap edge. Deformation of a  circular
gap into an eccentric one requires an $m=1$  disturbance.
By inspection of the form of the relative surface
density perturbation for  $Q_o=1.5$
  (see Fig. \ref{qm_density}), we see that edge modes are 
 associated with an  eccentric outer  gap edge. However, the
 inner gap edge remains fairly circular. 

In their non-self-gravitating disc-planet simulations, \cite{kley06}
found  eccentric discs can result for planetary masses $M_p > 0.003M_*$
fixed on circular orbit. Our simulations show that edge modes in a
self-gravitating disc can provide an additional perturbation to deform
the gap into an eccentric shape for lower mass planets.

 \begin{figure}
   \centering
   \includegraphics[width=0.25\textwidth,clip=true,trim=1cm 0cm 0cm 0cm]{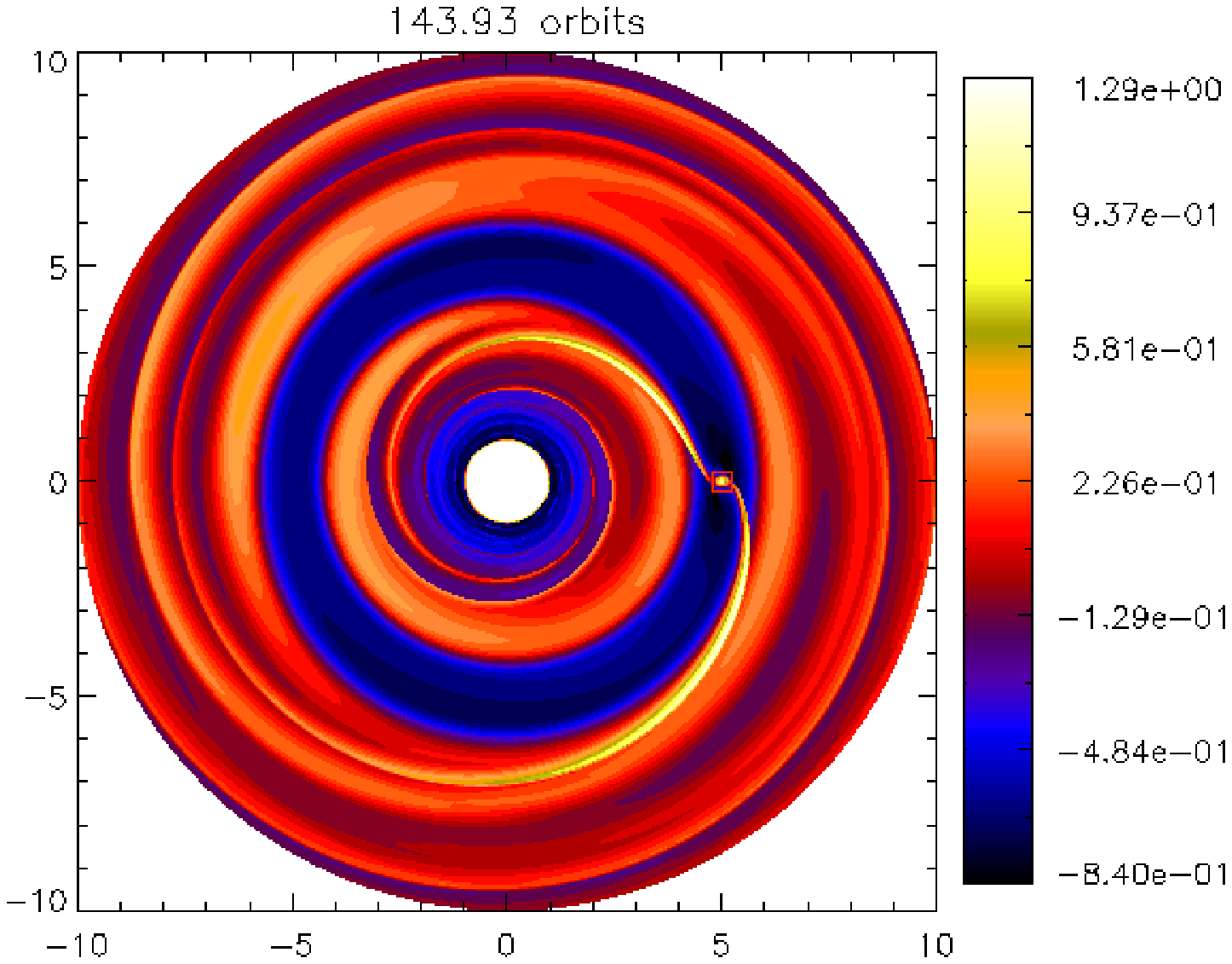}\includegraphics[width=0.25\textwidth,clip=true,trim=1cm 0cm 0cm 0cm]{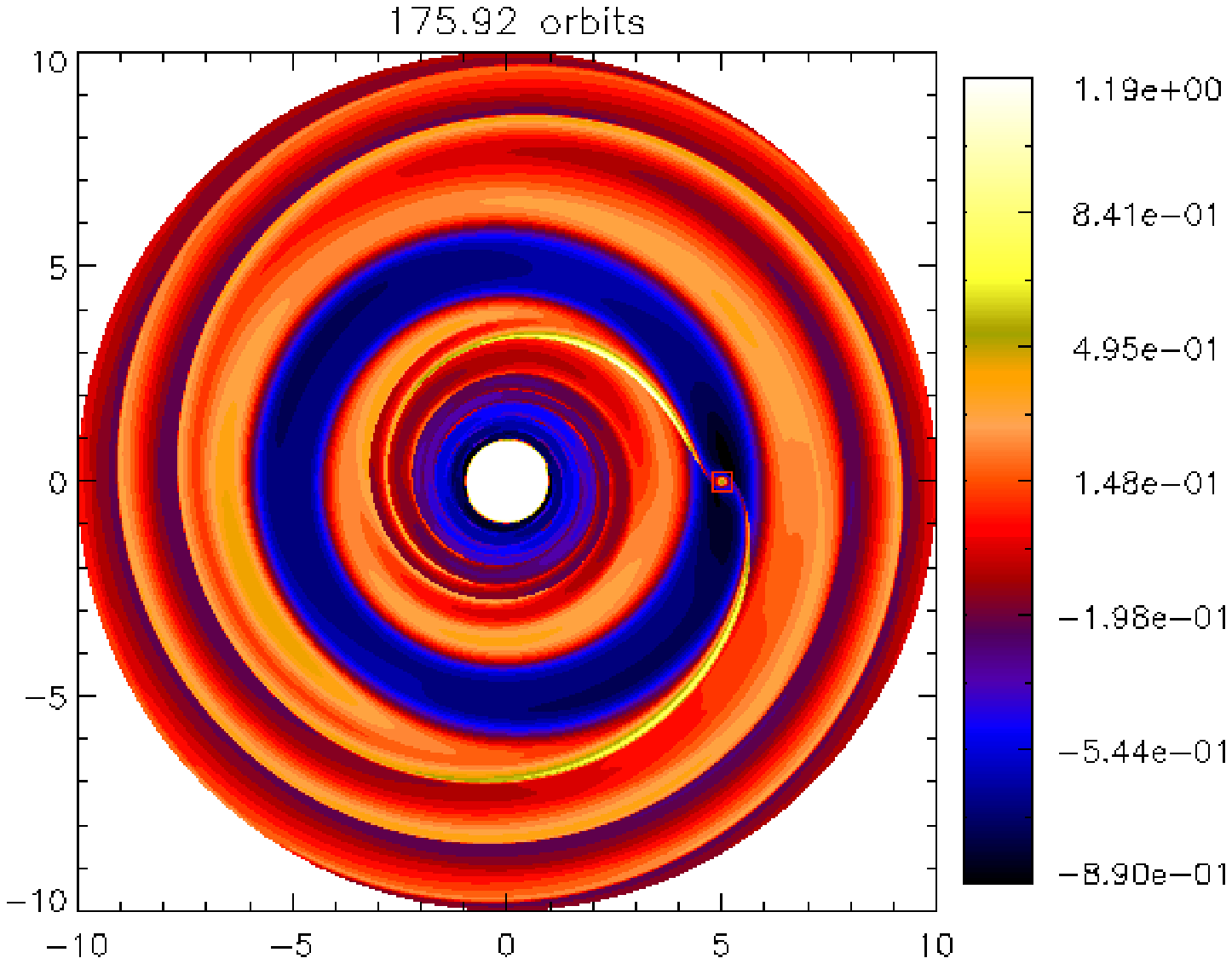} 
   \includegraphics[width=0.25\textwidth,clip=true,trim=1cm 0cm 0cm 0cm]{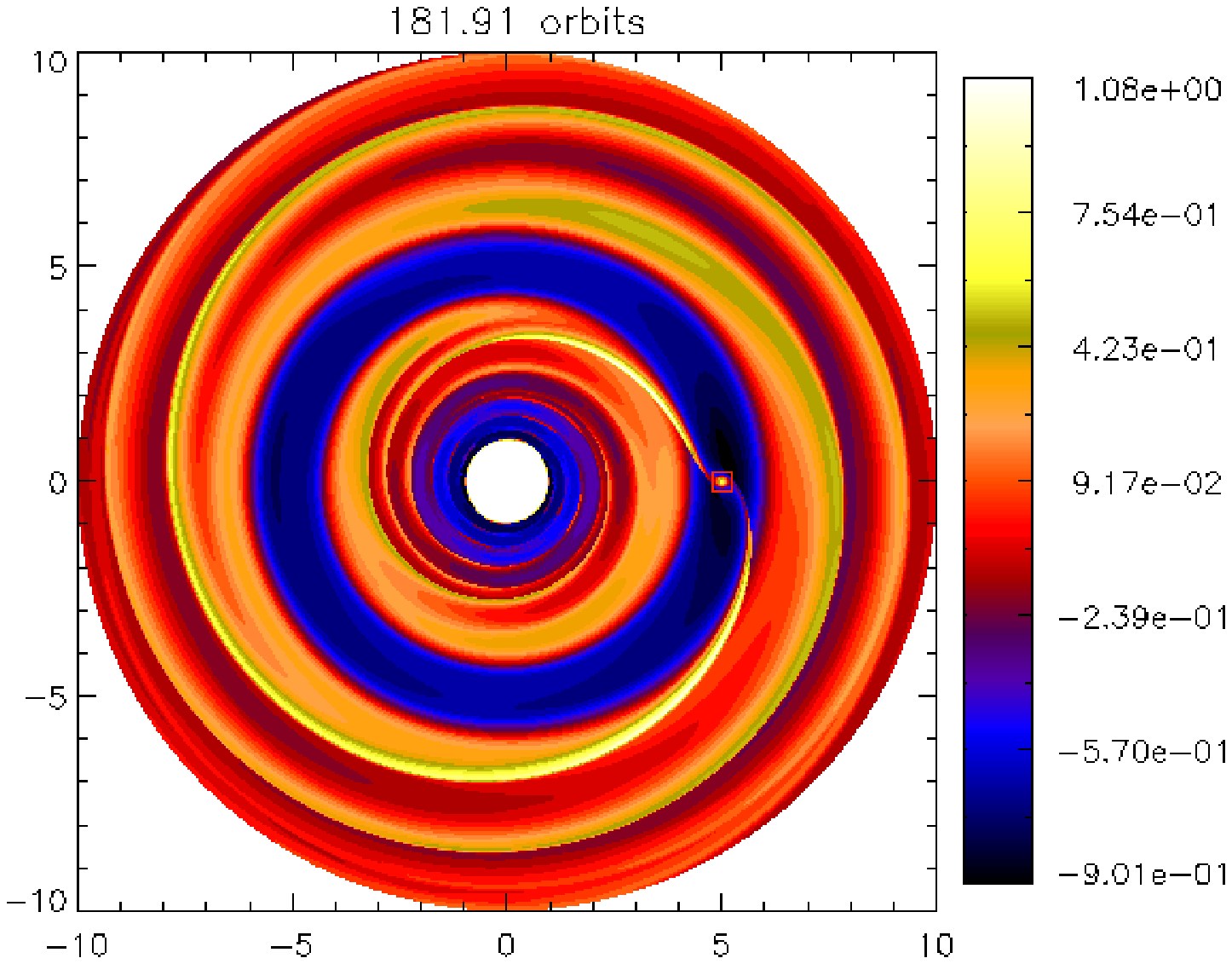} 
   \caption{Precession of an eccentric gap in the
     presence of an $m=2$ edge mode for a disc with  $Q_o=1.5.$ 
     \label{Qm1.5_ecc}}
 \end{figure}

\subsection{  Evolution of the gap structure  for  $Q_o=1.2$}

With increasing disc mass,  evolution of the gap profile away from its original form
takes place. Fig. \ref{Qm1.2} illustrates  the emergence of the $m=3$ edge
mode, expected from linear calculations, and subsequent
evolution of the gap profile. 
By $t\sim 42P_0$ the $m=3$ edge mode has already become non-linear,
whereas for  $Q_o=1.5$ the $m=2$ mode only begins to emerge at
$t\sim 44P_0,$ being consistent with linear calculations that the former has
almost twice the growth rate of the latter. In Fig. \ref{Qm1.2} 
there is a  $m=3$ spiral at $r<r_p$ in phase with the
mode for  $r>r_p.$ The gap narrows where they almost touch. These are
part of a single mode.
By  $t=44P_0,$   shocks  have formed   and  
extend continuously across the gap.  The   self-gravity of the disc
 has overcome the planet's gravity which is responsible for
gap-opening. Edge mode spirals can be more prominent than planetary wakes.

The evolution of the form  of the gap  is shown  in  the one-dimensional  averaged profiles
plotted in Fig. \ref{Qm1.2}. The quasi-steady  profile before the onset 
of the edge mode  instability is 
 manifest  at
$t\sim 40P_0.$ 
By $t=44P_0$, the original bump at $r=6$  has
diminished. This bump originates from the planet expelling
material as it opens a gap but is subsequently `undone' by the edge
mode as it grows and attempts to connect across the gap. This gap
filling effect is  possible as non axisymmetric 
perturbation  of the inner disc  may be induced via the outer  disc self-gravity.
Notice also in Fig. \ref{Qm1.2}
at $t=44P_0$ the increased surface density at $r\simeq7.5$ implying
radial re-distribution of material.
 By $t=120P_0$, the inner
bump ($r=4$) is similar to that produced by a  standard gap-opening planet. By
contrast, the outer bump cannot be maintained. The gap widens and
there is an overall increase in surface density for $r>6.5$. Note that
there is no additional  diminishing  of the outer edge bump from
$t=44P_0\to120P_0$.  
\begin{figure}
   \centering
   \includegraphics[scale=0.3,clip=true, trim=1cm 0cm 3.3cm
   0cm]{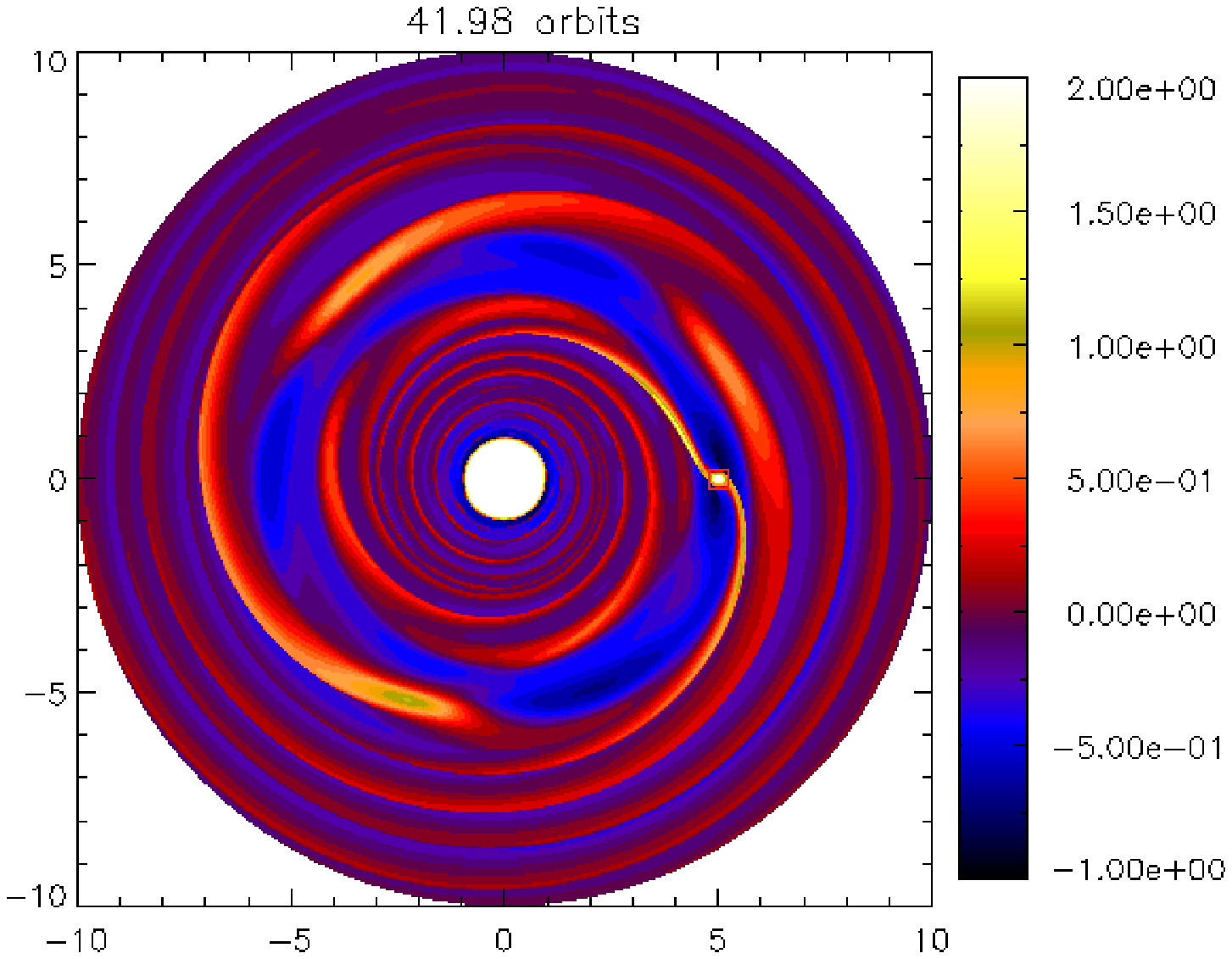}\includegraphics[scale=0.3,clip=true, trim=1cm 0cm 0cm 
   0cm]{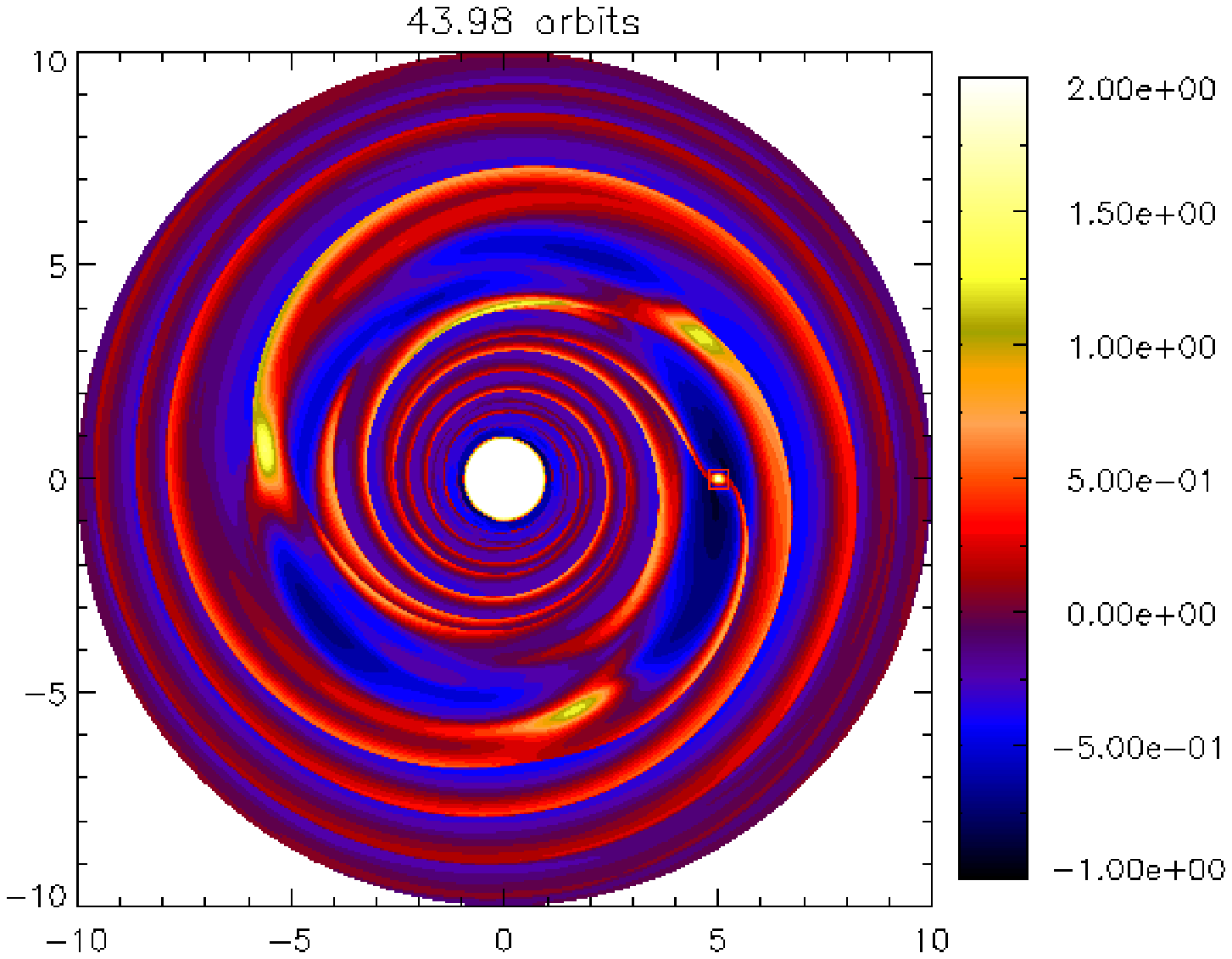} 
   \includegraphics[width=0.45\textwidth,clip=true,trim=0cm 0cm 0cm
   1cm]{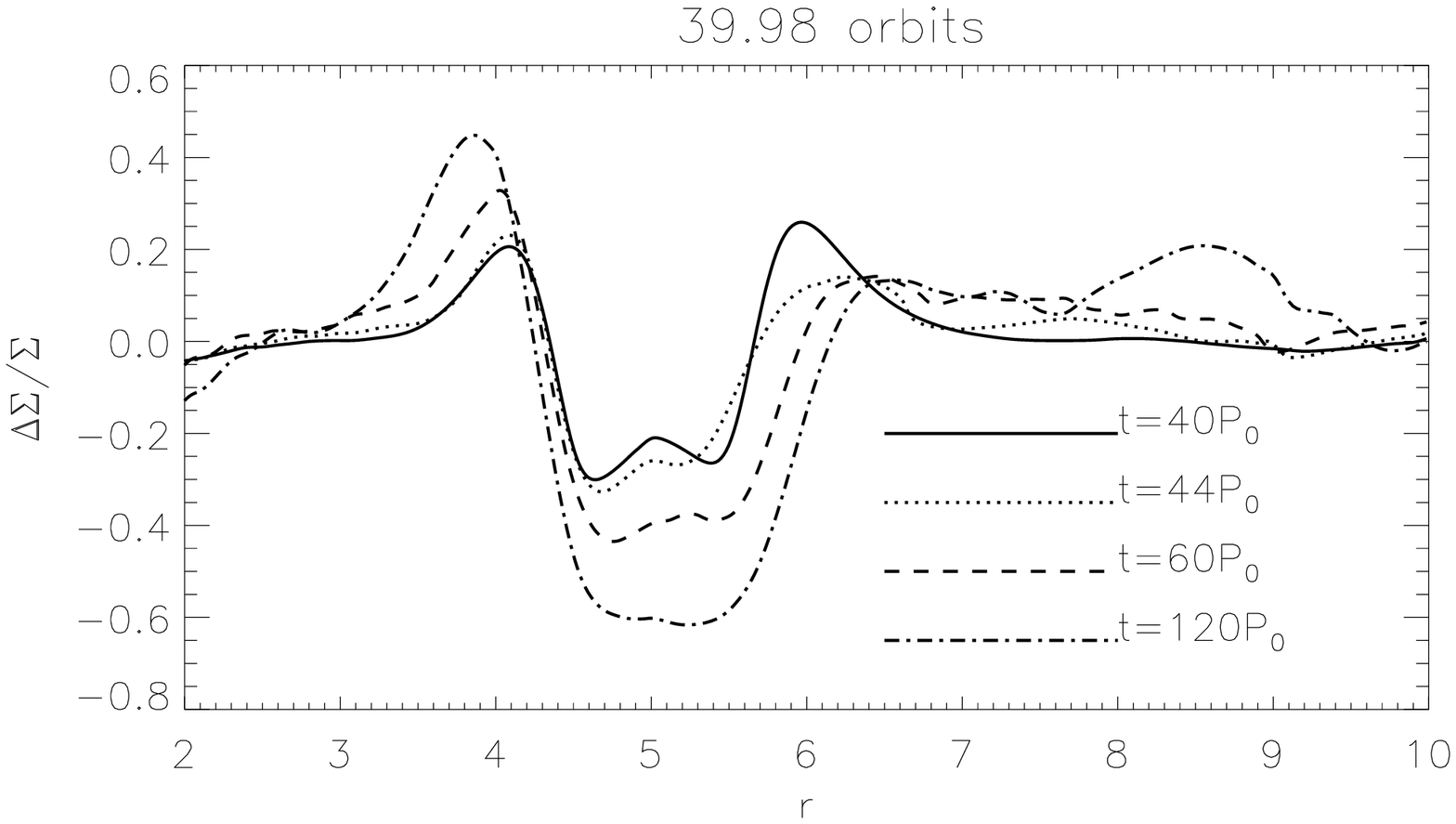}  
   \caption{ The disc model with  $Q_o=1.2$: emergence of the $m=3$ edge mode (top) and
     evolution of the gap profile (bottom). 
     \label{Qm1.2}}
 \end{figure}

Fig. \ref{Qm1.2_vortensity} illustrates
  the evolution of the vortensity profile within the
gap. The  development of edge modes temporarily reduces the
amplitude of the vortensity maxima set up by the perturbation of the planet.
This is particularly noticeable when the profiles at 
 $t=40P_0$ and  $t=44P_0$ are compared.  However, vortensity
is generated through material passing through shocks
 induced by the gravitational perturbation of
 the planet \citep{lin10}.
This provides a vortensity source that enables the amplitudes of
vortensity maxima to increase again, as can be seen from the profile  at $t=60P_0.$ 
The maxima remain roughly symmetric about the planet's location at
$r_p\pm 2r_h$.  Note a general increase in  the co-orbital vortensity level
 caused by  fluid elements repeatedly
passing  through  shocks. 

\begin{figure}
   \centering
   \includegraphics[width=0.45\textwidth,clip=true,trim=0cm 0cm 0cm 1cm]{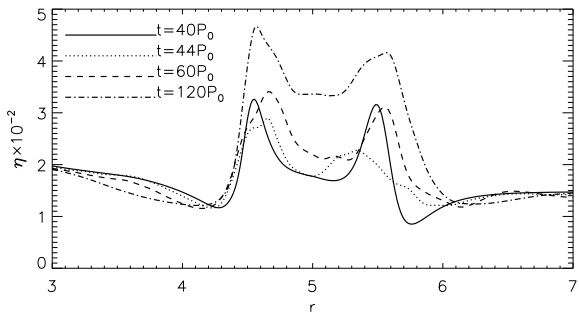}
   \caption{ The disc model with  $Q_o=1.2$: evolution of the
     azimuthally averaged vortensity $\eta$ profile in the region of the gap. 
     \label{Qm1.2_vortensity}}
 \end{figure}

Although  the $m=3$  mode emerges first and is predicted to be most unstable from
linear analysis, it does not persist. Linear theory can only predict the initially 
favoured mode. In the nonlinear regime, Fourier analysis shows that
$m=2$ becomes dominant for  $60P_0 \ga t \ga 145P_0$.  
We suspect this be due to  finite viscosity acting over this time-scale. 
Larger $m$ means  shorter  radial wavelengths, so
over given a time-scale, diffusion and hence stabilisation by viscosity is
more effective than smaller $m$. The evolution is  also 
complicated by the planetary wake which may gravitationally interact with edge modes. 
For $Q_o=1.2$ we found that $m=1$ becomes dominant at $t\ga145P_0$ when
there appears to be a coupling between the outer planetary wake and
the edge mode. 
\subsection{Mass transport}
The similarity between the effects induced by an  edge mode and an external
perturber (planet) is further illustrated   by the induced  mass transport.
Just as when a  planet opens a gap, the development of an  edge disturbance   results
in a  radial re-distribution of  mass.  We 
express the angular momentum transport  in terms of an $\alpha$
viscosity parameter, defined through  $\alpha = \Delta u_r\Delta u_\varphi/c_s^2$, where
$\Delta$ here denotes deviation from the  azimuthal average.

Angular momentum is also transmitted  through  gravitational torques.
However, in practice we found  this to be a  small effect  compared to
transport  due to Reynolds stresses. The discs here are not gravito-turbulent. 
The azimuthally averaged $\alpha$ parameters 
$\langle\alpha\rangle$  are shown in
Fig. \ref{masstrans}. 
The stable disc with $Q_o=2$ has
non-axisymmetric features entirely induced by the planetary
perturbation and $\langle\alpha\rangle$ is localised about the
planet's orbital radius giving a 
two-straw feature about $r_p$. This is typical of disc-planet interactions,
and it provides  a  useful case
  for comparison with  more massive discs to see the effect
of edge modes and their  spatial dependence.  

Increasing self-gravity  by adopting $Q_o=1.5$,
 we see that angular momentum  transport is
enhanced around $r_p$ and for  $r<4$, signifying the global nature of 
edge modes. Towards the inner boundary, $\alpha\simeq 10^{-3}$
being   twice as large as for  the case with  $Q_o=2.$
The   $Q_o=1.5$ case  also has enhanced
wave-like behaviour for  $r>7$ as compared to $Q_o=2.0.$ 

The $Q_o=1.2$ case is more dramatic, with $\langle\alpha\rangle$ reaching
a factor of 2--3 times larger than the imposed physical viscosity
   around the planet's location ($\nu/h^2 = 4\times10^{-3} $). 
This case displays a three-straw feature, with an additional 
trough at  about $r=5.5$, i.e. close to edge mode co-rotation, as if another
planet were placed there.
Although the $Q_0=1.5$ case also develops edge modes, there is no such 
 additional
trough at co-rotation. We suspect this may because the  $Q_o=1.2$ case  is more
unstable,  developing  an  $m=3$ mode ( which produces  3 additional 
effective coorbital  planets
placed at the gap edge), whereas the  $Q_o=1.5$ case  develops a less prominent
$m=2$ mode. 
However, compared to $Q_o=1.5$,  $Q_o=1.2$ has 
no enhanced transport away from the gap region.

Fig. \ref{masstrans} also shows the  evolution of  the disc mass as a function of time, $M_d(t)$. 
Viscous
evolution leads to mass loss ($Q_o=2.0$) and there is clearly enhanced
mass loss if edge modes develop ($Q_o=1.2,\,1.5$). For $t\leq 50P_0$
the curves are identical, this interval includes gap formation and
the emergence of edge modes. When edge modes become non-linear,
mass loss is enhanced ($t\simeq 50P_0$), but there is no significant
difference between $Q_o=1.2,\,1.5$, which is consistent with the
previous $\langle \alpha\rangle $ plots near the inner boundary. The $Q_o=1.2$ curve
is non-monotonic, though still decreasing overall, possibly because
of boundary effects, and mass loss is enhanced once more around
$t=125P_0$. As $Q_o$ is lowered, mass loss becomes less smooth,
showing the dynamical nature of edge modes.

\begin{figure}
   \centering
   \includegraphics[width=0.99\linewidth]{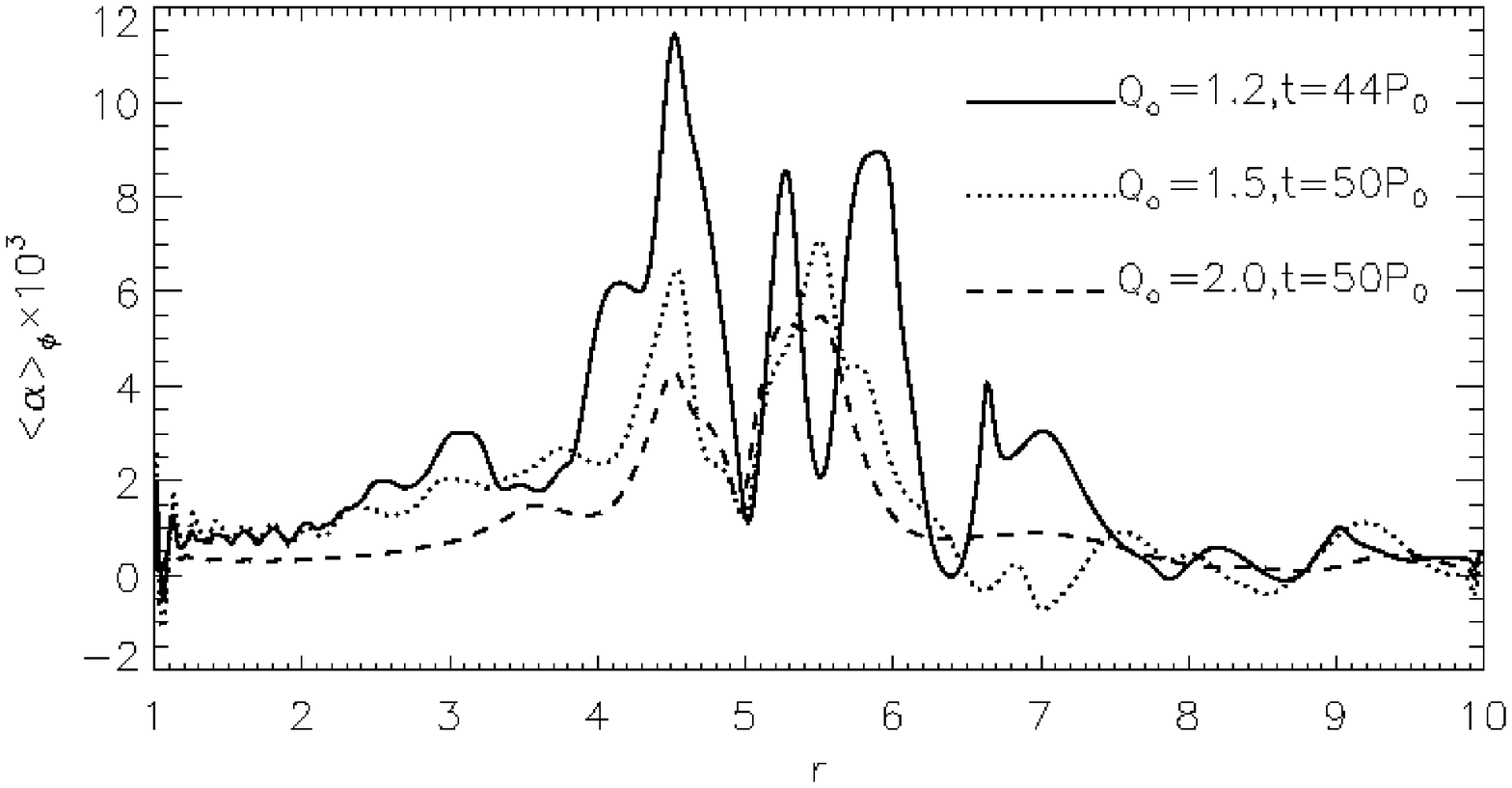}  
   \includegraphics[width=0.99\linewidth]{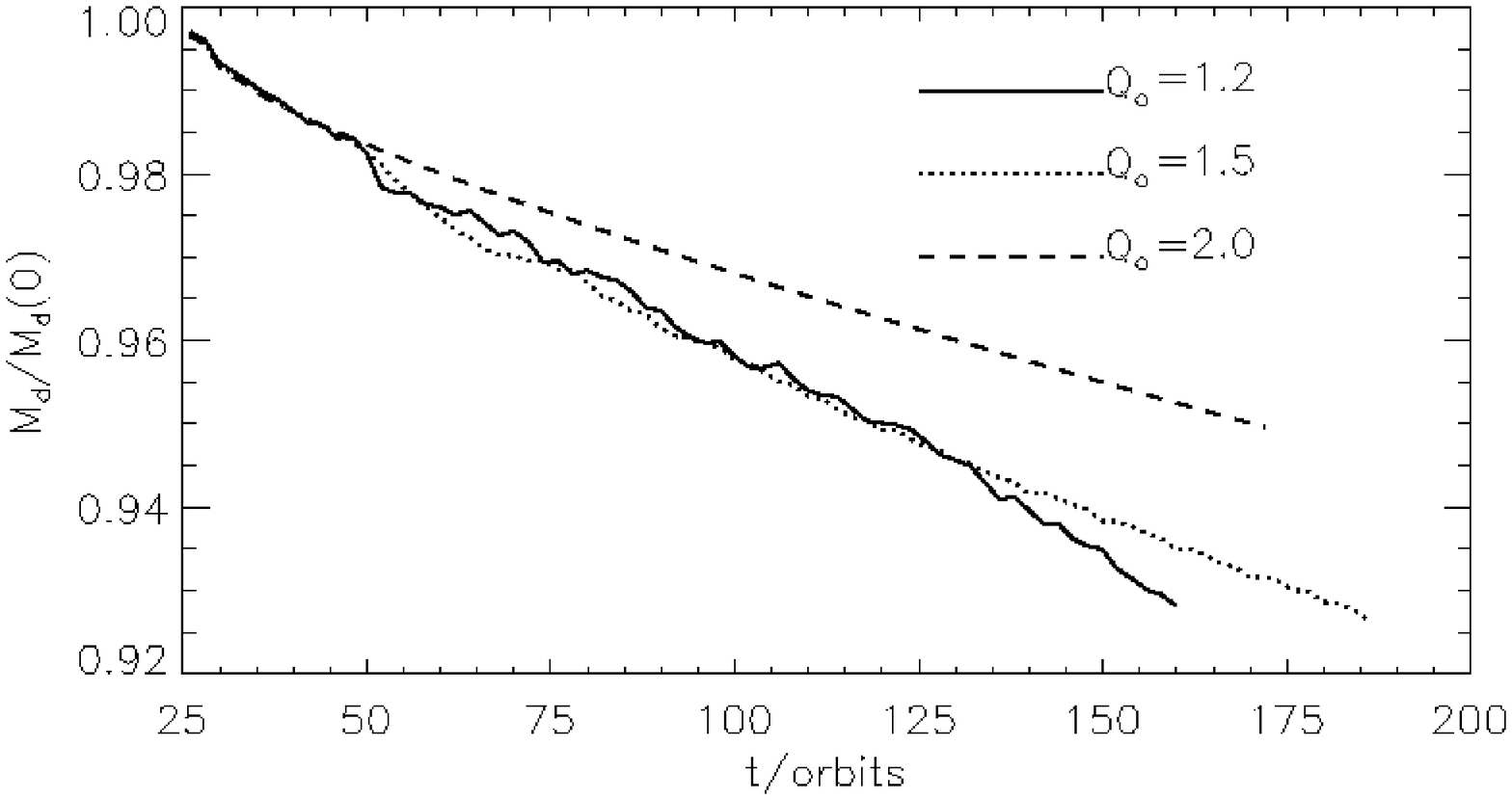} 
   \caption{Top: azimuthally averaged Reynolds stresses associated with edge modes 
     when they become non-linear ($Q_o=1.2,\,1.5$), the $Q_o=2$ case
     is shown as a stable case with no edge modes for
     comparison. Bottom: evolution of the total disc mass for the three cases, scaled by initial mass.  
     \label{masstrans}}
 \end{figure}

\subsection{Additional effects}
We briefly discuss effects of softening and viscosity on edge modes. To analyse mode
amplitudes, the surface density field is Fourier transformed in azimuth giving
$a_m(r)$ as the (complex) amplitude of the $m^{\mathrm{th}}$ mode. 
We then integrate over the outer disc $r>5$ to get $C_m =\int a_m dr$ 
and consider $A = C_m/C_0$.

\subsubsection{Softening}
In linear theory we found that edge modes cannot be constructed for
gravitational softening parameters that are too large. We have
simulated the $Q_o=1.5$ disc with
$\epsilon_{g0}=0.6,0.8,\,1.0$. \footnote{It should be remarked that increasing
$\epsilon_{g0}$ makes the Poisson kernel in simulations increasingly
non-symmetric, so the earlier analytical discussion is less applicable.}  

Fig. \ref{vsoft} shows the evolution of the running-time averaged $m=2$ amplitudes. 
The evolution for all cases is very similar for $t\leq
40P_0$. Unstable modes develop thereafter, with increasing growth rates
and saturation levels as $\epsilon_{g0}$ is lowered.
$\epsilon_{g0}=1$ displays either no growth or  a  much reduced growth rate. 
The gap for $Q_o=1.5,\,\epsilon_{g0}=1.0$ reaches a steady state
similar to the stable case with $Q_o=2.0,\,\epsilon_{g0}=0.3$.  This is
consistent with the fact that as softening weakens, edge disturbances
can no longer perturb the remainder of the disc via self-gravity, 
the residual non axisymmetric structure being due to the planetary perturbation. 

Interestingly, the decrease in
saturation level in going from $\epsilon_{g0}=0.3\to0.6$ is smaller than going from
$\epsilon_{g0}=0.6\to0.8$, which is in turn smaller than $\epsilon_{g0}=0.8\to1.0$,
despite a larger relative increase in softening in one pair than the next.  
This is suggestive of convergence for the nonlinear saturation amplitude  
as $\epsilon_g\to0$. This can be expected 
from convergence in linear theory (\S\ref{linear_convergence}).   
However,
  at $\epsilon_{g0}=0.3$ there are only 5 cells per softening
  length. To probe even smaller $\epsilon_{g0}$ requires prohibitively high
  resolution simulations.  
  Furthermore, since $\epsilon_g$ approximately accounts for 
  the disc's non-zero vertical extent, very small softening lengths 
  are not physically relevant to explore.  


\begin{figure}
   \centering
   \includegraphics[width=0.45\textwidth]{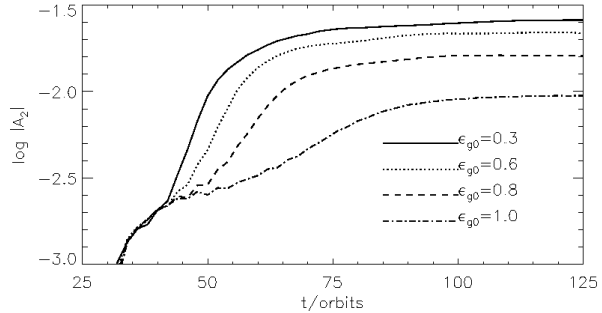} 
   \caption{Running-time average of the surface density $m=2$ Fourier amplitude, integrated over
     $r\in[5,10]$ and scaled by the axisymmetric component, 
     for the $Q_o=1.5$ disc, as a function of the  gravitational softening
     length. 
     \label{vsoft}}
 \end{figure}

\subsubsection{Viscosity}
An important difference between edge dominated  modes and the already
well-studied vortex modes in planetary gaps is the effect of
viscosity on them. The standard viscosity $\nu=10^{-5}$ prevents vortex mode
development, but we have seen in \S\ref{motivation} that such
viscosity values still allow the $m=2$ edge modes to develop. Vortices are
localised disturbances and more easily smeared out than global spirals
in a given time interval. Hence, vortex growth is inhibited more
easily by viscosity than spirals. 

We repeated the $Q_o=1.5$ runs  with a range of viscosities. Results are
shown in Fig. \ref{vvisc}. Generally, lowering viscosity 
increases the amplitudes of the non-axisymmetric modes.
For $\nu\leq 10^{-6}$, the $m=3$  mode emerges first 
(note that this is not in conflict with our linear calculations 
which used $\nu=10^{-5}$ in its basic state), rather than $m=2$.  
This is because lowering viscosity allows a sharper vortensity peak to develop in the background model, 
and thus has the same effect as increasing the disc mass. 
The latter can enable higher $m$ edge modes.

Unlike vortex modes, even with twice the standard viscosity
($\nu=2\times10^{-5}$) the edge mode develops and grows. It is only
is suppressed when $\nu\geq5\times
10^{-5}$. However, this is because no vortensity maxima could be setup
at the gap edges due to vortensity diffusion in the co-orbital region
 making an almost uniform vortensity distribution in the gap. Since
the necessary condition for edge modes is not achieved, no linear
modes exist. This differs  from the nature of the suppression of vortex modes, because in
that case vortensity extrema that are stable can still be setup. 
\begin{figure}
   \centering
   \includegraphics[width=0.45\textwidth]{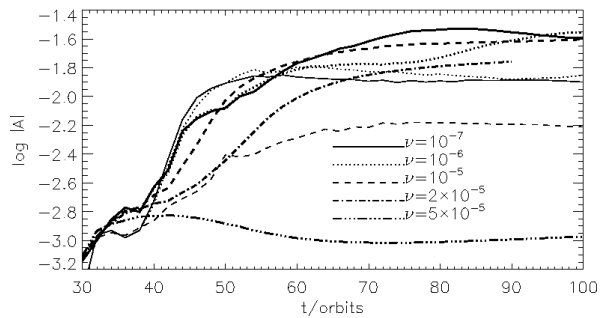} 
   \caption{Non-axisymmetric disc modes 
     for the $Q_o=1.5$ disc as a function of applied
     viscosity $\nu$. The Fourier amplitude has been integrated over $r\in[5,10]$
	and scaled by the axisymmetric amplitude and its running-time average plotted. 
     Thick lines indicate the $m=2$ mode and is
     plotted for all $\nu$. Thin lines indicate the $m=3$ mode and is
     plotted for $\nu=10^{-7}$---$10^{-5}$ only. 
     \label{vvisc}}
 \end{figure}

\section{Applications to disc-planet systems}\label{disc-planet}
In order to focus on the issue of the stability of the dip/gap
in the surface density profile induced by a giant planet,
the planet was held on a fixed orbit.
However, torques exerted by the non axisymmetric disc
on the planet will induce migration that could occur
on a short time scale \citep[eg.][]{peplinski08b}.
Such torques  will be significantly affected by the presence
of edge modes.
Accordingly, we now discuss  the torques
exerted by a  disc, in which edge modes are excited,  on the planet.

\subsection{Disc-planet torques}
The presence of large-scale edge mode spirals of comparable
amplitude as the planetary wake will significantly modify the torque 
on the planet, which in a stable disc is the origin of disc-planet 
torques. 
Here, we measure the disc-planet
torque but still keep the planet on fixed orbit.

Fig. \ref{Qm1.5_torques}(a) shows
the evolution of the 
torque per unit length  for the $Q_o=1.5$ fiducial case. 
The
torque profile at $t=40P_0,$ before the edge modes
have developed significantly,  shows the outer torque is larger in
magnitude than inner torque, implying inward  migration. This is a
typical result for disc-planet interactions and serves as a  reference. 
At $t=60P_0$ and $t=100P_0$, edge modes mostly modify the  torque contributions
exterior to the planet, though some disturbances are seen in the
interior disc ($r-r_p<-5r_h$, $r_h$ being the Hill radius). 
The original outer torque at $+r_h$, is reduced in magnitude as material
re-distributed to concentrate around the  co-rotation radius $r_c$ of the edge mode.
($r_c$ is $\sim2r_h$ away from the planet). 

Edge modes can both enhance or reduce disc-planet
torques associated with planetary wakes. Consider 
$r-r_p\in [3,5]r_h$ in 
Fig. \ref{Qm1.5_torques}(a).
 Comparing the situation at $t=60P_0$ to that at 
$t=40P_0$, the torque contribution from this region is seen to be  more
negative. This is because  an edge mode spiral overlaps the
planetary wake and therefore contributes an  additional negative torque on
the planet.
 However, at $t=100P_0$, an edge mode spiral is just upstream of the
planet,  exerting  a positive torque on the planet, hence the torque
contribution from  $r-r_p\in [3,5]r_h$ becomes positive. 

Differential rotation between edge modes and the planet produces
oscillatory torques, shown in Fig. \ref{Qm1.5_torques}(b). Unlike typical
disc-planet interactions, which produce inward  migration, the total
instantaneous torques in the presence of edge modes can be of either
sign. For a disturbance with $m$ fold symmetry the time interval between
encounters with the planet is $\Delta T = (2\pi/m)/\dd\Omega$, where
$\dd\Omega$ is the difference in angular velocity of the planet and
the disturbance pattern.  Approximating $\delta \Omega =
|\Omega_k(r_p) - \Omega_k(r_c)|$ with $r_c=5.46$ obtained from linear theory, we
obtain $\Delta T = 4.04P_0$. Indeed, the oscillation period in
Fig. \ref{Qm1.5_torques} is $\simeq 4P_0$. 
Both inner and outer torques oscillate, but since the edge mode is
more prominent in the outer disc, oscillations in the outer torque are
larger, particularly after mode saturation ($t>70P_0$).

\begin{figure*}
   \centering
\includegraphics[width=0.33\textwidth]{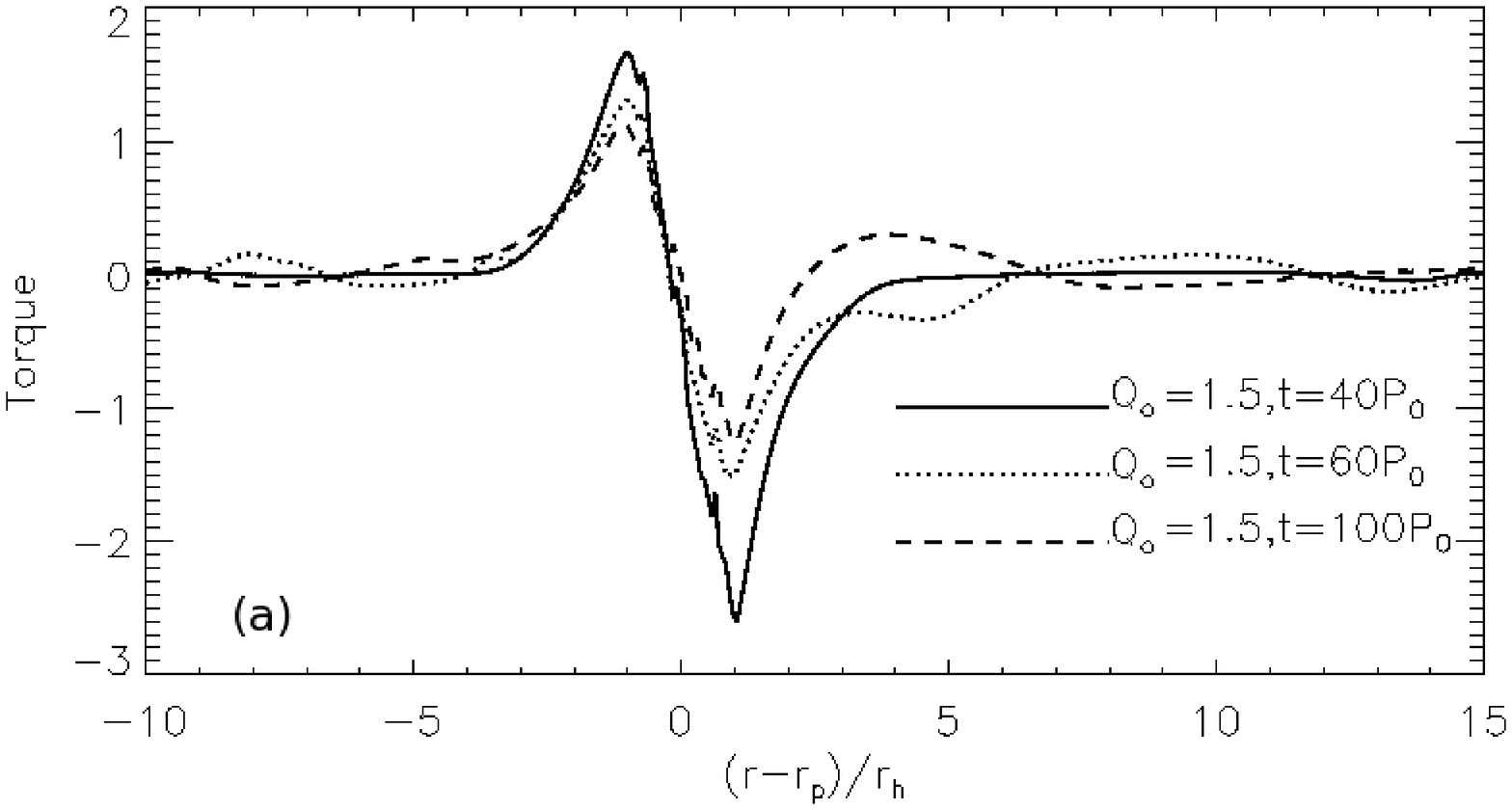}
\includegraphics[width=0.33\textwidth]{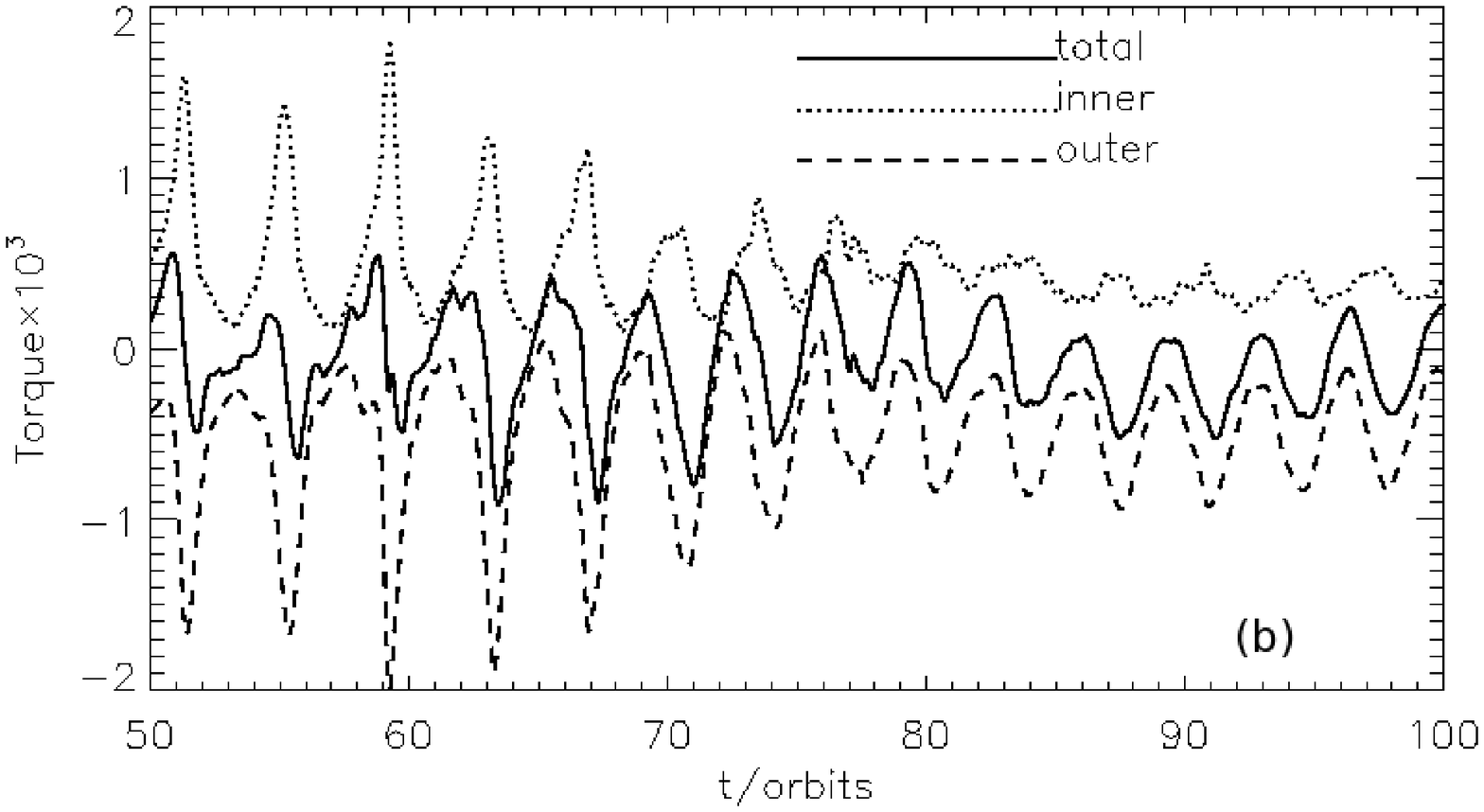}
\includegraphics[width=0.33\textwidth]{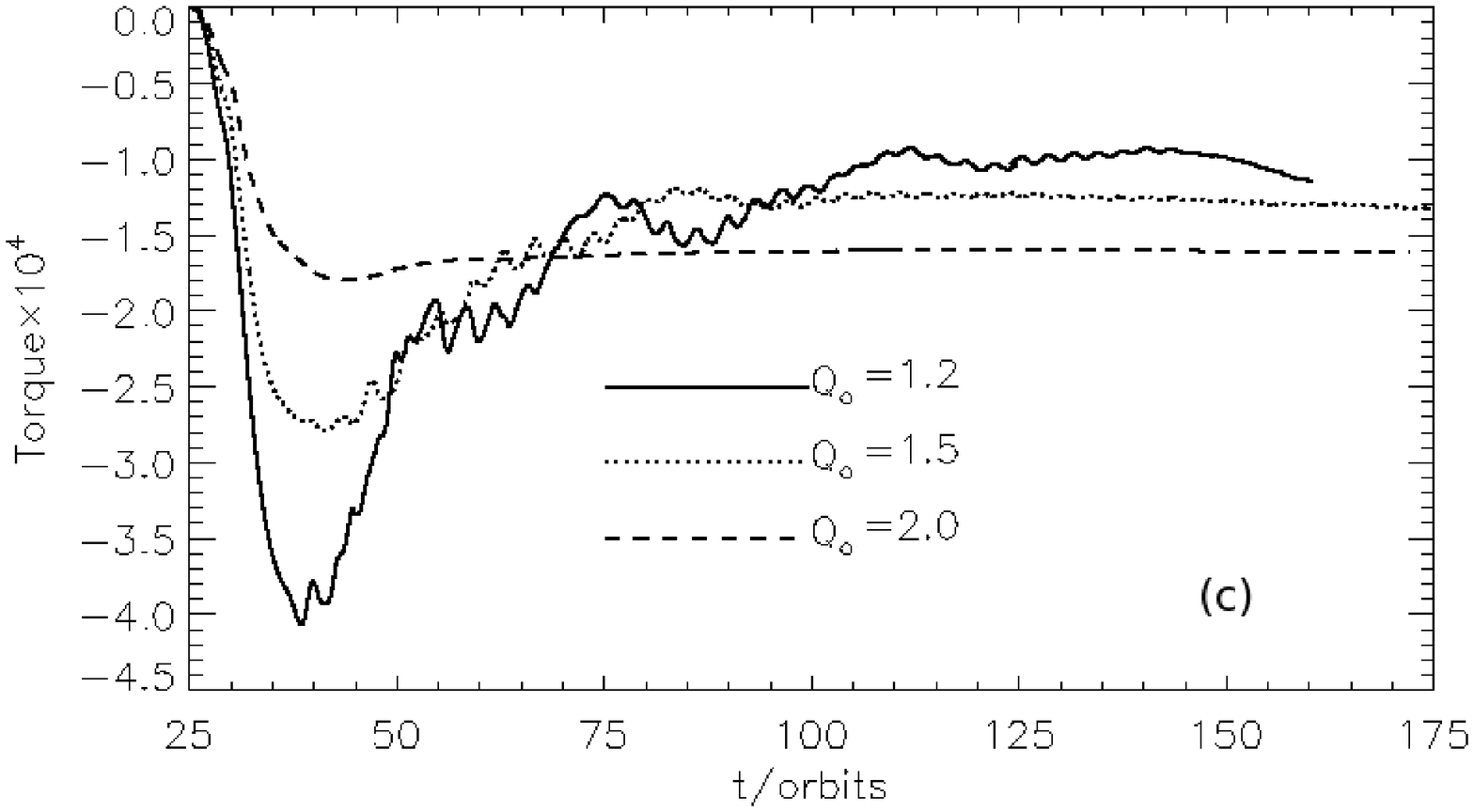} 
   \caption{(a) Disc  torque per unit radius acting on the planet.
      The length scale
     $r_h=(M_p/3)^{1/3}r_p$ is the Hill radius.
     (b) Time dependent evolution of the disc planet torques, the  contribution
        from the inner disc (dotted curve), the
     outer disc (dashed curve) and the  total torque (solid curve).
     (c) The running  time average of the total
      torque acting on the planet  as a function of $Q_o.$  
     \label{Qm1.5_torques}}
 \end{figure*}

We compare time-averaged torques in Fig. \ref{Qm1.5_torques}(c). In
all three cases,  on average a negative torque acts on the planet and its magnitude
largest at $t=40P_0$. 
Up to this point, the  torque becomes more negative as the disc mass increases, as expected.
However, development of edge modes makes the averaged torques more
positive, and eventually reverses the trend with $Q_o$.  While the disc
with $Q_o=1.5$ also attains steady torque values as in $Q_o=2$, the disc with
$Q_o=1.2$ has significant  oscillations
even after time-averaging and is remains  non-steady at the end of the run. 

Before the
instability sets in, torques arise from wakes induced by the planet,
and the outer (negative) torque is dominant. Edge modes concentrate
material into spiral arms, leaving voids in between. Therefore, except
when spiral arms  cross the planet's azimuth, the surface density in 
the planetary wake is reduced because it resides in the void in between edge mode
spirals. Hence the torque magnitude is
reduced. If the reduction in surface density due to edge mode voids is
greater than the increase in surface density scale produced 
by decreasing $Q_o,$ the  presence of edge modes will, 
on average make the torque acting on the planet more positive.



\subsection{Outward scattering by spiral arms}
If edge modes develop, planetary migration may be affected by them.
 Their association with gap edges inevitably affects co-orbital
flows. While a detailed  numerical study of migration is
deferred to future work, we highlight an interesting effect found when
edge modes are present. This is \emph{outward} migration induced through scattering
by spiral arms. 

We restarted some of the simulations described  above
 at $t=50P_0$  allowing the planet
to move in response to the gravitational forces due to the disc.
 The  equation of motion of the planet  is
integrated using a fifth order Runge-Kutta
integrator\footnote{\cite{lin10} found the same integrator to be 
 adequate for studying migration induced through
 scattering by vortices}. Fig. \ref{orbit3_rp} shows the orbital radius of the
planet in discs with edge modes present. 
 No obvious trend with
$Q_o$ is shown. This is because of oscillatory torques due to edge modes and
the partial gap associated with a Saturn mass planet, which is associated
with  type III migration \citep{masset03}. The initial direction of migration can
be inwards or outwards depending on the relative positioning  of the  edge mode
spirals with respect to the planet at the time
the planet was first allowed to move. 

Outward  migration is seen for $Q_o=1.3$ and $Q_o=1.5.$  
For $Q_o=1.5$, the planet migrates by $\Delta r = 2$, or 8.6 times its
initial Hill radius, within only $4P_0.$  This is essentially the result of  a
scattering event. 
Repeating this run with  tapering applied to  contributions 
to disc torques from
within $0.6r_h$ of the planet, we found outward  scattering still
occurs, but limited to $\Delta r = 1$ and the planet remains at
$r_p\simeq 6$ until $t=128P_0$. Hence, while the subsequent inward
migration seen for  $Q_o=1.5$ may be associated with
 conditions close to the planet,
the initial outward scattering is due to an exterior edge mode spiral.

For $Q_o=1.3$ the planet is scattered to $r_p\simeq 8$.  
It is interesting to note that the subsequent
inward migration for $Q_o=1.5$ and $Q_o=1.3$ stalls at $r\simeq6$, i.e. the
original outer gap edge. The planet remains there for sufficient time for
both gap and edge mode development and for $Q_o=1.3$ a second episode of
outward scattering occurs 
(it also occurs for $Q_o=1.5$ to a small extent around $t=70P_0$).   
Interaction with edge modes can affect
the disc well beyond the original co-orbital region of the planet. In the
case of outward  migration, it may promote gravitational activity in
the outer disc. However, boundary effects may be important after
significant outward scattering.  

\begin{figure}
   \centering
   \includegraphics[width=0.45\textwidth]{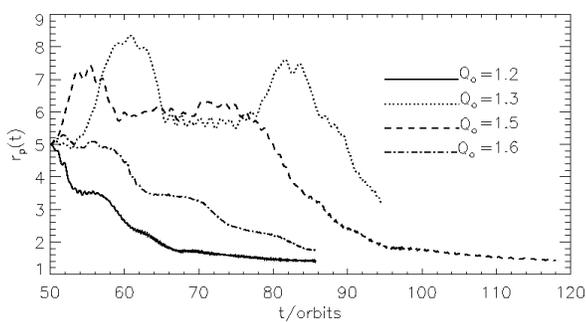} 
   \caption{Orbital radius evolution in discs with edge mode
     disturbances. The instantaneous orbital radius $r_p$ is shown as a
     function of time.
     \label{orbit3_rp}}
 \end{figure}

The spiral arm-planet interaction for $Q_o=1.5$ above is detailed in
Fig. \ref{Qm1.5_scatter}.
A spiral arm approaches the planet from the
upstream direction ($t=50.9P_0$) and exerts  positive torque on the planet,
increasing $r_p$. The gap is asymmetric in the azimuthal direction
with the surface density upstream  of the planet being larger
than that downstream of the planet ($t=51.5P_0$).
 As the spiral arm  passes through the
planetary wake ($t=51.5P_0$---$52.1P_0$) the gap surface density just
ahead of the planet builds up as an edge mode is setup across the
gap. Notice the fluid blob at $r=5,\,\varphi-\varphi_p=0.3\pi$. This
signifies material executing horseshoe turns ahead of the planet. 
At the later time $t=53.1P_0$ the planet has exited the original gap, in
 which the
average surface density is now higher than before the scattering.
 Material that was originally outside
the gap loses angular momentum and moves into the original gap. This
material is not necessarily that composing the spiral.

We remark that migration  through  scattering by spiral arms  differs from
 vortex
scattering \citep{lin10}. Vortices form about vortensity minima, which
lie further from the planet than vortensity maxima. We can assume
vortensity maxima are  stable in the case of vortex formation. This
means the ring of vortensity maxima must be disrupted in order for
vortices to flow across the planet's orbital radius for direct
interaction. Edge modes themselves correspond to a  disruption of
vortensity maxima, hence direct interaction is less hindered than in  the
vortex case. Furthermore, a vortex is a material volume of fluid. Vortex-planet 
scattering results in an  orbital radius change of both objects.
 A spiral pattern is generally not a material volume of fluid.  It can be seen
in Fig. \ref{Qm1.5_scatter} that the local $\mathrm{max}(\Sigma)$ in the spiral remains  at
approximately the same radius before and after the
interaction. The spiral as a whole does not move inwards. In this
case, the spiral first increases  the  planet's orbital radius, thereby
encouraging it to interact with the outer gap edge and scatter that
material inwards. 

\begin{figure}
   \centering
   \includegraphics[scale=0.4, clip=true, trim=0cm 1.85cm 0cm
   0cm]{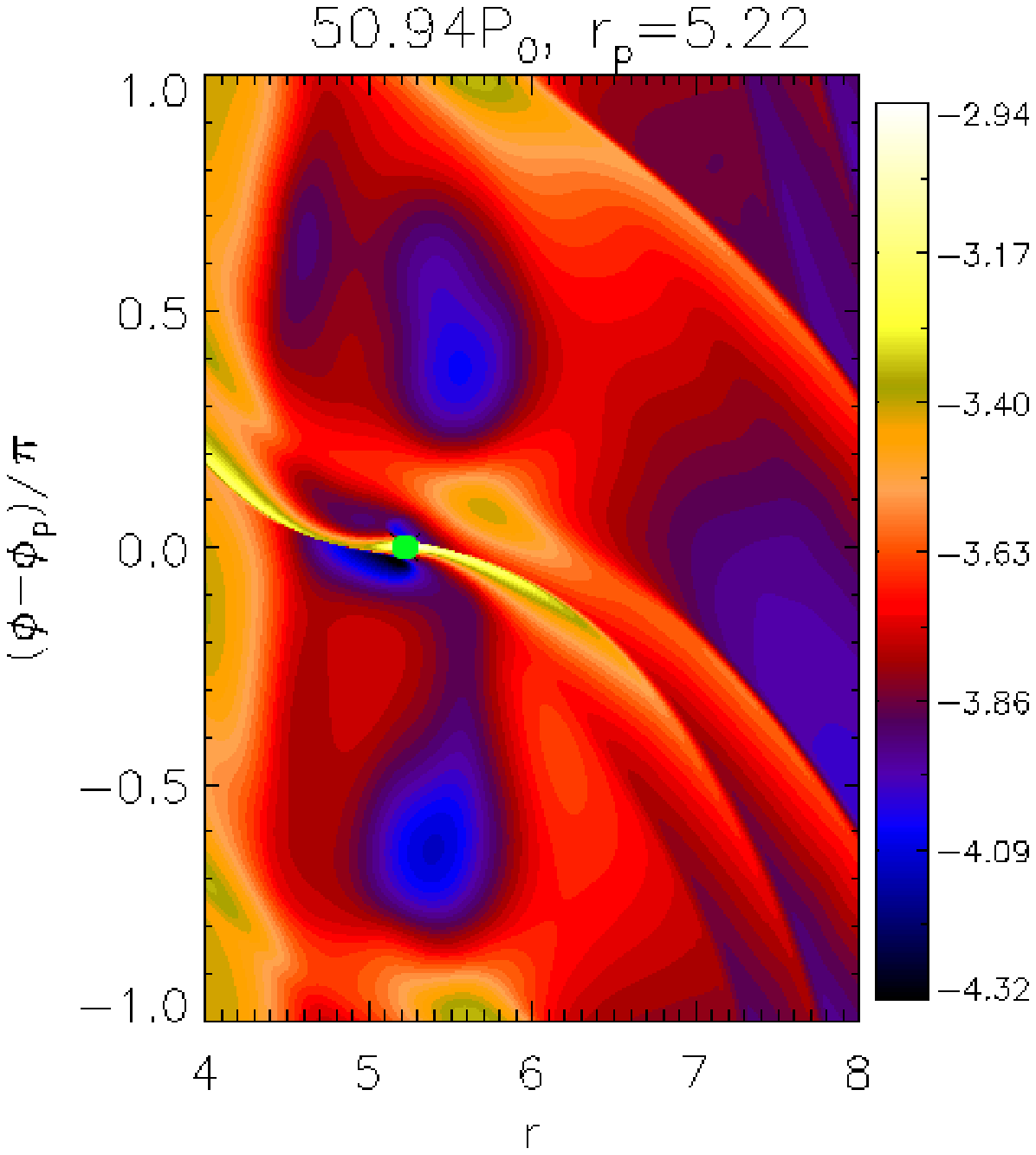}\includegraphics[scale=.4,clip=true,trim=2.2cm 1.85cm 0cm 0cm]{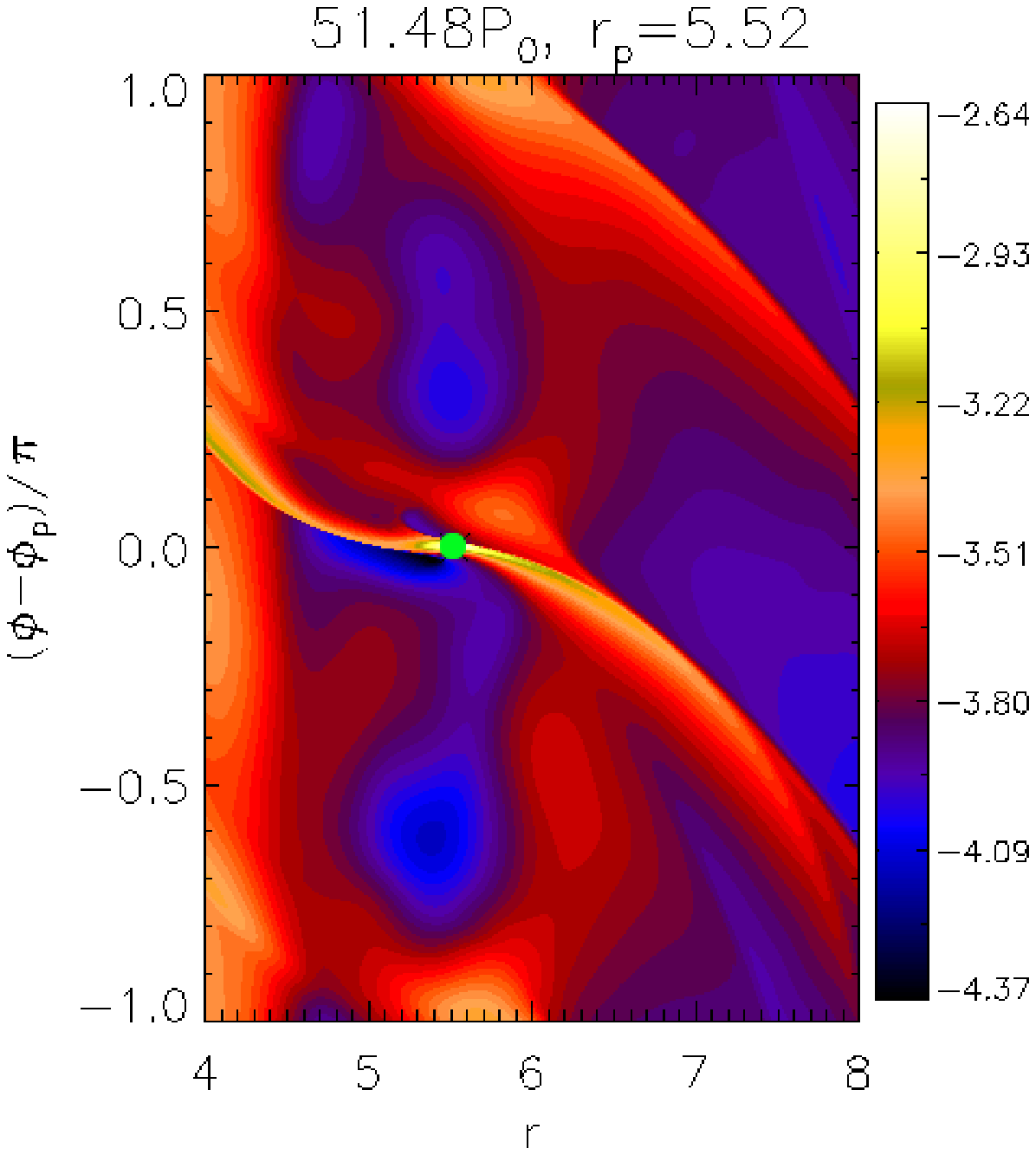}\\
   \includegraphics[scale=.4,clip=true,trim=0cm 0cm 0cm 0cm]{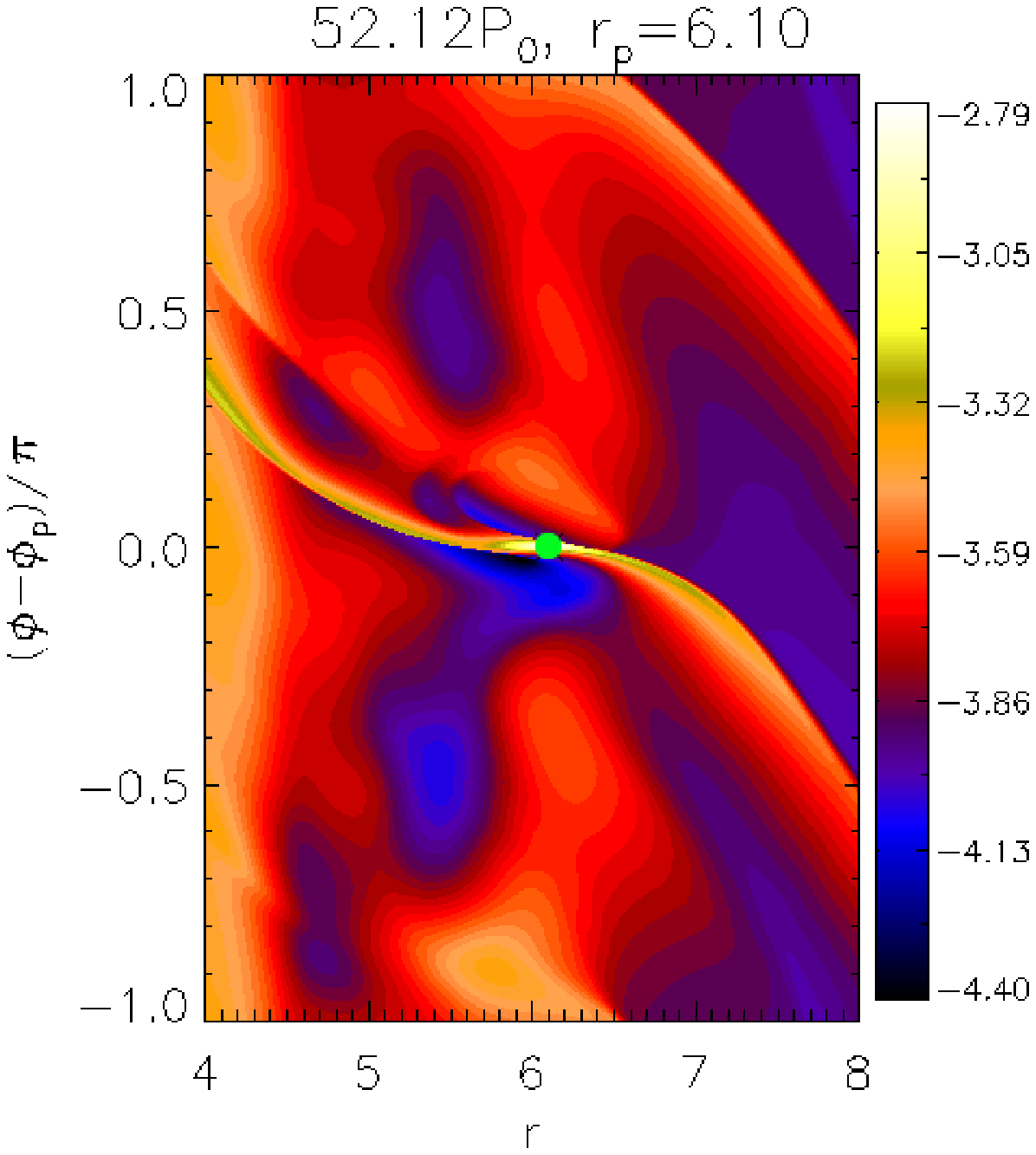}\includegraphics[scale=.4,clip=true,trim=2.2cm 0cm 0cm 0cm]{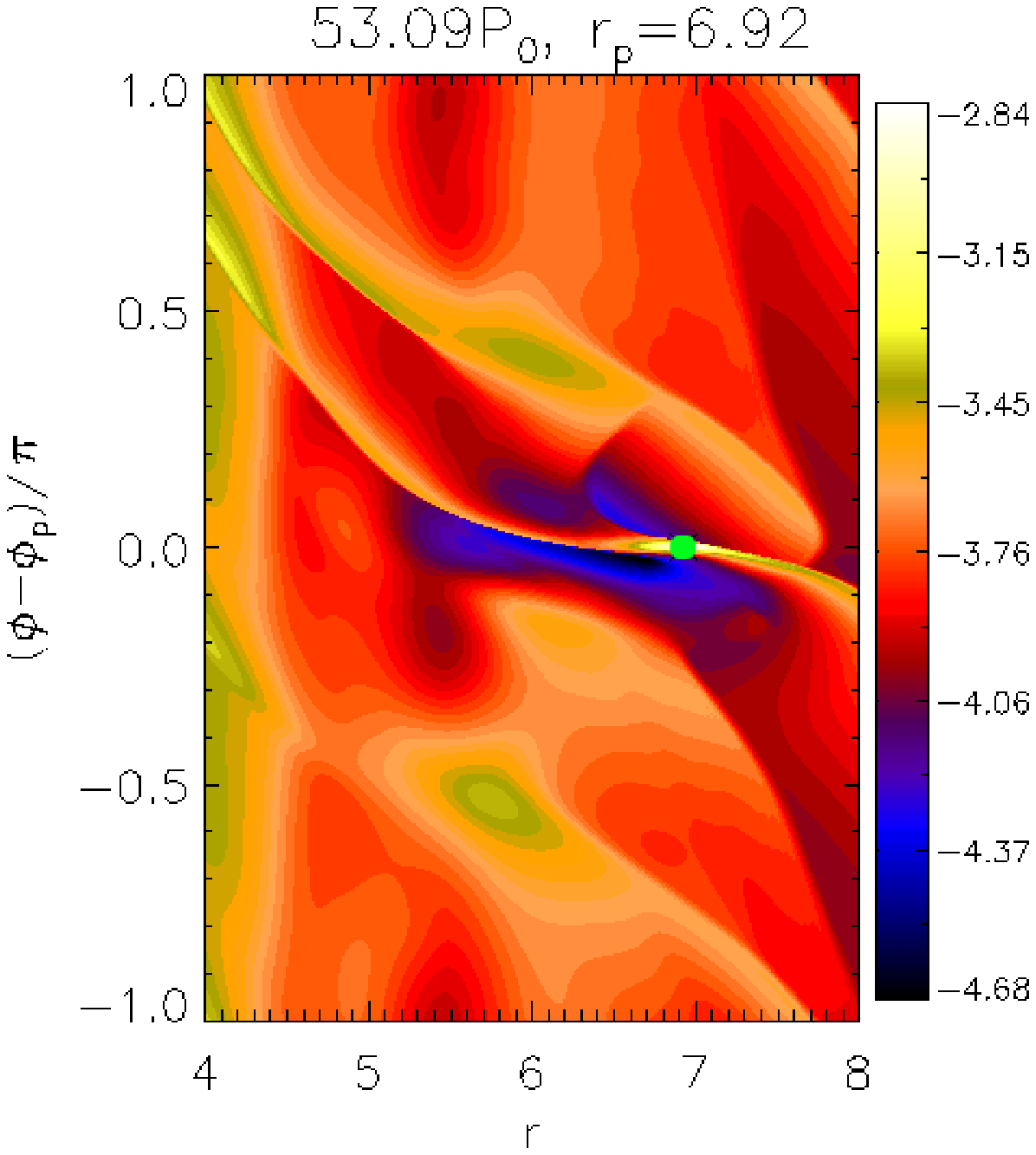} 
   \caption{Interaction between an edge mode spiral and the planet (green dot), 
	causing the latter to be scattered outwards. $\log\Sigma$ is shown. 
     \label{Qm1.5_scatter}}
 \end{figure}

\section{Summary and discussion}\label{conclusions}

We have studied  instabilities 
associated with a surface density gap opened by
 a giant planet embedded  in a massive 
protoplanetary disc. Vortex  producing instabilities,
associated with local vortensity minima, have previously 
been found to occur in weakly or non
self-gravitating discs. However,  edge modes  that are  associated with local
vortensity maxima and require sufficiently strong  self-gravity
were  found in the more massive discs that we have focused on in this paper.
These have very different properties in that they are global
rather than local and are associated with large scale spiral arms
as well as being less affected by viscous diffusion.

For our disc-planet models with fixed planet mass
  $M_p=3\times10^{-4}M_*$ and fixed disc aspect-ratio $h=0.05$ we
  found edge modes
  to develop for gap profiles with an average Toomre $Q\la 2$ exterior
  to the planet's orbital radius. In the unperturbed smooth disc, 
  this corresponds to $Q\la 2.6$ at the planet's location
  and $Q\la 1.5$ at the outer boundary, or equivalently a disc mass
  $M_d\ga 0.06M_*$.  For fixed $M_d$, nonlinear simulations show 
  edge modes develop readily for uniform kinematic
  viscosity $\nu\la 2\times10^{-5}$ and self-gravitational softening
  $\epsilon_g\la 0.8H$. 

A theoretical description  of edge modes was developed. The edge mode
is interpreted as a disturbance associated with an interior vortensity
maximum that requires self-gravity to be sustained, and which
 further perturbs the remainder of the disc through its
self-gravity causing the excitation of density waves.
 Linear calculations confirm this picture and also show
the expected strengthening of the instability
 when self-gravity is increased via the  disc mass
and the expected  weakening of the instability through  increasing
gravitational  softening.

Hydrodynamic simulations of disc-planet interactions were performed. 
We confirm earlier suggestions by \cite{meschiari08}, who found
  gravitational instabilities
associated with their prescribed gap profiles without a planet. Indeed, we found that
 edge modes can develop for
gaps self-consistently opened by introducing a planet
into the disc. However, our models showed their development \emph{during} gap
formation of a Saturn-mass planet, 
when the gap  only consists of a
$20$---$30\%$ deficit in surface density relative to the unperturbed
disc. 
This is much less shallow than the Jovian-mass planetary gaps 
considered in \cite{meschiari08},
 which are typically associated with a $90\%$ deficit. 
We found edge modes to exist for disc models
extending to a distance twice as large as the planets orbit with 
masses $M_d\ga 0.06M_*$, again this is less massive than 
required in \cite{meschiari08}, but  more massive than those typically
used in modelling protoplanetary discs. 

We remark that in
\citeauthor{meschiari08}'s disc model, the local $Q$ maximum 
occurs at the gap centre, presumably where the planet would lie.   
However, gap opening by the planet may disrupt the 
disturbance associated with this extremum, which is required to induce perturbations
in the smooth disc. It is then unclear if the edge mode could be setup. Furthermore,
since corotation lies at their gap centre, there would be no relative motion between spirals and
the planet of the kind seen in our simulations.  
The case of Jovian-mass planets will be explored
in a future work. We note in that case, significant disruption of the disc, leading to fragmentation, 
may occur \citep{armitage99,lufkin04}. 
%
%

Our simulations show edge modes with  $m=2$ and $m=3.$  They have
 surface density maxima localised near the outer gap edge, but the
disturbance they produce 
 extends throughout the entire disc, consistent with analytical 
expectation and linear calculations.
 One important difference between edge modes and
the vortex forming  modes in non self-gravitating discs
 is that the latter require low viscosity. The typical
viscosity values adopted for protoplanetary discs, $\nu=10^{-5}$, will
suppress vortex formation  but not edge modes associated with self-gravity.

We have also considered the effect of edge modes on planetary
migration. They produce large oscillations in the disc
torques acting on the planet,
 which can be positive or negative. The presence of edge modes
reverses the trend of the time 
averaged disc-on-planet torque as a function of
disc mass: the torque being more positive as disc mass increases.   
Direct interaction between the planet and  a
spiral arm associated with an  edge mode is
possible. These  should be more prominent in the disc 
 section with the lower
$Q.$ In practise this corresponds to $r>r_p,$  so the planet tends to
interact with the outer spirals,  which results  in a scattering of the planet 
outwards.

\subsection{Outstanding issues}


%


This work has been motivated by the wish to obtain
a better understanding of   disc-planet interactions
in massive discs. In particular a proper discussion
of phenomena such as type III migration requires
incorporation of self-gravity which we have undertaken here.
We have studied model discs which are unstable through  
the development of edge modes when a dip/gap
is produced by an embedded giant planet but which would be stable 
without the planet.
  
Our first calculations show that the  disc
torques acting on the planet 
in the presence of edge modes change sign and are  highly time-dependent. 
Thus  migration is  unpredictable. A statistical approach may be
ultimately required to assess the likelihood of reduced inward migration 
in practice.
Furthermore, development of edge modes during gap formation indicates 
their importance for planets before they reach a Jovian mass. This leads to
the possibility of type III migration, which is self-sustained and can
be in either direction.
Issues  such as what is the appropriate softening 
length to use in a two dimensional simulation  and
the treatment of material 
 inside the Hill radius, will be important for more thorough
understanding of planetary migration in the presence of large-scale
spiral arms in massive discs. 
Resolving these  requires improved numerical modelling  which  will be
presented in a future work.


\appendix
\section{Energy densities in linear theory}\label{energy}
In \S\ref{linear} we defined and discussed the thermal-gravitational
  energy  (TGE)  and the different
contributions to it. There, a factor $D^2$ was 
applied to the TGE per unit length and to  each contributing term 
in order to overcome numerical difficulties associated with Lindblad
resonances. For completeness, we provide  here a discussion of 
the behaviour of the  TGE and the various contributions to it considered
without the additional factor  $D^2.$

Fig. \ref{energy_den_original} shows very smilier features to the
corresponding  curves
discussed in \S\ref{energy_balance}. $\mathrm{Re}(\varrho)$ is negative around
co-rotation, again signifying a self-gravity driven edge mode. Beyond
$r\simeq 6.4$, $\mathrm{Re}(\rho)$ becomes positive due to pressure. The
  most extreme peak at $r\simeq 5.6$ coincides with the 
background vortensity edge (Fig. \ref{linear_basic}). The negative
contribution from the 
co-rotation term is larger in magnitude than that from the positive wave term,
resulting in $\mathrm{Re}(\rho)<0$. The co-rotation radius $r_c < 5.6$ 
making the shifted frequency $\sbar_R(5.6) = m[\Omega(5.6) -
\Omega(r_c)]<0$. Also at $r=5.6$, $d(\eta^{-1})/dr > 0$, resulting in 
$\mathrm{Re}(\rho_\mathrm{corot}) <0$ at this point.




$\mathrm{Re}(\rho_\mathrm{wave})$ and $\mathrm{Re}(\rho_\mathrm{corot})$ have 
relatively small spurious bumps around $r=7.2$  that are associated with
the outer Lindblad resonance and are numerical.
The eigenfunctions $W,\, \Phi'$ are well-defined
without singularities there, but evaluation of
$\rho_\mathrm{corot}$  and $\rho_\mathrm{wave}$ involves division by
$D$, which can amplify numerical errors at Lindblad
resonances where $D\to 0$. 
Integrating $\mathrm{Re}(\rho)$ over  $[5,10]$, we find
$U<0$, the TGE is negative, which means
gravitational energy dominates over the  pressure contribution  for  $r\geq 5$. 
Integrating the contributions to the  TGE separately, we find
\begin{align*}
  & U_\mathrm{corot}\equiv\mathrm{Re}\int_5^{10} \rho_\mathrm{corot} dr\simeq -0.94|U|,\\
 & U_\mathrm{wave}\equiv\mathrm{Re}\int_5^{10} \rho_\mathrm{wave} dr\simeq -0.077|U|.
\end{align*}
The integration range includes the OLR and thus the
spurious bumps in $\rho_\mathrm{wave}$ and
$\rho_\mathrm{corot}.$ However,  we still find that
$U$ approximately equals $ U_\mathrm{corot} + U_\mathrm{wave}$. The correct
energy
balance is still maintained despite  being subject to
possible numerical error due to the diverging factor $1/D.$  
The above means that, aside from the spurious bumps, 
 the energy  density values for $r\in[5,10]$ may still be used to
interpret energy balance. We find  $|U_\mathrm{corot}/ U|\sim 0.9$, so
as before, the TGE  is predominantly accounted for by  the vortensity term.  


\begin{figure}
  \centering
  \includegraphics[width=0.45\textwidth]{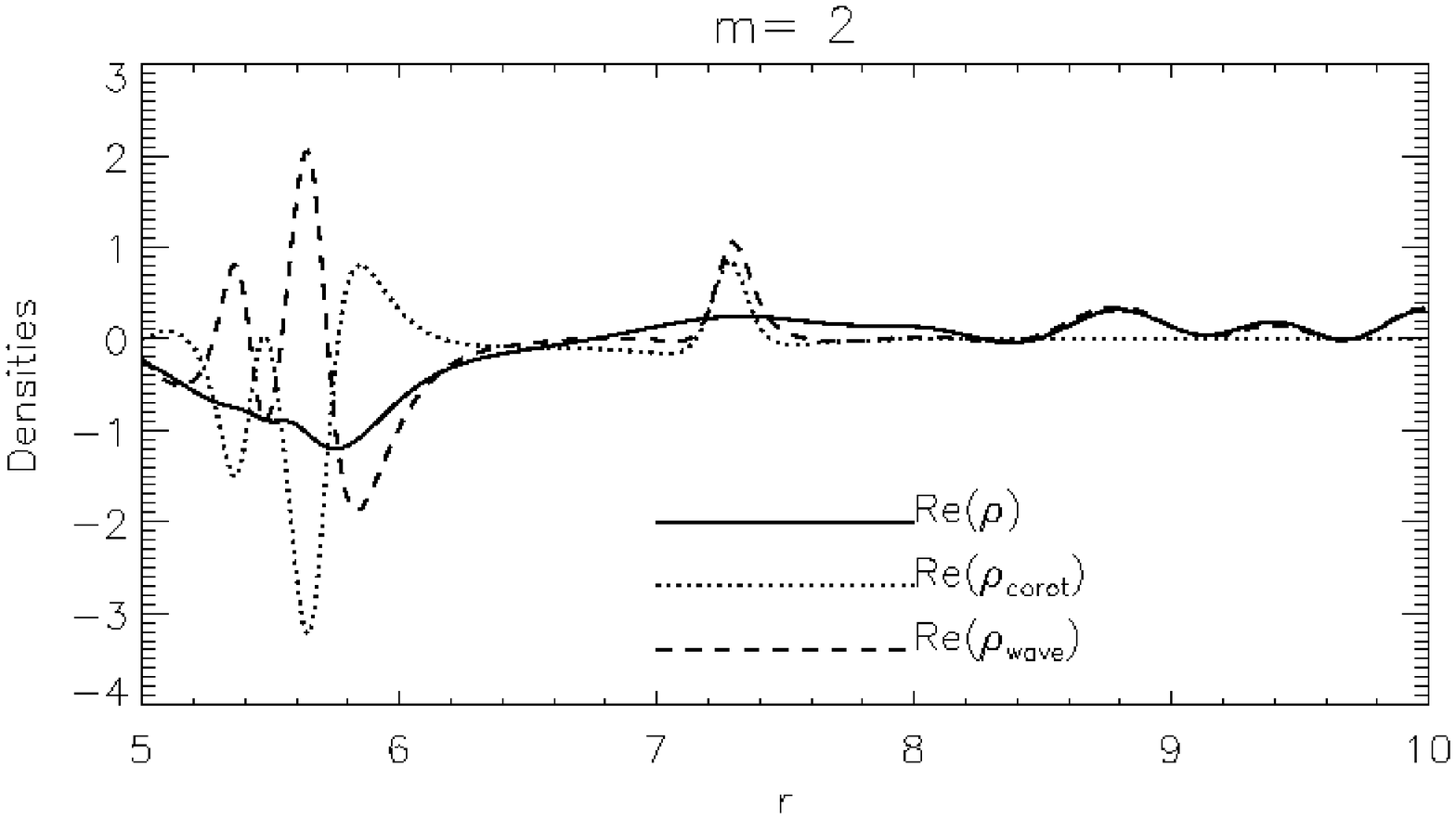}
  \caption{ The thermal and gravitational 
     energy (TGE)  per unit length  (solid) computed from 
    eigensolutions for the fiducial case $Q_o=1.5$ from linear
    theory. The  contributions from the  vortensity term (dotted line) and other
    terms (dashed line), defined by Eq. \ref{totalenerg}, are also shown.  
    The relatively small 
    bumps near $r=7$ are numerical.  However as they are the
    same magnitude for both contributions, incorporating them
      does  not affect conclusions
    concerning the overall energy balance in the system. 
\label{energy_den_original}}
\end{figure}



\end{document}